\documentclass[aps,prd,twocolumn,showpacs,eqsecnum,amsmath,nofootinbib,showkeys]{revtex4}
\usepackage[T1]{fontenc}
\usepackage[utf8]{inputenc}
\usepackage{lmodern}
\usepackage[english]{babel}
\usepackage{xcolor}
\usepackage{graphicx,amssymb,amsthm,amsmath}
\usepackage{hyperref}
\usepackage{physics}
\usepackage{mathrsfs}
\usepackage[scr=rsfs]{mathalfa}

\def \d {{\rm d}}

\newcommand{\energy}{\mathscr{E}}
\newcommand{\angmom}{\mathscr{L}}

\usepackage{comment}

\begin{document}

\title{Orbital Motion in Spacetimes Influenced by the Presence of Scalar and Electromagnetic Fields}


\author{J. Horák}
\email{horak@astro.cas.cz}
\affiliation{Astronomical Institute, Academy of Sciences, Bo\v{c}n\'{\i}~II 141\,31 Prague, Czech Republic}

\author{T. Tahamtan}
\email{tahamtan@utf.mff.cuni.cz}
\affiliation{Institute of Theoretical Physics, Faculty of Mathematics and 
Physics, Charles University, V~Hole\v{s}ovi\v{c}k\'ach 2, 180~00 
Prague 8, Czech Republic}

\author{T. Hale}
\email{tomas.hale@utf.mff.cuni.cz}
\affiliation{Institute of Theoretical Physics, Faculty of Mathematics and 
Physics, Charles University, V~Hole\v{s}ovi\v{c}k\'ach 2, 180~00 
Prague 8, Czech Republic}

\author{G. T\"{o}r\"{o}k}
\email{gabriel.torok@physics.slu.cz}
\affiliation{Institute of Physics, Silesian University in Opava, Bezru\v{c}ovo n\'{a}m. 13, CZ-74601 Opava, Czech Republic}

\author{A. Kotrlov\'{a}}
\email{andrea.kotrlova@physics.slu.cz}
\affiliation{Institute of Physics, Silesian University in Opava, Bezru\v{c}ovo n\'{a}m. 13, CZ-74601 Opava, Czech Republic}

\author{E. \v{S}r\'{a}mkov\'{a}}
\email{eva.sramkova@physics.slu.cz}
\affiliation{Institute of Physics, Silesian University in Opava, Bezru\v{c}ovo n\'{a}m. 13, CZ-74601 Opava, Czech Republic}

\begin{abstract}
The study investigates orbital motion of test particles near compact objects described by solutions involving massless scalar fields, electromagnetic fields, and nonlinear electrodynamics. Specifically, we analyze orbital dynamics in the Janis-Newman-Winicour, Janis-Newman-Winicour-Maxwell, Schwarzschild-Melvine, and Bonnor-Melvin spacetimes, comparing the results with those obtained for the Schwarzschild and Reissner-Nordström solutions. We examine the stability of circular orbits and the behavior of epicyclic frequencies under varying physical parameters. Our analysis shows that in certain cases the central object transitions into a naked singularity. Deviations from classical Schwarzschild and Reissner-Nordström solutions reveal conditions for the existence of multiple photon orbits or marginally stable orbits. In some instances, the geometry allows the presence of two photon orbits—one stable and one unstable—with an interesting connection to the region of stable orbits. We find that at lower intensities, the effects of the scalar field and electromagnetic fields are comparable and seemingly interchangeable. However, for a sufficiently strong scalar field, its influence becomes dominant, leading to the emergence of a distinct region of stable orbits near the naked singularity. These effects are illustrated within the framework of optical geometry using embedding diagrams.
\end{abstract}

\date{\today}

\maketitle

\section{Introduction}
\label{sec:Introduction}
Fast variability is one of the prominent observational features of accreting compact objects. Apart from chaotic fluctuations, Fourier analysis of the X-ray light curves of several low-mass X-ray binaries (LMXRBs) in our Galaxy often reveals distinct and relatively coherent high-frequency quasi-periodic oscillations (QPOs) superimposed on the broadband noise spectra (see also \cite{Klis2000, Remillard+McClintock2006, Ingram+Motta2019} for reviews). The frequencies of these oscillations are of the order of a few hundred Hz (peaking at around 1.2 kHz in some neutron star sources), which roughly corresponds to the orbital periods of matter in the vicinity of the central compact object. The majority of the astrophysical community is convinced that the origin of QPOs is related to orbital motion, supported by spectroscopic arguments and the observed scaling of QPO frequencies with the mass of accreting objects \cite{2000MNRAS.316..923G,2004ApJ...609L..63A,Gol-etal:2019}.

The QPOs are most commonly detected either as a single oscillation or as a pair of two oscillations (referred to as `upper' and `lower' QPOs according to their frequencies). For neutron star (NS) sources, the two QPO frequencies follow remarkably stable correlations specific to a given source, while the overall distribution of simultaneous detections of two QPOs tends to cluster around frequencies in ratios of small integers, such as 3:2, 4:3, or 5:4 \cite{Belloni+2005}.  The clustering most likely arises as a result of the weakness of (one or both) QPOs outside the limited frequency range \citep{Abramowicz+Bulik+Bursa+Kluzniak2003, Belloni+2005, Torok+2008b, Torok+2008a, Boutelier+2010}. The NS QPO frequencies also appear to correlate well with the frequencies of broad features in the low-frequency part of the Fourier power density spectra \citep{Klis2000, Remillard+McClintock2006, Ingram+Motta2019}.

Small integer ratios, especially the 3:2 ratio, have also been reported for black hole Galactic microquasars \citep{Remillard+McClintock2006}, where QPO detections also seem to reveal frequencies specific to a given source. The black hole QPOs are rather elusive weak features, but the sources are likely to have frequency correlations similar to those observed in NS sources \citep{Belloni+2012,Motta+2014,Varniere+Rodriguez2018}.

Both the high-frequency values and the remarkable stability of the QPO frequency correlations suggest that, based on a correct description of the orbital motion of accreted matter, QPO observations can serve as a very useful probe of the spacetime geometry of compact objects. Most QPO models involve either the orbital motion of individual test particles (associated with `blobs' in the disk) or the collective motion of the fluid along nearly circular paths in a strong gravitational field. Generally, three different frequencies are involved in the nearly circular test particle geodesic motion: the Keplerian orbital frequency $\Omega_\mathrm{K}$ (sometimes referred to as $\omega_\phi$) and two epicyclic frequencies $\omega_r$ and $\omega_\theta$ (radial and vertical, alternatively referred to as $\kappa$ and $\Omega_\perp$) corresponding to the oscillations of a particle around a strictly circular equatorial orbit in two perpendicular directions. In the spherically symmetric Newtonian gravitational field described by the $\Phi\propto r^{-1}$ potential, all the three frequencies coincide, $\omega_r = \omega_z = \Omega_\mathrm{K}$, leading to closed planar orbits (ellipses). In the Schwarzschild spacetime, the radial epicyclic frequency is lower than the orbital frequency and their difference $\omega_\mathrm{p} = \Omega_\mathrm{K} - \omega_r$ gives the frequency of the pericentre precession. The latter two frequencies still coincide due to spacetime spherical symmetry.  In more general scenarios with axisymmetric spacetimes, however, all the frequencies can be different.

In the relativistic precession model \cite{Stella+Vietri1999, Motta+2014}, the frequencies of the twin-peak QPOs are identified with combinations of the orbital and epicyclic frequencies of a blob moving around the central object inside the accretion flow. The lower and upper QPO frequencies are given by the pericentre precession frequency $\omega_\mathrm{p}$ and the orbital frequency $\Omega_\mathrm{K}$, respectively.   

Other models involve the collective motion of the fluid rather than the geodesic motion of individual particles. Rezolla \cite{Rezzolla+2003} suggested that QPOs represent different global oscillation modes of a geometrically thick torus that may be located in the innermost part of the accretion flow. He identifies the lower and upper QPOs with the fundamental mode of the radial oscillations of the torus and its first overtone, respectively. Alternatively, Blaes et al. \cite{Blaes+2006} have proposed a scenario in which the QPOs are represented by the fundamental mode of the vertical oscillations and the breathing mode of the torus. The authors showed that this combination gives a 3:2 frequency ratio almost independently of the fluid angular momentum distribution and the polytropic index of the gas. This is true even for tori of considerable thickness. In addition, for small (``slender'') tori, both the shape of the torus and the frequencies of the oscillation modes are determined solely by the values of the orbital and epicyclic frequencies at the center of the torus and the polytropic index of the gas, other spacetime properties do not enter \cite{Abramowicz+2006}. 

In all the models mentioned so far, the values of the observed frequencies depend on the actual position of the spot or torus, which is a free parameter of the model. Relatively stable frequencies of QPOs in black-hole candidate sources can only be explained if there is some additional mechanism that ensures that the position of the source of the variability is always the same. In this context, an interesting idea has been proposed by Abramowicz \& Kluzniak \cite{Abramowicz+Kluzniak2001} that resonance phenomena could be responsible for the observed QPO variability. More specifically, the QPOs have been attributed to a nonlinear resonance between the radial and vertical epicyclic modes of the accretion torus \cite{Abramowicz+2003}. Although their frequencies vary with the position of the torus, the oscillations reach substantial amplitudes only at radii where the two modes are in resonance. The authors also argue why the 3:2 resonance is the strongest among the epicyclic modes \cite{Kluzniak+Abramowicz2002, Horak2008}.    

On the other hand, within the class of diskoseismic models that relate QPOs to global oscillations of geometrically thin accretion disks, there is increasing evidence that waves trapped in the disk can reach substantial amplitudes and contribute to the observed high-frequency variability \cite{Kato2001}. Among the proposed candidates are p-modes excited by the corotational instability \cite{Lai+Tsang2009, Horak+Lai2013} or g-modes excited in nonlinear interactions with other oscillation modes or waves present in the disk \cite{Kato2004, Ferreira+Ogilvie2008}. In the innermost region of disks surrounding Kerr black holes, both the axisymmetric and non-axisymmetric p-modes are typically trapped between the inner edge of the disk at the innermost stable circular orbit (ISCO) and the Lindblad resonances, where the oscillation frequency observed in the comoving reference frame coincides with the local radial epicyclic frequency. On the other hand, in the case of axisymmetric g-modes a wave can be trapped between two Lindblad resonances separated by the wave-propagation region due to the non-monotonic behavior of the radial epicyclic frequency \cite{Okazaki+1987}. Since a typical wavelength of the waves is comparable to the thickness of the disk, the frequencies of the lowest-order trapped g-modes are always near the maximum of the radial epicyclic frequency, independently of the detail properties of the disk \cite{Nowak+Wagoner1992, Wagoner+2001}.  Although a precise mechanism giving rise to a sequence of multiple QPOs in harmonic ratios is not yet well established, thin disk oscillation modes may still be a viable candidate for those sources where a single QPO is observed. 

The crucial quantities that determine numerical values of the observed QPOs in all the models mentioned so far are the three geodesic frequencies. Although the models were originally presented with the values corresponding to the Kerr metric describing the gravitational field of a rotating black hole, there is now a growing interest in other, more exotic spacetimes, opening up the possibility of testing alternative theories of gravity and/or collapse scenarios. The final stage of the gravitational collapse of massive stars can lead to black holes or naked singularities, depending on the initial conditions. Whether or not naked singularities exist, it is worth finding and discussing the differences in observational properties between these exotic objects and black holes. In this context, the behavior of epicyclic frequencies in the spacetime of various, sometimes quite exotic, objects has been studied, including Kerr naked singularities \cite{Torok+Stuchlik2005}, braneworld models of neutron stars \cite{Kotrlova+2008}, regular non-minimal magnetic black holes \cite{Rayimbaev+2021}, rotating traversable wormholes \cite{Deligianni+2021a, Deligianni+2021b}, Simpson-Wisser wormholes \cite{Stuchlik+Vrba2021a, Stuchlik+Vrba2021b}, non-semi-quantum commutatively inspired black holes \cite{Rayimbaev+2022}, Schwarzschild black holes in $f(R)$ gravity \cite{DeFalco+2023}, or hairy black holes in Horndeski gravity \cite{Rayimbaev+2023}. Recently, a comprehensive study of the various properties of orbital motion in a wide variety of alternative spacetimes has been published by Shahzadi et al \cite{Shahzadi+2023a}. On the other hand, the spacetime may also deviate from Kerr because of the presence of another gravitating object in the vicinity. Deviations of epicyclic frequencies due to accretion disk gravity have been studied in ref.~\cite{Semerak+Zacek2000a, Semerak+Zacek2000b}.  

 \subsection*{Aims and scope of this paper}
 In this paper, we derive the epicyclic frequencies for two well-known solutions describing naked singularities arising from the Einstein's massless scalar field theory. In the vanishing scalar field limit, these two solutions coincide with the Schwarzschild solution describing a black hole. We show that the presence of the scalar field strongly affects the properties of the orbital motion. Furthermore, shifting our attention to nonlinear electrodynamics, we derive explicit formulae for the epicyclic frequencies in the vicinity of a black hole with a magnetic charge, considering a popular model of nonlinear electrodynamics with the square-root Lagrangian \cite{Tahamtan2020, Tahamtan-RAG19}. Although this solution includes an electromagnetic field, it is very different from the well-known Reisner-Nordstr\"{o}m solution, namely due to the presence of a single event horizon. Finally, we study a static and axially symmetric magnetic solution describing a Schwarzschild black hole immersed in an external magnetic field.
 
The outline of the paper is as follows. In Sec.~\ref{sec:epicyclic-motion}, we introduce all the necessary quantities by presenting a quick derivation of the orbital frequencies in a general spherically-symmetric spacetime. These formulae are then applied in the following Secs.~\ref{sec:scalar-field} and \ref{sec:Einstein-Maxwell-scalar-field} to non-rotating black holes with massless and Einstein-Maxwell scalar fields. In Sec~\ref{sec:nonlinear-electrodynamics}, we explore the epicyclic oscillations within nonlinear electrodynamics, while Sec~\ref{sec:magnetic-field} is devoted to static black holes immersed in an external magnetic field. This is followed by the discussion and summary presented in Sec~\ref{sec:conclusion}.

\section{Circular and epicyclic motion}
\label{sec:epicyclic-motion}
The expressions for orbital frequencies valid in the Schwarzschild and Kerr spacetimes have been derived in the past by several authors using the geodesic equation\cite{Shikorov1973, Aliev+Galtsov1981, Okazaki+1987, Kato1990}. Formulas applicable to a general stationary spacetime have been introduced by Abramowicz \& Klu\'{z}niak\cite{Abramowicz+Kluzniak2005}. There is also an alternative derivation by Kerner et al. \cite{Kerner+2001}, which is derived from the geodesic deviation equation. 

Here, we outline a very quick derivation of the general expressions valid in spherically symmetric spacetimes, in order to introduce the necessary notation. We employ geometrical units where $c = G = 1$; both times and lengths are therefore measured in the units of mass. We consider nearly circular geodesics in a general spherically symmetric spacetime described by the metric
\begin{eqnarray}
    \label{ourmetric}
    \d s^2&=&-f\,\d t^2+\frac{\d r^2}{f} +
    R^2 \left(\d \theta^2 +{\sin^2{\theta}}\,\d \phi^2\right),
\end{eqnarray}
where the metric function $f(r)$ is positive in a domain of interest and $R(r)$ is the circumferential radius. The trajectory of a free test particle $x_\alpha(\tau)$ (where $\tau$ is the proper time) satisfies the geodesic equation 
\begin{equation}
    \frac{\d u_\alpha}{\d\tau} + \frac{1}{2}g^{\mu\nu}{}_{,\alpha} u_\mu u_\nu = 0,
    \label{eq:geodetics}
\end{equation}
where $u_\alpha = \d x_\alpha/\d\tau$ is the four-velocity of the particle. Due to spherical symmetry, our spacetime geometry admits equatorial circular orbits described by $r=r_0=\mathrm{const}$ and $\theta=\pi/2$. Consequently, the only non-zero covariant components of the particle four-velocity are $u_t$ and $u_\phi$. Both $u_t$ and $u_\phi$ are conserved along the orbit, $u_t$ is related to the total energy of the particle per particle rest mass, $u_t = -\energy_\mathrm{K}$, and $u_\phi$ is the angular momentum per particle rest mass, $u_\phi = \angmom_\mathrm{K}$. The subscript `K' relates to Keplerian (i.e.\ geodesic) orbits.  
The normalization condition $u^\mu u_\mu=-1$ and the geodesic equation (\ref{eq:geodetics}) imply the relations
\begin{align}
    g^{tt}\,\energy_\mathrm{K}^2 + 
    g^{\phi\phi}\,\angmom_\mathrm{K}^2 &= -1,
    \label{eq:u-normalization}
    \\
    g^{tt}{}_{,r}\energy_\mathrm{K}^2 + 
    g^{\phi\phi}{}_{,r}\angmom_\mathrm{K}^2 &= 0,
    \label{eq:circular-geodesic-equation}
\end{align}
from which $\energy_\mathrm{K}^2$ and $\angmom_\mathrm{K}^2$ immediately follow,
\begin{equation}
    \energy_\mathrm{K}^2 = 
    \frac{g_{\phi\phi,r}}{\left(\frac{\partial\tilde{r}^2}{\partial r}\right)},
    \quad
    \angmom_\mathrm{K}^2 = -\tilde{r}^4
    \frac{g_{tt,r}}{\left(\frac{\partial\tilde{r}^2}{\partial r}\right)}.
    \label{eq:EKLK}
\end{equation}
The quantity 
\begin{equation}
    \tilde{r} = \sqrt{-\frac{g_{\phi\phi}}{g_{tt}}} = \frac{R}{\sqrt{f}}
    \label{eq:radius-of-gyration}
\end{equation}
plays an important role in our analysis. As will be shown in the following paragraphs, it has several physical meanings. The time-like geodesics correspond to positive and finite $\angmom_\mathrm{K}^2$ and $\energy_\mathrm{K}^2$, and thus to $\tilde{r}^2$ that increases with $r$.

Denoting $\ell_\mathrm{K}\equiv \mathcal{L_\mathrm{K}}/(-\energy_\mathrm{K}) = -u_\phi/u_t$ the angular momentum per energy of the particle, we find 
\begin{align}
    \ell_\mathrm{K} = \sqrt{-\frac{g^{tt}{}_{,r}}{g^{\phi\phi}{}_{,r}}}.
    \label{eq:ellK}
\end{align}
As we will demonstrate, this quantity (and its radial derivative) has important consequences for the stability of the circular motion.

Closely related to the Keplerian angular momentum is the Keplerian angular velocity
\begin{equation}
    \label{OmegaK}
    \Omega_\mathrm{K} = \frac{\d\phi}{\d t} = \frac{u^\phi}{u^t} = 
    -\frac{g^{\phi\phi}}{g^{tt}}\ell_\mathrm{K} = 
    \sqrt{-\frac{g_{tt,r}}{g_{\phi\phi,r}}}\,,
\end{equation}
which describes the orbital motion as seen by a distant observer. 

In the case of photons, the normalization condition $u_\mu u^\mu=0$ translates to the vanishing right-hand side of (\ref{eq:u-normalization}). In combination with geodesic equation (\ref{eq:circular-geodesic-equation}), we find the relation $\Omega_K\ell_K= 1$ and 
\begin{align}
    \frac{\partial \tilde{r}}{\partial r} = 0
    \label{eq:photon orbit}
\end{align}
for the circular \emph{photon orbit} radius. The photon orbits therefore correspond to radii of infinite $\energy_\mathrm{K}$ and $\angmom_\mathrm{K}$.

\subsection{Radius of gyration}
\label{ssec:intro-gyration}
In Newtonian dynamics, the angular velocity and the specific angular momentum of a particle in circular motion are related by $\ell=(r\sin\theta)^2\Omega$. Here, the factor $r\sin\theta\equiv r_\mathrm{gyr}$ is sometimes called the \textit{radius of gyration}. This concept can be easily generalized to general relativity in the case of static spacetimes, where the relation between the specific angular momentum $\ell=-u_\phi/u_t$ and the angular velocity $\Omega=u^\phi/u^t$ of a particle at the circular orbits (not necessarily geodesics) allows one to uniquely introduce the radius of gyration \cite{Abramowicz+1993} using the same relation
\begin{equation}
    \ell_\mathrm{K} = r_\mathrm{gyr}^2\Omega_\mathrm{K}.
    \label{eq:ell-Omega-relation}
    \nonumber
\end{equation}
In our case, $r_\mathrm{gyr}=\tilde{r}$, as follows from relations (\ref{eq:ellK}) and (\ref{OmegaK}). The radius of gyration plays an important role in the theory of relativistic accretion flows in static spacetimes. The surfaces of constant $\tilde{r}$ coincide with the von Zeipel cylinders. In the case of a differentially-rotating body made of a polytropic perfect fluid, the angular velocity and specific angular momentum are constant at these surfaces (see Ref.~\cite{Abramowicz1971} for a relativistic version of the von Zeipel's theorem). In addition, a vector normal to these surfaces gives the direction of a suitably defined centrifugal acceleration, $a_\alpha = (\tilde{v}^2/\tilde{r})(\nabla_\alpha\tilde{r})$ (see Ref.~\cite{Abramowicz+1993} for more details).

\subsection{Epicyclic frequencies}
\label{ssec:intro-epicyclic-frequencies}
Let us now introduce a small perturbation to the circular geodesics, $x_\alpha \rightarrow x_\alpha + \xi_\alpha$. The four velocity will change as $u_\alpha\rightarrow u_\alpha + \delta u_\alpha$ with $\delta u_\alpha = \d\xi_\alpha/d\tau$. We assume that the perturbation keeps the angular momentum unchanged, therefore one may write $\delta u_\phi = 0$. From the normalization condition for the four-velocity, we immediately find that one may also write $\delta u_t = 0$ up to the linear order in $\xi$. The perturbed orbit is also geodesic and thus it has to satisfy equation (\ref{eq:geodetics}). Up to the linear order in $\xi$, we find
\begin{equation}
    \frac{\d^2 \xi_\alpha}{\d\tau^2} + 
    \frac{1}{2}\energy_\mathrm{K}^2\left(g^{tt}{}_{,\alpha\beta} + 
    \ell_\mathrm{K}^2 g^{\phi\phi}{}_{,\alpha\beta}\right) g^{\beta\gamma}\xi_\gamma = 0.
\end{equation}
Here, due to the symmetry with respect to the equatorial plane, the odd-order derivatives of the metric tensor with respect to $\theta$ vanish. The equation therefore describes two uncoupled harmonic oscillations in $r$ and $\theta$ directions with frequencies
\begin{equation}
    \omega^2_x= \frac{\energy_\mathrm{K}^2}{2g_{xx}}
    \left(g^{tt}{}_{,xx} + \ell_\mathrm{K}^2 g^{\phi\phi}{}_{,xx}\right),
    \label{eq:omegax}
\end{equation}
where $x=r,\theta$. These \textit{epicyclic frequencies} are measured by the observer comoving with the particle (note that the proper time $\tau$ is measured along the unperturbed circular orbit). Equation (\ref{eq:omegax}) can also be written in a more familiar form as
\begin{equation}
    \omega_x^2 = \left(\frac{\partial^2\mathcal{U}}{\partial x_\ast^2}\right)_\ell,
    \quad
    \mathcal{U} = -\frac{1}{2}\ln\left(-g^{tt} - \ell^2g^{\phi\phi}\right),
\end{equation}
where $\mathcal{U}$ is the effective potential as defined in Ref.~\cite{Abramowicz+Kluzniak2005}, and $x_\ast$ is the locally measured radial or meridional distance, $\dd x_\ast = \sqrt{g_{xx}}\dd x$. The derivative is taken assuming a constant value of the specific angular momentum, $\ell=\ell_\mathrm{K}(r_0)$. This corresponds to an unperturbed circular orbit and its value follows from the condition
\begin{equation}
    \left(\frac{\partial\mathcal{U}}{\partial X}\right)_\ell = 0
\end{equation}
(see Ref.~\cite{Abramowicz+Kluzniak2005} for more details). The frequencies $\Omega_x$ measured by a distant observer are related to $\omega_x$ by the `red-shift' factor $\d t/\d\tau = u^t$, therefore
\begin{equation}
    \Omega_x^2 = \frac{g_{tt}^2}{2g_{xx}}\left(g^{tt}{}_{,xx} + \ell_\mathrm{K}^2 g^{\phi\phi}{}_{,xx}\right).
    \label{eq:Omegax}
\end{equation}

In spherically symmetric spacetimes, $g^{tt}$ is a function of $r$ only, and $g^{\phi\phi} = g^{\theta\theta}/\sin^2\theta$. Therefore, we immediately find that $\Omega_\theta = \Omega_\mathrm{K}$. This result has a simple interpretation -- due to the spherical symmetry, all the geodesic orbits are planar, hence the particle has to make exactly one vertical oscillation per revolution.

\subsection{Signatures of curved spacetimes}
\label{ssec:into-optical-geometry}
The existence of a difference between the radial epicyclic frequency and the Keplerian orbital frequency in a static and strictly spherically symmetric gravitational field is a manifestation of general relativity. In fact, the resulting apsidal precession remains one of the first experimental tests of this theory. Many authors have attributed this difference to various aspects of general relativity. An interesting interpretation has been proposed by Abramowicz and Klu\'{z}niak \cite{Abramowicz+Kluzniak2003}, in which this inequality arises from geometrical properties of the space-like hyper-surfaces perpendicular to the time-like Killing vectors in the optical geometry introduced by Abramowicz, Carter and Lasota \cite{Abramowicz+Carter+Lasota1988}. 

\begin{figure*}
    \includegraphics[width=0.47\textwidth]{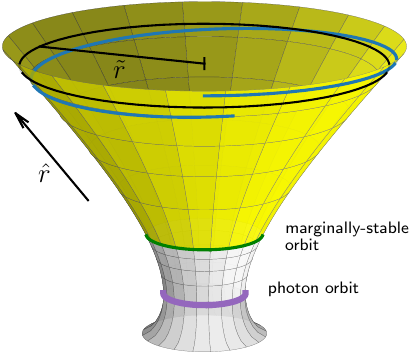} 
    \hfill
    \includegraphics[width=0.47\textwidth]{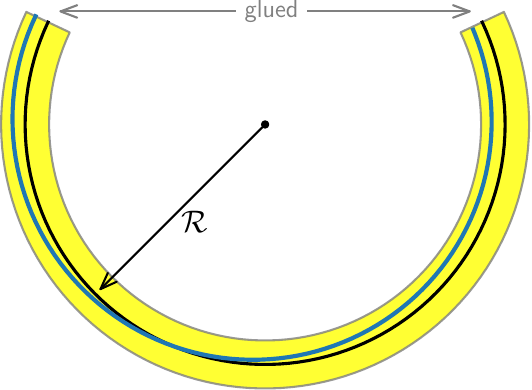}
    \caption{Equatorial cut of the optical space corresponding to the square-root model of the nonlinear electrodynamics with $\alpha=1/2$ that will be described in greater details in Sec.~\ref{sec:nonlinear-electrodynamics}. The \textit{left} panel shows the embedding diagram of the region $r\geq (\sqrt{5}-2)M$. The photon orbit at $r_\mathrm{ph}=3M/\alpha$ shown by the purple line in the throat corresponds to a circular geodesic in the optical space. Regions of stable and unstable circular orbits are indicated by yellow and gray color, respectively. The blue line in the upper part of the plot shows a perturbed circular orbit (the unperturbed one with $r=7 r_\mathrm{ph}$ is marked by the black circle). The plot also shows geometrical meaning of the circumferential ($\tilde{r}$) and geodesic ($\hat{r}$) radii, the latter one measures a radial distance on the embedded surface. The curvature radius $\mathcal{R}$ becomes apparent when `unfolding' the region around the orbit as shown in the \textit{right} panel of this figure.}
    \label{fig:sqrt-embedding}
\end{figure*}

The optical geometry becomes apparent after conformal rescaling of the spatial part of the metric (\ref{ourmetric}) to
\begin{equation}
    \d s^2 = f\left(-\d t^2 + \d h^2\right).
\end{equation}
with $\d h^2$ being a line element describing the three-dimensional optical space. The name `optical geometry' refers to the fact that the light rays correspond to geodesics also in this space and thus obey the Fermat's principle as can be seen from the above equation. It is also a convenient way how to separate the gravitational and inertial effects in a way analogical to the Newton's mechanics (see again the Ref.~\cite{Abramowicz+Carter+Lasota1988}). In particular, it is possible to show that the relativistic orbital dynamics of particles in the optical space obeys the very same equation as in Newton's mechanics in space that is curved in a specific way \cite{Abramowicz+Kluzniak2003}. This stresses a direct connection between the curvature and apsidal precession of the orbiting particles. In the following, we express the line element as
\begin{equation}
    \d h^2 = \d\hat{r}^2 + 
    \tilde{r}^2\left(\d\theta^2 + \sin^2\theta\,\d\phi^2\right)
\end{equation}
 with
\begin{equation}
    \hat{r} = \int^r \frac{\dd r}{f},
    \quad
    \tilde{r}^2 = \frac{R^2}{f}.
\end{equation}
The coordinate $\hat{r}$ measures a proper radial distance in the optical space and will be further referred to as the \textit{geodesic} radius. The second quantity, $\tilde{r}$, coincides with the radius of gyration introduced in Sec.~\ref{ssec:intro-gyration}, however, one can see its another meaning: as it measures the circumferences of the circles of constant $r$ and $\theta$ in the optical space. That is why the same quantity will be also referred to as the \textit{circumferential} radius. 

While in the flat space the circular orbit is associated with just a single radius, there are three different ways how to define radius of the circular orbit in a curved space. In addition to the geodesic and circumferential radii just mentioned, there is also a \textit{curvature} radius $\mathcal{R}$ measuring how far is the orbit from being a geodesic in the optical space. While the geodesics correspond to $\mathcal{R}=\infty$, the curvature radius of the time-like orbits can be calculated according to
\begin{equation}
    \mathcal{R} = \tilde{r}\,\frac{\d\hat{r}}{\d\tilde{r}}.
    \label{eq:opt-curvature-radius}
\end{equation}
The geometrical meaning of the three orbital radii is apparent in Fig.~\ref{fig:sqrt-embedding}, showing an example of the embedding diagram of the optical space corresponding to the equatorial plane in the case of a Schwarzschild black hole surrounded by a nonlinear electromagnetic field described by the square root model. Orbital properties of this spacetime are studied in great details in Sec.~\ref{sec:nonlinear-electrodynamics}. 

Abramowicz and Kluzniak \cite{Abramowicz+Kluzniak2003} found an interesting relation between the radial epicyclic and orbital frequency involving just these three radii,
\begin{equation}
    \Omega_r^2 = \left(\frac{\tilde{r}^2}{\mathcal{R}^2}
    \frac{\mathrm{d}\mathcal{R}}{\mathrm{d} \hat{r}}\right)\Omega_\mathrm{K}^2.
    \label{eq:Abramowicz-Kluzniak-radii}
\end{equation}
In Euclidean space, all the three radii coincide, leading to the $\Omega_r^2 = \Omega_\mathrm{K}^2$ equality. Measuring the $\Omega_r/\Omega_\mathrm{K}$ ratio in physical situations described above would provide a direct clue as to how these radii are related, as well as to the intrinsic properties of the curved spacetimes itself.

\subsection{Stability of circular orbits}
Finally, using equation (\ref{eq:ellK}), expression (\ref{eq:Omegax}) for the radial epicyclic frequency can be rewritten as
\begin{equation}
    \Omega_r^2 = g^{rr}\, 
    \frac{\d \left(-g_{tt}\right)}{\d r}\,
    \frac{\d\ln\ell_\mathrm{K}}{\d r}.
    \label{eq:Omegar}
\end{equation}
This equation describes a special case of the well-known Rayleigh criterion for geodesic motion -- the orbital geodetic motion is stable (i.e.\ $\Omega_r^2 > 0$) when the angular momentum increases outward. The radii $r_\mathrm{ms}$ of the marginally stable circular orbits, separating the regions of stable and unstable radial epicyclic oscillations, are given by the condition
\begin{equation}\label{rms-Eq}
    \frac{\d\ell_\mathrm{K}}{\d r} = 0.
\end{equation}
On the other hand, for large radii, both the orbital and epicyclic frequencies should obey their common Newtonian limit $\sim \sqrt{m/r^3}$, where $m$ is the ADM mass. Therefore, if a spacetime contains a marginally-stable circular orbit, the radial epicyclic frequency $\Omega_r$ is necessarily non-monotonic with a maximum somewhere in the region of stable orbits. If the compact object is surrounded by a thin accretion disk, this property will guarantee the existence of trapped axisymmetric g-mode oscillations that are one of the candidates for the observed high-frequency variability.   

Alternatively, the condition for existence of stable Keplerian orbits can be formulated in terms of optical geometry. As we have already shown, the condition for existence of a time-like circular geodesic is such that  the circumferential radius increases with increasing radial coordinate. Additional stability condition follows immediately from relation (\ref{eq:Abramowicz-Kluzniak-radii}). Therefore, stable Keplerian orbits may exist in those parts of the spacetime where both the circumferential and curvature radii increase with increasing geodesic radius. 

On the other hand, photon orbits correspond to extremal circumferential and infinite curvature radii. Depending on whether the circumferential radius takes its minimum or maximum, the photon orbit is either unstable or stable. The unstable photon orbits plays an important role in gravitational lensing, shaping the observed images. The stable `anti-photon' orbits are important for the stability of the spacetime with respect to gravitational radiation. It has been shown that gravitational waves trapped near these orbits decay extremely slowly and may even reach substantial amplitudes in the nonlinear regime. Sufficient concentration of gravitational waves may then induce gravitational collapse and formation of small black holes \cite{Keir2016}. 

\section{Scalar Field}
\label{sec:scalar-field}
Solutions to Einstein's equations with a scalar-field source provide a very useful tool for understanding relativity due to the simplicity of the source. Recently, it has become evident that fields of this type indeed do exist (Large Hadron Collider) and play a fundamental role in the standard model of particle physics. At the same time, scalar fields feature in models of dark energy and dark matter.

The famous static spherically symmetric solution of gravitating massless scalar field (minimally coupled to gravity) is called the Janis, Newmann and Winicour (JNW) solution \cite{JNW1}. It was originally discovered by Fisher \cite{Fisher} in the late forties and then independently rediscovered by Janis, Newman and Winicour \cite{JNW1} in the late sixties. Although the JNW solution is static it has received significant attention from different fields ranging from those involving quantum gravity (\cite{Svitek+2020}) up to astrophysical phenomena and optical properties concerning the JNW spacetime,  e.g., gravitational lensing and relativistic images \cite{JNW-lensing1, JNW-lensing2}, accretion processes \cite{JNW-accretion1, JNW-accretion2, Gyulchev+2019}, and also other observational phenomena \cite{JNW-shadow, Mirza+2023}. Recently in \cite{Azizallahi+2024}, radial epicyclic frequencies for rotating metrics including generalization of JNW were studied.

For the naked-singularity or singular horizon solutions it is important to establish potential observational differences compared to the Schwarzschild black hole and similar standard solutions. The JNW spacetime sourced by a  standard scalar field represents a good candidate for such study.  Additionally, studying the influence of gravitating scalar field on the orbital motion is interesting on its own.  It was shown in Ref.~[44] that the regularity of horizon which is spoiled by scalar field in spherically symmetric static scalar spacetimes is not improved by including nonlinear electrodynamics or other sources satisfying certain condition. 

First, we present the metric functions for the metric ansatz \eqref{ourmetric} corresponding to the JNW solution 
\begin{align} \label{JNW}
    f(r) &= \left [\frac {r-M}{{r}+M}\right]^{\frac{1}{\mu}}, \nonumber\\
    R(r) &= \sqrt{\frac{r^2-M^2}{f(r)}}
\end{align}
with the radial scalar field
\begin{equation}
    \varphi(r)=\frac{C_{0}}{2\,M}\ln{\left\{\frac{r -M}{r+ M}\right \}}\,,
\end{equation}
where the $\mu$ parameter reads $\mu=[1-C_{0}^2/(2M^2)]^{-1/2}$ (which means that $\mu \geq 1$ and $C_0<\sqrt{2}\,M$).
Obviously, if we write $C_{0}=0$ then we have $\mu=1$ and the solution coincides with the Schwarzschild metric with a shift in the radial coordinate $r+M \rightarrow r_\mathrm{Schw}$. 

The event horizon, located at $r_\mathrm{H}=M$, is a singular point because the Kretschmann scalar diverges here. In other words, this spacetime represents a naked (null) singularity. 

If we expand the $f(r)$ and $R(r)$ metric functions for a large $r$, we find
\begin{equation}
    R(r)\approx r,
    \quad
    f(r)\approx 1 - \frac{2M}{\mu r},
\end{equation}
from which we obtain the ADM mass to be 
\begin{equation}\label{ADM-JNW}
    m=\frac{M}{\mu}=\sqrt{M^2-\frac{C^2_0}{2}}.
\end{equation}
Since $r_\mathrm{H}=M$ (horizon), if one keeps the position of the horizon fixed it is clear that by increasing $C_0$  (the scalar-field charge) the ADM mass decreases. This behavior agrees with the Misner-Sharp Mass behavior, which is demonstrated in the Appendix \ref{MS}. As expected, the Misner-Sharp mass and ADM mass are equivalent at spatial infinity in asymptotically flat and spherically symmetric spacetimes. 

\begin{figure*}
	\includegraphics[width=\textwidth]{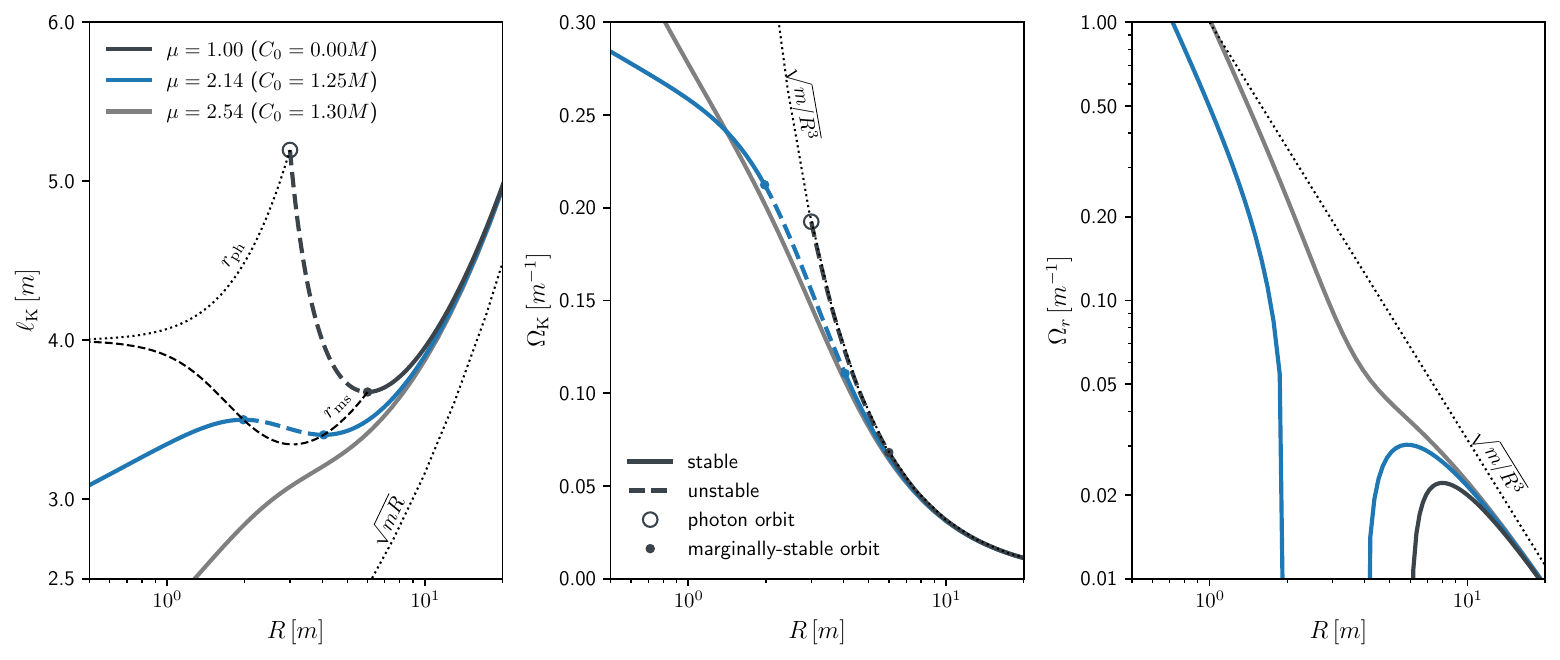}
    \caption{Keplerian angular momentum $\ell_\mathrm{K}$(\textit{left}), Keplerian orbital velocity $\Omega_\mathrm{K}$(\textit{middle}) and the radial epicyclic frequency $\Omega_r$(\textit{right}) in JNW spacetime as functions of the metric function $R$. All quantities are normalized by the ADM mass $m$. The value $\mu=1$ (black curves) corresponds to the Schwarzschild spacetime with the  photon orbit and marginally stable orbit located at $R=3m$ and $R_\mathrm{ms}=6m$. The position of the marginally stable orbit coincides with the minimum of $\ell_\mathrm{K}$ and zero of $\Omega_r$. The Keplerian orbital motion is possible only at $R > R_\mathrm{ph}$, in between $R_\mathrm{ph}$ and $R_\mathrm{ms}$, the motion is unstable. The corresponding part of the curves is shown by the dashed line. As $\mu\rightarrow 2$, the photon orbit approaches the singularity and $\ell_\mathrm{K}$ approach a constant value. For $2<\mu<\sqrt{5}$, there are two marginally stable orbits that correspond to extremes of $\ell_\mathrm{K}$ and no photon orbits. Keplerian orbital motion is possible everywhere outside the singularity, although it is unstable in between the marginally stable orbits. Consequently, the radial epicyclic frequency is imaginary in this region. Finally, for $\mu>\sqrt{5}$, there are neither photon nor marginally stable orbits. Keplerian orbital motion is stable in the entire space outside the singularity. The angular momentum monotonically increases and the radial epicyclic frequency is positive everywhere outside the singularity.}
    \label{fig:jnw-lf}
\end{figure*}

\subsection{Photon orbits}
The circumferential radius $\tilde{r}$ for this spacetime reads
\begin{equation}
    \tilde{r} = \left(r + M\right)^{(\mu+2)/2\mu}\left(r - M\right)^{(\mu-2)/2\mu}.
    \label{eq:JNW-circumferential-radius}
\end{equation}
Using expressions (\ref{eq:EKLK}), we find the squared energy and angular momentum (per particle mass) as
\begin{align}
    \energy_\mathrm{K}^2 &= \frac{\left(\mu r - M\right)f}{\mu r - 2M},
    \\
    \angmom_\mathrm{K}^2 &= \frac{\left(r^2 - M^2\right)M}{\left(\mu r - 2M\right)f}.
\end{align}
Both quantities are positive when $r>2M/\mu$. The singularity corresponds to the  location of the photon circular geodesics,
\begin{equation}
    r_\mathrm{ph} = \frac{2M}{\mu}\,,
    \label{eq:JNW-photon-orbit}
\end{equation}
and agrees with the minimal circumferential radius given by expression (\ref{eq:JNW-circumferential-radius}). When $\mu=1$, one gets $r=2M$ and recovers the Schwarzschild's photon orbit in our shifted coordinates. Scalar field tents to push the orbit closer to the singularity. When $\mu>2$ ($C_0>\sqrt{3/2}M$) the existence of the photon orbit is fully suppressed by the scalar field. In the following, we will restrict ourselves to radii $r>r_\mathrm{ph}$ (when $\mu < 2$), or $r > M$ (otherwise).

\subsection{Properties of the orbital motion}
The profile of the Keplerian angular momentum follows from Eq.~(\ref{eq:ellK}) and reads
\begin{equation}
    \ell_\mathrm{K} = \frac{1}{f}\sqrt{\frac{M\,(r^2-M^2)}{\mu r-M}}\,.
    \label{eq:jnw-lK}
\end{equation}
Clearly, at large radii, the angular momentum obeys the Newtonian formula $\ell_\mathrm{K} = \sqrt{mr}$. 

There are two radii at which $\ell_\mathrm{K}$ potentially diverges: 
\begin{equation}
    r_1=M,\qquad r_2=\frac{M}{\mu}.
    \nonumber
\end{equation}
While the first one, $r_1$, corresponds to the horizon, the second one is always below the horizon because $\mu\geq 1$. Boundaries between the regions of stable and unstable radial epicyclic oscillations are determined by the marginally stable orbits whose radii correspond to local extremes of the Keplerian angular momentum distribution. Using expression \eqref{rms-Eq}, we find that there are potentially two such radii 
\begin{equation}
    \label{eq:jnw-isco}
    r_\mathrm{ms}^\pm =\frac{M}{\mu}(3 \pm \sqrt{5-\mu^2}).
\end{equation}
The same result also follows from minimizing the curvature radius of the orbit (see Eq.~\ref{eq:opt-curvature-radius}),
\begin{equation}
    \mathcal{R} = \frac{\mu R^2}{\mu r - 2M}.
\end{equation}
For $1\leq\mu\leq2$ (corresponding to $0\leq C_0\leq\sqrt{3/2}$), the Keplerian angular momentum has just a single minimum at $r^{+}_\mathrm{ms}$ outside the horizon.  When $r>r_\mathrm{ms}^{+}$, the angular momentum increases with increasing $r$ implying stability of circular orbits, while inside $r_\mathrm{ms}^{+}$, it is a decreasing function of $r$ and all particle orbits there are unstable with respect to radial perturbations. The angular momentum distribution terminates at the photon orbit, where $\ell_\mathrm{K}$ takes a finite value. 

An interesting situation occurs when $2<\mu<\sqrt{5}$ ($\sqrt{3/2}<C_0<\sqrt{8/5}$). In that case, $\ell_\mathrm{K}$ has two extrema outside the horizon, located at $r_\mathrm{ms}^{-}$ and $r_\mathrm{ms}^{+}$. Consequently, there are two marginally stable orbits limiting a narrow region of unstable orbits between them. In the vicinity of the singularity angular momentum increases as $\propto(r-M)^{(\mu-2)/2\mu}$.

Finally, when $\mu>\sqrt{5}$ (i.e. when $C_0 > \sqrt{8/5}M$), the Keplerian angular momentum increases monotonically everywhere outside the horizon and all the circular geodesics are stable. Near the singularity it has the same asymptotics as in the previous case.

The radial profiles of the Keplerian angular momentum are shown in the left panel of Fig.~\ref{fig:jnw-lf} for representative cases of each type of behavior. In the plot, the angular momentum is normalized by the ADM mass $m$ as it is the quantity most easily accessible by a distant observer. As a representative of the radial coordinate, we have chosen the metric function $R$ that gives the proper length of the orbit divided by $2\pi$. The topmost curve corresponds to Schwarzschild spacetime, the middle curve represents the JNW spacetime with two marginally stable orbits with a region of unstable Keplerian orbits between them and the bottom curve represents the case of the spacetime with stable orbits everywhere outside the singularity.

In order to calculate the Keplerian angular velocity observed by a distant observer for the JNW solution, we use Eq.~(\ref{OmegaK}). We find
\begin{equation}
    \Omega_\mathrm{K}=f(r)\sqrt{\frac{M}{(\mu r-M)(r^2-M^2)}}\,.
    \label{eq:jnw-OmegaK}
\end{equation}
Radial profiles of the Keplerian angular frequency as functions of $R$ are shown in the middle panel of Fig.~\ref{fig:jnw-lf}. The frequencies are again normalized by the ADM mass $m$. At large radii, the expression (\ref{eq:jnw-OmegaK}) recovers the Newtonian behavior, $\Omega_\mathrm{K}\approx\sqrt{m/r^3}$. Closer to the singularity the angular frequency always increases with decreasing $r$. For $\mu<2$, the frequency curves terminate at the photon orbit, for $\mu>2$, they extend towards the singularity with the asymptotics $\Omega_\mathrm{K}\propto (r-M)^{-(\mu-2)/2\mu}$.

The last quantity is the radial epicyclic frequency. Using equation (\ref{eq:Omegar}), we get
\begin{align}
    \Omega_r & 
    = \frac{\Omega_\mathrm{K}}{M}\sqrt{\left(r-r_\mathrm{ms}^{-}\right)
    \left(r-r_\mathrm{ms}^{+}\right)} 
    \nonumber \\
    &
    =f(r)\sqrt{\frac{M\left[M^2(\mu^2+4)-6\mu M r+\mu^2r^2\right]}
    {\mu^2(\mu r-M)(r^2-M^2)^2}}
    \,.
    \label{eq:jnw-Omegar}
\end{align}
The radial epicyclic frequency vanishes at the marginally stable orbits given by Eq.~(\ref{eq:jnw-isco}), and, at large distances, it decays as $\Omega_\mathrm{K}\sim\sqrt{m/r^3}$. Consequently, for $\mu<\sqrt{5}$ there is a maximum of $\Omega_r$ between the (outer) marginally stable orbit and infinity. Existence of such maxima has important consequences for the oscillations of thin accretion disks. The astrophysical implications of it are discussed in details in Sec.~\ref{sec:conclusion}. For $2<\mu<\sqrt{5}$, the orbital motion becomes stable again below the inner marginally stable orbit and the singularity is approached with the asymptotics $\Omega_r\propto (r-M)^{(2-\mu)/2\mu}$. Finally, well above $\mu=\sqrt{5}$, the radial epicyclic frequency decreases monotonically with increasing $r$.

\section{The Einstein--Maxwell and Scalar Field}
\label{sec:Einstein-Maxwell-scalar-field}
In this section, we explore how the results of the previous section change when additional strong electromagnetic field is present. We consider another exact naked-singularity solution, where the sources of gravity are mass, and the scalar and electromagnetic (Maxwell) fields. This `Einstein-Maxwell-scalar field' solution, (sometimes also referred to as the `Janis-Newman-Winicour-Maxwell` -- JNWM) has been found by several authors \cite{JNW2, Penney1969, Teixeira1976, Banerjee} and represents a rather rare example of exact analytic solution of Einstein's equations, with three nonlinearly coupled fields of three different kinds. As such it is very useful for exploring relative impact  of the three fields on the stability of the orbital motion. In what follows, we show that the electromagnetic field does not change much its qualitative properties. In this respect, the orbital dynamics is still dominated by the Einstein and scalar field. Nevertheless, the electromagnetic field brings a new effect of the two photon orbits appearing in the spacetime exactly at level of the scalar field strength where their existence is suppressed in the original JNW solution. In a narrow range of radii this limits existence of Keplerian circular orbits. 

First, we briefly describe the structure of the solution. We assume the electromagnetic vector potential in the following form:
\begin{equation}
    \mathbf{A} = \psi(r)\,\mathbf{d}t.
    \nonumber
\end{equation}
The Maxwell equations then imply
\begin{equation}
    \psi = q\,\int\frac{\d r}{R^2},
    \nonumber
\end{equation}
where $R$ is the metric function from the metric ansatz \eqref{ourmetric}. The metric functions are given by:
\begin{align}
    f(r) &= \frac{4M^2\nu^2 x}{\left(C_1x - C_2\right)^2}, 
    \quad
    x \equiv \left(\frac{r-M}{r+M}\right)^{\nu/2},
    \label{JNW-Maxwell}\\
    R(r) &= \sqrt{\frac{r^2-M^2}{f(r)}}
    \label{JNW-Maxwell-R}
\end{align}
and the scalar field is described by the potential
\begin{equation}
      \varphi(r) = \frac{C_{0}}{2\,M}\ln{\left\{\frac{r -M}{r+ M}\right \}}.
\end{equation}
Here, $C_{0}$, $C_1$, and $C_2$ are integration constants with dimensions of mass, and the dimensionless parameter $\nu$ is related to the strength of the scalar field by $\nu = 2\sqrt{1-C_{0}^2/(2M^2)}$. Note that $\nu$ is limited to the range $0 < \nu \leq 2$ and therefore there is an upper limit for $C_0$, $C_0 < \sqrt{2}\,M$. Additionally, the $C_1$ and $C_2$ constants are constrained by the Einstein--Maxwell equations to satisfy $C_1 = 4\,q^2/C_2$, where $q$ is interpreted as a charge (see Ref.~\cite{Tahamtan2020} for more details). Unless $C_0 = 0$, the event horizon located at $r_\mathrm{H} = M$ becomes a singular point because the Kretschmann scalar diverges there. In other words, this spacetime represents a naked (null) singularity just like the JNW solution does.

After inserting the expression for $R$ from \eqref{JNW-Maxwell-R} and integrating, the electromagnetic potential reads
\begin{equation}
    \psi = -\frac{4\,M\,\nu\,q}{C_1\,\left(C_1\,x-C_2\right)}+\mathrm{const}.
\end{equation}

Finally, in the $r \gg M$ limit, we find
\begin{align}
    R(r) &\approx \frac{\sqrt{C_1 - C_2}}{2\nu M}\, r,
    \\
    f(r) &\approx \frac{4M^2\nu^2}{\left(C_1 - C_2\right)^2}
    \left[1 + \frac{M\nu}{r}\frac{C_1 + C_2}{C_1 - C_2}\right].
\end{align}
Thus, $\nu$, $C_1$, and $C_2$ are constrained by another relation:
\begin{equation}
    \left(C_1 - C_2\right)^2 = 4M^2\nu^2,
\end{equation}
and the ADM mass is:
\begin{equation}
    m = \sqrt{\left(\frac{\nu M}{2}\right)^2 + q^2} = \sqrt{M^2 + q^2 - \frac{C^2_0}{2}}.
    \label{eq:JNWM-AMD}
\end{equation}

Similar to the JNW case \eqref{ADM-JNW}, by keeping the position of the horizon fixed, $r_H = M$, increasing the scalar field charge $C_0$ leads to a decrease in the ADM mass. Close to the upper limit for $C_0$ (note that $C_0 < \sqrt{2}M$), the ADM mass approaches the charge $q$. A similar behavior is present for asymptotically large radii $r$ for the Misner-Sharp mass, as shown in Appendix \ref{MS}.

The spacetime is therefore fully described by only three parameters: the mass $M$, the strength of the scalar field $C_0$, and the charge $q$, which gives rise to the electromagnetic field. The values of the two other integration constants then follow from simple relations, $C_{1,2} = 2m \mp \nu M$ (note that $|C_2| > |C_1|$).

When $q$ vanishes, the solution \eqref{JNW-Maxwell} approaches the previously mentioned JNW solution \eqref{JNW}. On the other hand, in the limit of a vanishing scalar field, the solution coincides with the Reissner-Nordström (RN) solution. To recover the standard form of the RN metric, the following change in the radial coordinate should be applied:
\begin{align}
    r_\mathrm{RN} &= -\frac{1}{4M}\left[r\,(C_1 - C_2) - M\,(C_1 + C_2)\right]
    \nonumber \\
    &= \frac{\nu}{2}\, r + m.
\end{align}
Then, with $\nu = 2$, the metric function $f$ takes the standard form:
\begin{equation}
    f = 1 - \frac{2\,m}{r_\mathrm{RN}} + \frac{q^2}{r_\mathrm{RN}^2}
    \nonumber
\end{equation}
and at the same time
\begin{equation}
    R = r_\mathrm{RN}.
\end{equation}
Clearly, in this case, depending on $m > q$, $m < q$, and $m = q$, several scenarios may occur: the existence of two horizons, one extremal horizon, or no horizon at all (a naked singularity). However, if we retain definition of $m$ given by \eqref{eq:JNWM-AMD}, we arrive at the standard two-horizons scenario and the naked singularity case by replacing $M$ with $iM$. The extremal case occurs for $M = 0$. This shows that the ADM mass plays the role of the RN mass in the limit of a vanishing scalar field.

Naturally, when both $q$ and $C_0$ vanish, the Schwarzschild solution is recovered in the shifted radial coordinate $r = r_\mathrm{Schw} - M$. As shown in the following sections, the properties of orbital motion remain similar in many respects to the JNW solution because of the influence of the scalar field, impact of which is dominant compared to the electromagnetic field.

\subsection{Photon orbits}
The radius of gyration follows directly from expression (\ref{eq:radius-of-gyration}). In the equatorial plane, we have
\begin{equation}
    \tilde{r} = \frac{\sqrt{r^2 - M^2}}{f(r)}.
    \label{eq:radius-of-gyration-JNW-Maxwell}
\end{equation}
Using expressions (\ref{eq:EKLK}), we find squared energy and angular momentum per unit mass,
\begin{align}
    \energy_\mathrm{K}^2 &= \left(\frac{y-1}{y-2}\right)f,
    \label{eq:JNWM-EK2}
    \\
    \angmom_\mathrm{K}^2 &= \frac{R^2}{y-2},
    \label{eq:JNWM-LK2}
\end{align}
where
\begin{equation}
    y(r) \equiv -\frac{2r}{\nu M}\left(\frac{C_1 x - C_2}{C_1 x + C_2}\right).
\end{equation}
For vanishing charge, $y=r/m$. The photon orbits correspond to $y=2$. Time-like circular geodesics exist only when both $\energy_\mathrm{K}^2$ and $\angmom_\mathrm{K}^2$ are positive, i.e., when $y > 2$. Near the singularity at $r=M$, the function $y(r)$ first decreases as 
\begin{equation}
    y(r)\approx \frac{2}{\nu}\left[\frac{r}{M} - \frac{2C_1}{C_2}\left(\frac{r-M}{2M}\right)^{\nu/2}\right],
    \nonumber
\end{equation}
reaching a minimum at a finite radius $r_\mathrm{min}$ and then starts to increase, approximately following
\begin{equation}
    y(r)\approx \frac{r}{m} + \left(\frac{q}{m}\right)^2
    \nonumber
\end{equation}
at large radii. Depending on the values of $y(M)$ and $y(r_\mathrm{min})$, one of the following four cases occurs:
\begin{enumerate}
    \item $y(M) < 2$: This situation arises when $\nu > 1$, or equivalently $C_0 < \sqrt{3/2}M$. In this case, $y(r)$ reaches the value $y=2$ at a single point. Consequently, the spacetime contains a single photon orbit outside the singularity. Circular orbits of unaccelerated test particles exist only above this orbit. Inside, particles necessarily have nonzero radial component of the velocity. This situation is similar to the case of the RN black-hole spacetime.
 
    \item $y(M) > 2$ but $y(r_\mathrm{min}) < 2$: In this case, $y(r)$ intersects the $y=2$ line at two points, on either side of the minimum $r=r_\mathrm{min}$. Consequently, there are two photon orbits outside the singularity. Test particles may orbit the central singularity circularly only inside the inner photon orbit or outside the outer one. In between, no circular time-like geodesics are possible.

    \item $y(M) > 2$ and $y(r_\mathrm{min}) = 2$: In this case, $y(r)$ just touches the $y=2$ line at a single point $r_\mathrm{ph}^\ast$ that coincides with the minimum $r_\mathrm{min}$. Time-like circular geodesics exist in the whole domain outside the singularity (except for this particular radius).
 
    \item $y(r_\mathrm{min}) > 2$: In this case, $y(r)$ is always above the $y=2$ line. Consequently, no photon orbits exist in the spacetime, and time-like circular geodesics are possible throughout the entire domain $r>M$.
\end{enumerate}

\begin{figure}
    \includegraphics[width=0.48\textwidth]{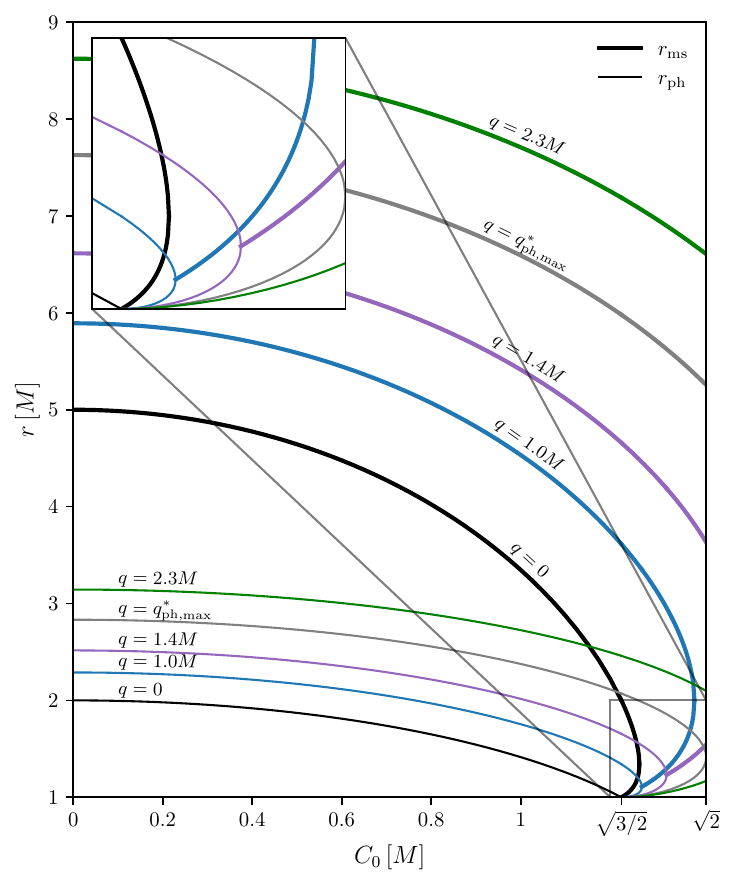}
    \caption{Position of photon orbits (thin lines) and marginally stable orbits (thick lines) as functions of the scalar-field strength $C_0$ for several values of the electric charge $q$. Depending on $C_0$ and $q$, the spacetime may contain either zero, one, or two orbits of both kinds.}
    \label{fig:jnwm-orbits}
\end{figure}

The cases 2-4 occur for rather strong scalar fields, when $\nu<1$, corresponding to $\sqrt{3/2}M < C_0 < \sqrt{2}M$. In the following, we elaborate the limit case 3 in more detail. The conditions for this case can be summarized as
\begin{align}
   y = -\frac{2r}{\nu M}\left(\frac{C_1 x - C_2}{C_1 x + C_2}\right) &= 2,
   \nonumber \\
   y' = \frac{y}{r} - \frac{q^2 x r}{\left(C_1x + C_2\right)^2\left(r^2-M^2\right)} &= 0.
   \nonumber
\end{align}
Solving these two equations for $x$ and $r$ gives
\begin{align}
   r = r_\mathrm{ph}^\ast = \sqrt{2 - \nu^2} M,
   \quad
   x = \frac{\sqrt{2-\nu^2} - \nu}{\sqrt{2 - \nu^2} + \nu}\,
   \left(\frac{C_2}{C_1}\right).
   \nonumber
\end{align}
Obviously, this solution is compatible with our definition of $x$ only when
\begin{equation}
   \frac{C_2}{C_1} = \frac{\sqrt{2-\nu^2}+\nu}{\sqrt{2-\nu^2}-\nu}\left(\frac{\sqrt{2-\nu^2}-1}{\sqrt{2-\nu^2}+1}\right)^{\nu/2} \equiv \gamma
   \nonumber
\end{equation}
Using $C_{1,2} = 2m\mp \nu M$ and expression (\ref{eq:JNWM-AMD}) for the ADM mass $m$, we arrive at the critical value of the charge related to this case,
\begin{equation}
   q = \frac{\nu\sqrt{\gamma}}{\gamma - 1} M \equiv q^\ast_\mathrm{ph}(\nu),
   \quad
   (\nu < 1).
   \label{eq:jnwm-q_ph}
\end{equation}

As soon as the charge of the singularity exceeds this value and $\nu < 1$, the spacetime given by the JNWM solution contains two photon orbits. Obviously, $q^\ast_\mathrm{ph}$ vanishes as $\nu$ approaches 1 (or equivalently as $C_0$ approaches $\sqrt{3/2}M$). At the same time, the radius of the single photon orbit $r_\mathrm{ph}^\ast$ approaches $M$. This is consistent with the fact that there are no photon orbits in the JNW spacetime for scalar fields stronger than $C_0 = \sqrt{3/2}M$. When $C_0 \to \sqrt{2}M$, $q_\mathrm{ph}^\ast$ approaches the value $q_\mathrm{ph, max} = \frac{2}{2^{3/2} + \ln(3 - 2^{3/2})}M \approx 1.877M$. Spacetimes with greater charges than this value host two photon orbits throughout the range $\sqrt{3/2}M < C_0 < \sqrt{2}M$.

\begin{figure}
    \includegraphics[width=0.48\textwidth]{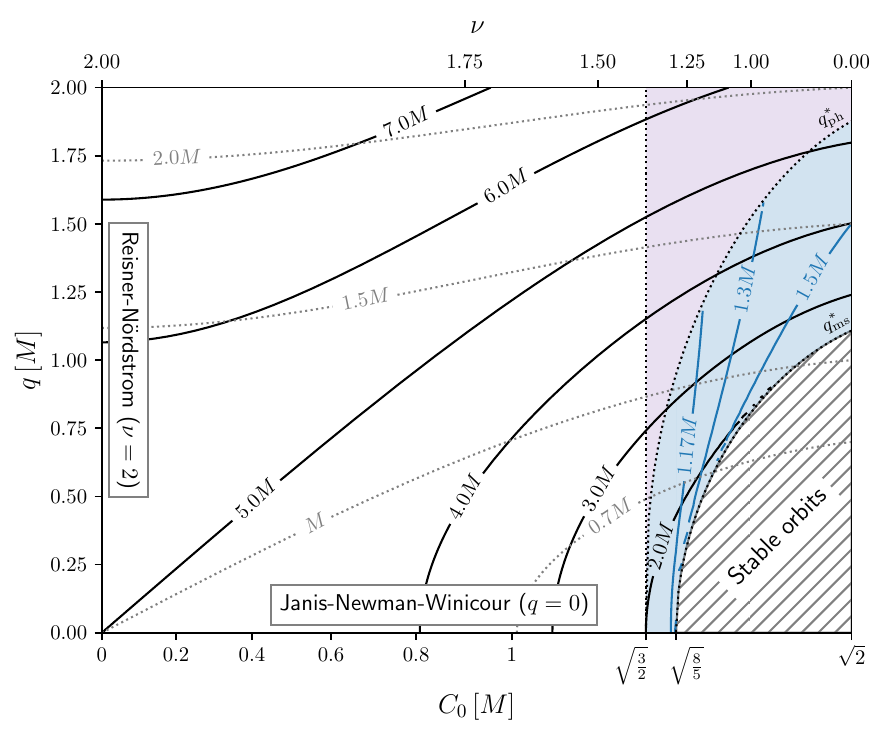}
    \caption{Occurrence of the photon and marginally stable orbits in the JNWM spacetime as functions of the charge and scalar field. When $C_0 < \sqrt{3/2}M$, the spacetime contains a single photon orbit and a single marginally stable orbit. The radius of the latter is shown by black contours in the plot. For a combination of $q$ and $C_0$ in the purple region, there are two photon orbits and a single marginally stable orbit. This domain is separated from the neighboring blue region, where two marginally stable orbits and no photon orbit exist, by the $q_\mathrm{ph}^\ast(\nu)$ curve given by equation (\ref{eq:jnwm-q_ph}). The blue contours in the latter region indicate radii of the inner marginally stable orbits. Finally, the dashed area in the lower right corner shows the region where neither marginally stable nor photon orbits exist and all circular geodesics are stable. This region is separated from the previous one by the curve $q_\mathrm{ms}^\ast(\nu)$, shown as a bent thin solid line starting at $C_0 = \sqrt{8/5}M$. The gray dotted contours in the plot indicate constant ADM mass $m$.}
    \label{fig:jnwm-parameter-space}
\end{figure}

\subsection{Marginally-stable orbits}
Using the formulas (\ref{eq:ellK}) and (\ref{OmegaK}) and with the aid of expressions (\ref{JNW-Maxwell}), (\ref{JNW-Maxwell-R}), (\ref{eq:radius-of-gyration-JNW-Maxwell}), (\ref{eq:JNWM-EK2}) and (\ref{eq:JNWM-LK2}), we find rather simple expressions for the Keplerian angular momentum and angular velocity,
\begin{equation}
    \ell_\mathrm{K} = \frac{\tilde{r}}{\sqrt{y-1}},
    \label{eq:jnwm-lK-OmegaK}
    \quad
    \Omega_\mathrm{K} = \frac{1}{\tilde{r}\sqrt{y-1}}.
\end{equation}
The time-like circular orbits correspond to $y>2$. The radial epicyclic frequency follows from equation (\ref{eq:Omegar}). After some algebra, we obtain
\begin{equation}
    \Omega_r = \Omega_\mathrm{K}\left[
    \frac{2r^2}{fR^2}\frac{(y-1)(y-2)}{y^2} 
    -\frac{\dd \ln y}{\dd \ln r}
    \right]^{1/2}
    \label{eq:jnwm-Omegar}
\end{equation}
with
\begin{equation}
    \frac{\dd \ln y}{\dd \ln r} = 1 - \left(\frac{q}{R}\right)^2.
    \nonumber
\end{equation}
Positions of the marginally stable orbits correspond to zeros of the radial epicyclic frequency. Putting $\Omega_r^2=0$ however leads to a much more complicated relation between $r$ and $x$ and $y$ than a simple condition $y=2$, which was analyzed analytically in the case of the photon orbits. This is why we have chosen a numerical approach. In short, we found a similar qualitative behavior to the previous case. Depending on the scalar-field strength and the charge, the spacetime may host either zero, one, or two marginally stable orbits. In addition, there is also an interesting link between the presence of the two types of orbits.

Our findings are summarized in~Figs.~\ref{fig:jnwm-orbits} and \ref{fig:jnwm-parameter-space}. Keeping the JNWM mass and charge constant and gradually increasing $C_0$ from zero, the scalar field slightly alters the positions of the single photon and marginally stable orbits, pushing both closer to the singularity. This is evident in the leftmost parts of Fig.~\ref{fig:jnwm-orbits} and \ref{fig:jnwm-parameter-space}.

At $C_0 = \sqrt{3/2}M$, the spacetime begins to host two photon orbits, with the newly appearing one emerging at the radius $r = M$. As $C_0$ increases further, the two photon orbits converge, while the single marginally stable orbit outside of them continues moving inward. This corresponds to the purple region in Fig.~\ref{fig:jnwm-parameter-space}.

If $q > q_\mathrm{ph,max}^\ast$, this situation persists until $C_0 = \sqrt{2}M$. For lower $q$, the two photon orbits eventually merge before $C_0$ reaches $\sqrt{2}M$. The scalar field parameter $\nu$, for which this occurs, satisfies $q_\mathrm{ph}^\ast(\nu) = q$. The merging point corresponds to case 3 discussed above, where $y = 2$ and $y^\prime = 0$ at the merging radius. From relation (\ref{eq:jnwm-Omegar}), it follows that $\Omega_r^2$ vanishes at the same radius. Consequently, as $C_0$ increases further, a new, inner marginally stable orbit emerges from exactly the same location where the two photon orbits previously merged. At the same time, the outer marginally stable orbit continues to move inward.

At this stage, corresponding to the blue region in Fig.~\ref{fig:jnwm-parameter-space}, the spacetime contains two marginally stable orbits and no photon orbit. However, at even higher scalar-field strength, the two marginally stable orbits might eventually merge, leaving the spacetime entirely stable for time-like circular geodesics (dashed region in Fig.~\ref{fig:jnwm-parameter-space}). Alternatively, they might gradually converge, maintaining a region of unstable circular orbits between them until the maximum value $C_0 = \sqrt{2}M$ is reached.

Similarly to the previous case of two photon orbits, we define a critical charge $q_\mathrm{ms}^\ast(\nu)$, at which the marginally stable orbits merge for a fixed value of $\nu$. We determined this charge by numerically solving the equation $\Omega_r^2 = 0$, with $\Omega_r^2$ given by equation (\ref{eq:jnwm-Omegar}). Evidently, $q_\mathrm{ms}^\ast(\nu) \to 0$ as $\nu \to 2/\sqrt{5}$ (or $C_0 \to \sqrt{8/5}M$), corresponding to the JNW spacetime with $\mu = \sqrt{5}$. In the opposite limit $\nu \to 0$ (or $C_0 \to \sqrt{2}M$), $q_\mathrm{ms}^\ast$ approaches $q_\mathrm{ms,max}^\ast = 1.085M$. Consequently, entirely stable orbits may exist only in the JNWM spacetimes with charges lower than this value.

\begin{figure*}
	\includegraphics[width=\textwidth]{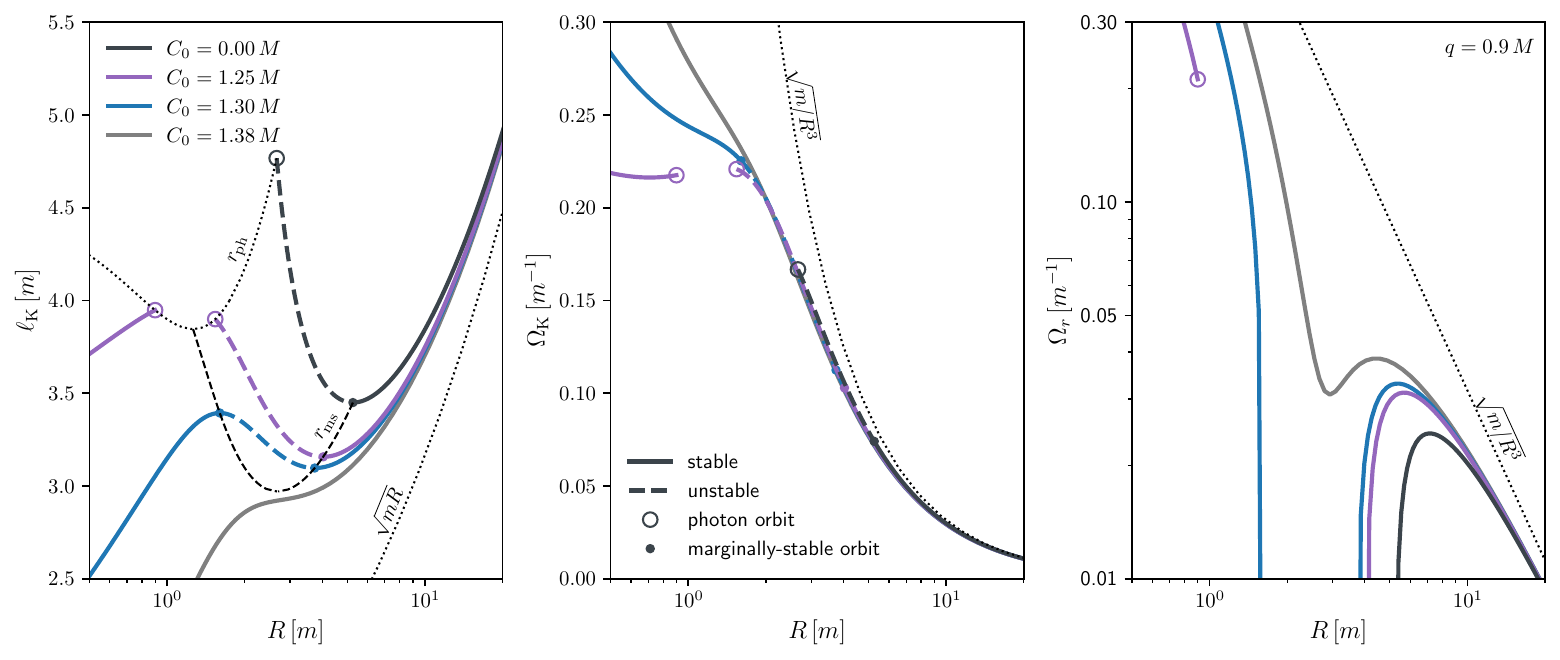}
    \caption{Keplerian specific angular momentum (\textit{left}), Keplerian angular velocity (\textit{middle}) and the radial epicyclic frequency (\textit{right}) normalized by the ADM mass $m$ as functions of the metric function $R$. Four cases corresponding to the same electric charge $q=0.9M$ and different scalar-field strength ($C_0=0, 1.25M, 1.30M$ and $1.38M$) are shown. Each of them corresponds to a different region in the parameter space shown in Fig~\ref{fig:jnwm-parameter-space}. The dotted line show asymptotics ($\sqrt{mR}$ in the case of the angular momentum and $\sqrt{m/R^3}$ in the case of the geodesic frequencies) at large distances.}
    \label{fig:jnwm-l-omegas}
\end{figure*}

\subsection{Geodesic orbital motion}
In this section, we analyze the behavior of specific angular momentum, as well as the Keplerian and radial epicyclic frequencies, for the four distinct cases identified in the previous section. The specific angular momentum and angular velocity are determined by equation (\ref{eq:jnwm-lK-OmegaK}), while the radial epicyclic frequency is described by expression (\ref{eq:jnwm-Omegar}).

It is clear that since $\tilde{r}\approx r$ and $y\approx r/m$ at large distances, all three expressions conform to the Newtonian limits. Specifically, the angular momentum increases as $\ell_\mathrm{K}\approx\sqrt{mr}$, while both angular frequencies decrease as $\sqrt{m/r^3}$ with the ADM mass $m$. However, closer to the singularity, the behavior becomes more sensitive to the relative strengths of the gravitational, scalar, and electromagnetic fields.

In the following, we fix the value of the charge to $q=0.9M$. Then just by varying the scalar field in the whole range $0\leq C_0\leq\sqrt{2}M$, we smoothly pass through all different regions of the parameter space shown in Fig.~\ref{fig:jnwm-parameter-space}. The profiles of the specific Keplerian angular momentum, Keplerian angular velocity, and epicyclic frequency are shown in the three panels of Fig.~\ref{fig:jnwm-l-omegas}. All the quantities are normalized by the ADM mass $m$, as this is the quantity with clear physical meaning for the distant observer. For the same reason, we have chosen the metric function $R$ as a convenient representation of the radial coordinate since it expresses the proper length of the orbit (measured by the local static observer) divided by $2\pi$. In addition, in the limit of vanishing scalar field, $R$ coincides with the RN radial coordinate $r_\mathrm{RN}$, and $m$ becomes the RN mass. It is also good to note that for non-vanishing scalar field $R\rightarrow 0$ as one approaches the singularity at $r=M$.  

The case of vanishing scalar field ($C_0=0$) corresponds to the uppermost curve in the angular-momentum plot. The spacetime coincides with that of the charged RN black hole with ADM mass $m=1.35M$ and $q/m=0.67$. Consequently, there is one unstable photon orbit at $R = 2.664\,m$ and one marginally stable orbit at $R=5.264\,m$. Circular geodesic motion is possible outside the photon orbit, the motion is stable against radial perturbations outside the marginally stable orbit. The corresponding Keplerian angular frequency and radial epicyclic frequency are shown in the middle and right panels of Fig.~\ref{fig:jnwm-l-omegas}.

Increasing the scalar field above $C_0=\sqrt{3/2}M$, one enters the purple region of Fig.~\ref{fig:jnwm-parameter-space}, where two photon orbits and one marginally stable orbits exist. A particular example of this case shown in the left panel of Fig.~\ref{fig:jnwm-l-omegas} by a second upper curve, corresponds to $C_0=1.25M = 1.23\,m$ and $q=0.9M=1.89\,m$ with ADM mass $m=1.01\,M$. The two photon orbits are located at $R=0.900\,m$ and $1.539\,m$ and the marginally stable orbit is at $4.036\,m$. While the outer photon orbit is unstable, the inner photon orbit is stable. The curves in all three panels have two branches. The outer branch starting at the outer photon orbit describes a similar structure of the orbits as in the case of the vanishing scalar field: first unstable orbits just outside the photon orbit, then marginally stable orbit and stable orbits further away. The inner branch between the singularity and the inner photon orbit contains time-like geodesics that are entirely stable.  

Increasing the scalar field further to $C_0=1.30\,M$, we have entered the blue region of Fig~\ref{fig:jnwm-parameter-space} with two marginally stable circular orbits and no photon orbits. A particular case shown in the figure corresponds to ADM mass $m=0.98\,M$, $C_0=1.32\,m$, and $q=0.92\,m$. The marginally stable orbits are located at $R=1.603\,m$ and $3.741\, m$. The time-like circular geodesics exist at all radii above the singularity, although they are unstable with respect to radial perturbations in between the two marginally stable orbits. This is qualitatively similar to the case of the JNW spacetime with $\mu$ between $2$ and $\sqrt{5}$.

Finally, for scalar field strength near the maximum $C_0=\sqrt{2}\,M$, and $q<q_\mathrm{ms,max}^\ast$, one enters the region of the parameter space, where the spacetime is entirely stable for all time-like geodesics outside the horizon (dashed region in Fig~\ref{fig:jnwm-parameter-space}). Fig.~\ref{fig:jnwm-l-omegas} shows the particular case of $C_0=1.38M=1.49\,m$ and $q=0.97\,m$ with ADM mass $m=0.93M$ (the lowest curve in the angular momentum plot). The properties of such spacetime are qualitatively similar to the JNW case with $\mu>\sqrt{5}$.

\begin{figure}
	\includegraphics[width=0.48\textwidth]{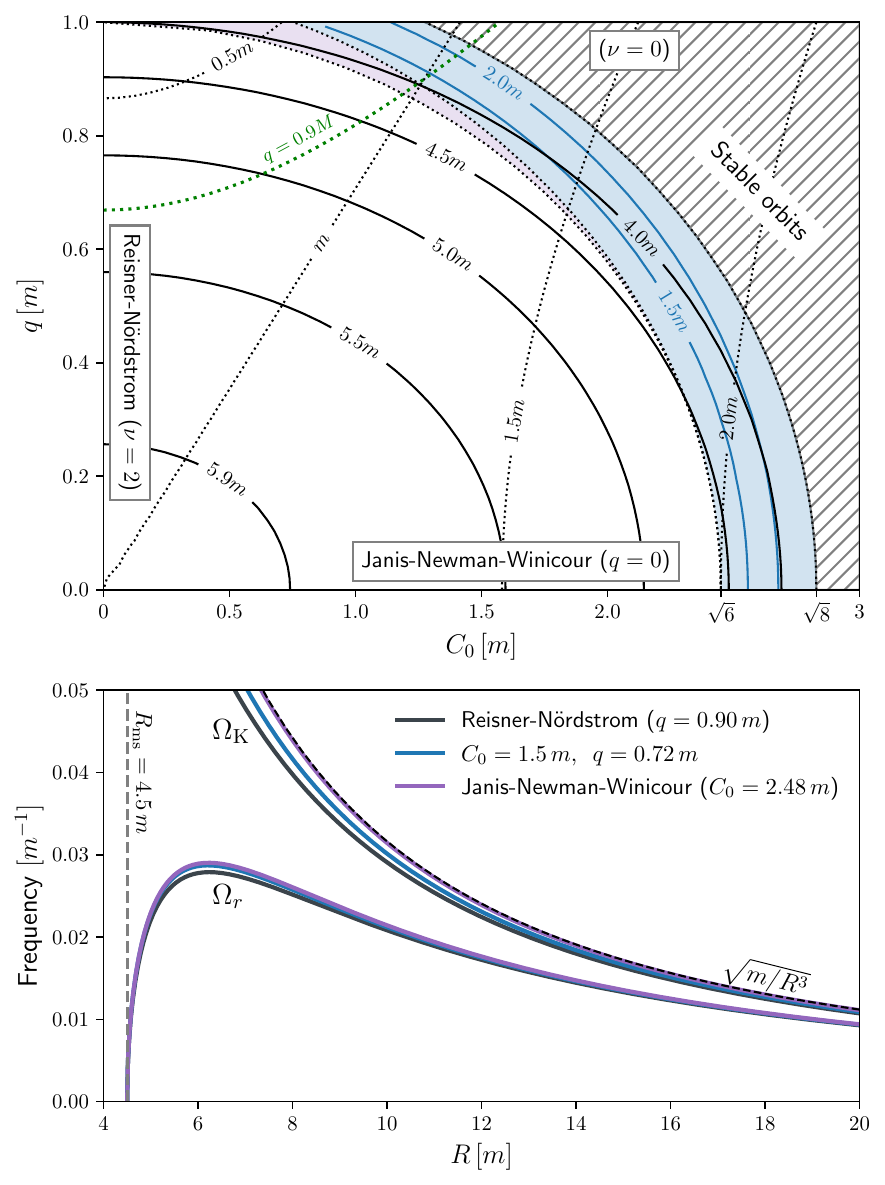}
    \caption{\textit{Top}: The parameter space of the JNWM spacetime for charge and scalar-field strength expressed in units of the ADM mass $m$. The solid black contours show positions of the (outer) marginally stable orbits measured by the metric function $R$ and normalized by $m$. The shaded region corresponds to parameters for which there are either two photon orbits and one marginally stable orbit (purple area) or two marginally stable and no photon orbits (blue area). The dashed area corresponds to parameters for which the singularity is entirely surrounded by stable orbits. The gray dotted contours show levels of constant JNWM mass $M$, and the green dotted curve shows the level of constant charge in terms of $M$. The cases shown in Fig.~\ref{fig:jnwm-l-omegas} correspond to 4 points on this curve. \textit{Bottom}: Radial profiles of the Keplerian angular frequency and radial epicyclic frequency for three JNWM spacetimes giving the same position of the marginally stable orbit $R_\mathrm{ms}=4.5\,m$. The frequencies are normalized by the ADM mass $m$ and the metric function $R$ normalized by $m$ is chosen as the radial coordinate.}
    \label{fig:jnwm-degeneracy}
\end{figure}

The qualitative behavior summarized above shows an interesting influence of the combined scalar and electromagnetic fields on the existence and stability of the circular orbital motion. The main effect of the scalar field is the change of the Keplerian angular momentum in the vicinity of the singularity when its strength exceeds $\sqrt{3/2}M$ (or when  $\nu<1$). The angular momentum is an increasing function of the radial coordinate in this region implying stability of the orbits. In addition, the radial epicyclic frequency rises when approaching the singularity suggesting an increasingly strong restoring force towards the unperturbed circular orbit. Consequently, only particles with vanishing angular momentum moving along strictly radial trajectories may reach the singularity. When $C_0>\sqrt{8/5}M$, the angular momentum is a monotonically increasing function of the radial coordinate in the entire space outside the singularity. In a narrow range of $C_0$ between $\sqrt{3/2}$ and $\sqrt{8/5}M$, there exists an intermediate region where the Keplerian circular orbits exist, but they are unstable with respect to the radial perturbations. 

The presence of the electromagnetic field does not change the reversal of the Keplerian angular momentum near the singularity. The critical value of the scalar field strength remains $\sqrt{3/2}M$ as in the JNW spacetime. However the charge shifts and extends the size of the intermediate region, where the Keplerian circular motion is unstable, as can be seen from the position and size of the blue region in the parameter space in  Fig.~\ref{fig:jnwm-parameter-space}. In addition, the charge brings up a new effect of the presence of two photon orbits that further enlarge the gap in the distribution of stable circular geodesics in the vicinity of the singularity. For a JNWM spacetime with charge and scalar field corresponding to the purple region of the parameter space, the Keplerian angular momentum has essentially two parts. The outer part shows qualitatively similar behavior to the RN or even Schwarzschild spacetime: with decreasing radii, initially stable Keplerian orbits become unstable after crossing the marginally stable orbit and finally impossible after the (outer) photon orbit is reached. Due to a combined influence of the scalar and electromagnetic field, a new region of stable circular geodesics appears just in the vicinity of the singularity reaching up to the inner photon orbit. Implications of this structure for accretion processes on such objects, as well as possible observational consequences of the presence of the two photon orbits in the spacetime, are discussed in Section~\ref{sec:conclusion}.

Finally, we comment on the shape of the parameter space when the scalar field strength and electric charge are normalized by the ADM mass $m$. The plot is shown in the upper panel of Fig~\ref{fig:jnwm-degeneracy} with contours showing locations of the marginally stable orbits $R_\mathrm{ms}$ measured in terms of the metric function $R$ and expressed in units of $m$ as well. Possible values of the normalized charge are limited to $(q/m)^2<1$, while $C_0/m$ is allowed to reach arbitrarily large values. Nevertheless, the unstable orbits may be found only when $C_0/m \leq \sqrt{8}$ (corresponding to $\mu\leq\sqrt{5}$ in the JNW spacetime). The region of two photon orbits is mostly pronounced for strong charge and weak scalar field. 

The elliptic shape of the contours of $R_\mathrm{ms}$ suggests that both types of fields push the marginally stable orbit closer to singularity and that a lack of the field of one type may be to some extent compensated by the field of the other type. This is shown in the bottom panel of Fig.~\ref{fig:jnwm-degeneracy}, where the orbital frequencies normalized by $m$ are plotted for three different values of the charge and scalar-field strength along the contour $R_\mathrm{ms}=4.5\,m$. Indeed the curves are almost indistinguishable. This suggests that even if the ADM mass could be well constrained and departures of orbital motion from the one predicted by Schwarzschild spacetime detected, it would be very hard to decide which of the two fields is responsible for measured deviations. The situation becomes different with qualitative change at $C_0=\sqrt{3/2}M$, when the scalar field is strong enough to allow the presence of another region of stable orbits in the vicinity of the singularity. Observational consequences of this effect are discussed in Sec~\ref{sec:conclusion}.

\section{Nonlinear Electrodynamics}
\label{sec:nonlinear-electrodynamics}
Various theories of nonlinear electrodynamics (NE) with different Lagrangians have been proposed \cite{Sorokin2021,  Boillat1970}. The most promising one is the Born-Infeld model (BI, \cite{BornInfeld}), which in the weak-field limit gives the linear Maxwell theory while the limit of strong field (considering only magnetic charge) implies the Lagrangian in the form $~ \sim \sqrt{F_{\mu \nu}F^{\mu \nu}}$ with $F_{\mu \nu}$ being the electromagnetic tensor.

In this section, we study the so-called ``square-root'' model with the Lagrangian given by $\angmom=-\sqrt{F}$. Details of obtaining the magnetic charge solution of this specific model can be found in \cite{Tahamtan2020}. The electromagnetic invariant $F=F_{\mu\nu}F^{\mu \nu}$ is assumed to read 
\begin{equation} 
    \label{F-invariant}
    F=\frac{2\,q^2_{m}}{R^4},
\end{equation}
where $q_\mathrm{m}$ is the magnetic charge. The corresponding spherically-symmetric solution of Einstein's equations with the square root model is then described by line element (\ref{ourmetric}) with the $f(r)$ and $R(r)$ metric functions given by the following simple expressions:
\begin{align}
    \label{f-NE}
    f(r) &= \alpha-\frac{2\,M}{r}, \\
    \label{R-NE}
    R(r) &= r.
\end{align}
Here, $\alpha=1-q_{m}\sqrt{2}$.
The Schwarzschild solution is recovered in the limit of $q_m=0$, corresponding to  $\alpha=1$, as should be expected in the absence of a source of the electromagnetic field. Obviously, for positive values of $\alpha$, $f(r)$ has one root, $r_0$, which can be considered an event horizon. The $\alpha$ parameter should be positive and is related to the solid angle deficit/excess (it can be bigger or smaller than unity depending on the sign of the magnetic charge).

Obviously, this spacetime described by the metric functions (\ref{f-NE}) and  (\ref{R-NE}) is not asymptotically flat. The $\alpha$ parameter is related to the solid angle deficit. The spacetime geometry is similar to the one outside the core of a so-called global monopole, a spacetime defect usually considered to be sourced by a self-coupling triplet of scalar fields whose original $O(3)$ symmetry is spontaneously broken to $U(1)$. The global monopole has been discussed in detail in the literature along with many applications; some of the original work can be found in \cite{Letelier-1979, Barriola, global1} and more recent work in \cite{Tahamtan-quantum:2014, Tahamtan-Boosting}. 

Moreover, although the square root model has a (magnetic) charge, the corresponding spacetime solution differs a lot from the linear Maxwell model, namely from the Reissner–Nordström solution. As we see from expression \eqref{f-NE}, the spacetime contains a single horizon, while the Reissner–Nordström solution has two of them. In addition, their global asymptotics disagree as well.

\subsection{Photon orbit}
Photons in NE obey the NE-Maxwell equations. This results in birefringence of the two propagating modes. The only exceptions are the standard Maxwell electrodynamics and BI. In the geometric optics approximation, rays of two linearly independent polarizations can be described by null geodesics in an effective geometry \cite{Novello:1999pg}. In restricted $\mathscr{L}(F)$ theories, there exists a polarization, for which the effective metric reduces to the spacetime metric. As we're interested only in the motion of massive particles, the optical space corresponding to their relativistic orbital dynamics is thus given precisely by the null geodesics of only the spacetime metric.

The radius of gyration follows from equation (\ref{eq:radius-of-gyration}). In the equatorial plane, we have
\begin{equation}
    \tilde{r} = \frac{r}{\sqrt{\alpha - 2M/r}}.
    \label{eq:nline-circumferential-radius}
\end{equation}
Then using expressions (\ref{eq:EKLK}), we find the squared energy and angular momentum on the Keplerian circular orbit per unit particle mass
\begin{align}
    \energy_\mathrm{K}^2 &= \frac{\left(\alpha - 2M/r\right)^2}{\alpha - 3M/r},
    \\
    \angmom_\mathrm{K}^2 &= \frac{Mr}{\alpha - 3M/r}\,.
\end{align}
Clearly, both quantities change signs at the radius
\begin{equation}
    r_\mathrm{ph} = \frac{3M}{\alpha}
\end{equation}
that corresponds to the photon orbit. The time-like circular geodesics are located outside this radius, at $r>r_\mathrm{ph}$, where both $\mathscr{L}_\mathrm{K}^2$ and $\mathscr{E}_\mathrm{K}^2$ are positive.

\subsection{Properties of the orbital motion}
\begin{figure}
    \includegraphics[width=0.47\textwidth]{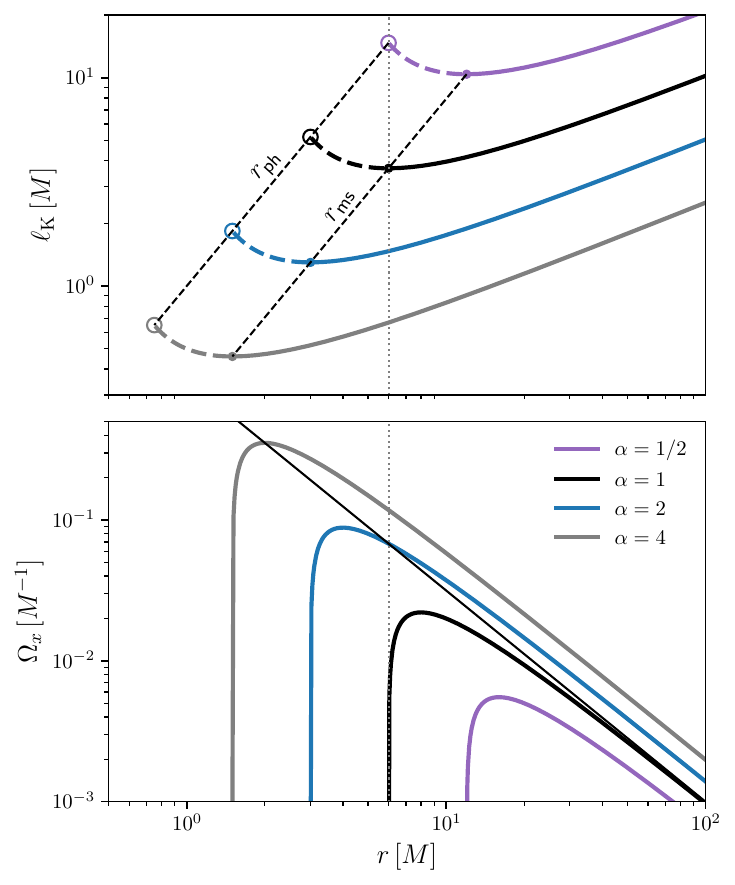}
    \caption{\textit{Top}: Keplerian specific angular momentum $\ell_\mathrm{K}$ for the square root Lagrangian for different values of the parameter $\alpha$, where $\alpha=1$ corresponds to Schwarzschild spacetime. The marginally-stable orbits are denoted by a cross, and the vertical dotted gray line marks $r=6M$. \textit{Bottom}: (1) $x=K$: The formula for the Keplerian orbital velocity $\Omega_\mathrm{K}$ for the square root Lagrangian \ref{OK-NE}, as a function of the radial coordinate, is independent of $\alpha$ and is displayed as a straight black line. (2) $x=r$: The radial epicyclic frequency $\Omega_r$ for different values of $\alpha$ is shown in terms of solid colored curves.}
    \label{fig:sqrt-lwK}
\end{figure}

The ratio of $\mathscr{L}_\mathrm{K}$ and $\mathscr{E}_\mathrm{K}$ gives the specific Keplerian angular momentum
\begin{equation}
    \ell_\mathrm{K}=\pm\frac{\sqrt{Mr}}{\alpha-\frac{2M}{r}}\,.
\end{equation}
As mentioned, $\alpha=1$ corresponds to the Schwarzschild case. As one can see from the plots shown in Fig.~\ref{fig:sqrt-lwK}, the Keplerian angular momenta behave similarly also for different values of $\alpha$.  The minimum of the angular momentum profile giving a position of the marginally stable orbit occurs at
\begin{equation}
    r_\mathrm{ms} = \frac{6M}{\alpha}.
\end{equation}
It is therefore just shifted by the same factor $1/\alpha$ as in the case of the photon orbit. As in the Schwarzschild case, the Keplerian orbits are unstable inside this radius. 

In order to obtain the Keplerian angular velocity for the NE solution, we use \eqref{OmegaK}, which gives us the same formula like in the Schwarzschild case,
\begin{equation}
    \label{OK-NE}
    \Omega_\mathrm{K}=\sqrt{\frac{M}{r^3}}\,.
\end{equation}
Finally, from (\ref{eq:Omegar}), we find the radial epicyclic frequency $\Omega_{r}$ to be
\begin{equation}
    \Omega_r=
    \Omega_\mathrm{K}\sqrt{\alpha - \frac{6M}{r}} =
    \frac{\sqrt{M(\alpha r - 6M)}}{r^2}.
    \label{eq:nline-omegar}
\end{equation}
Based on the above expression, we do not expect the behavior of $\Omega_r$ to differ much from that in the Schwarzschild case either, see Fig. \ref{fig:sqrt-lwK}. Note that contrary to the previous cases with a scalar field, the radial coordinate of our metric coincides with $R$ metric function. However, for different values of $\alpha$, even the asymptotic values of the radial epicyclic frequency differ from the Keplerian frequency by a factor of $\sqrt{\alpha}$. The reason is the solid angle deficit which becomes apparent when expressing the curvature radius, $\mathcal{R}$, and the geodesic radius, $\hat{r}$, in the optical space as functions of the radial coordinate $r$,
\begin{align}
    \mathcal{R} &= \frac{r}{\alpha - 3M/r},
    \label{eq:nline-curvature-radius}
    \\
    \hat{r} &= \int^r\frac{\mathrm{d}r}{\alpha - 2M/r}.
    \label{eq:nline-geodesic-radius}
\end{align}
The circumferential radius $r_\ast$ is given by expression (\ref{eq:nline-circumferential-radius}). Clearly, at large distances, both geodesic and curvature radii behave as $r/\alpha$, while the circumferential radius behaves as $r/\sqrt{\alpha}$ implying the circumference of the orbits to asymptotically follow $C = \sqrt{\alpha}\times(2\pi r_\ast)$. Therefore, the asymptotic solid-angle deficit is $\Delta\Omega \approx 4\pi(1 - \alpha)$. When expressions (\ref{eq:nline-circumferential-radius})--(\ref{eq:nline-geodesic-radius}) are used in the Abramowicz and Kluzniak formula (\ref{eq:Abramowicz-Kluzniak-radii}), we recover expression (\ref{eq:nline-omegar}). 

The equatorial cut of the optical space corresponding to the NE solution embedded in the three-dimensional Euclidean space has been already shown in Fig~\ref{fig:sqrt-embedding}. The asymptotically conical shape of the surface nicely illustrates the deficit of the solid angle and related absence of the asymptotical flatness. The photon orbit shown by the purple line is the only circular geodesic at the surface, corresponding to an infinite curvature radius. The areas of the stable and unstable orbits are indicated by the yellow and gray color, respectively. While in the region of the unstable orbits the curvature radius decreases with increasing $r$, it increases in the region of the stable orbits. The marginally stable orbit separating the two regions corresponds to the minimum of $\mathcal{R}(r)$ in agreement with equation (\ref{eq:Abramowicz-Kluzniak-radii}).

\section{Schwarzschild-Melvin spacetime}
\label{sec:magnetic-field}
Magnetic fields are significant on many different astrophysical scales, from the elusive intergalactic magnetic fields to those associated with active galactic nuclei and even stellar-mass compact objects.

One of the interesting exact solutions of the Einstein-Maxwell equations is the Schwarzschild-Melvin (static and axially symmetric) solution, a Schwarzschild black hole immersed in a strong external electromagnetic field. This solution has been generated by Ernst \cite{Ernst} utilizing a Harrison-type transformation \cite{Stephanietal:book} applied to the standard Schwarzschild metric. The  Schwarzschild-Melvin line element reads \cite{Marcello}
\begin{align}
    \label{Schw-Melvin}
    \d s^2 =& N\left[-f\,\d t^2+{f}^{-1}{\d r^2}+r^2 \d {\theta}^2 \right] 
    \nonumber \\
    &+\frac{1}{N}\,r^2\,{\sin^2\theta} \,\d\phi^2\,, 
\end{align}
where the functions $N$ and $f$ are given by
\begin{eqnarray}
    \label{f-N-Melvin}
    N&=&\left(1+\frac{B^2\,r^2{\sin{\theta}}^2}{4}\right)^2, \nonumber\\
    f&=&1-\frac{2\,M}{r}. \nonumber
\end{eqnarray} 
The constant $B$ determines the strength of the magnetic field. The black-hole horizon is located at $r=2M$, independently of the magnetic field strength. In the absence of the mass parameter, $M=0$, the resulting spacetime, which is referred to as the ``Bonnor--Melvin Universe'' \cite{Bonor-Melvin, Melvin1, Melvin2, Tahamtan-Melvin}, consists of parallel magnetic lines of force remaining in equilibrium under their mutual gravitational attraction. Since the magnetic field contributes to gravity even at large distances, neither Bonnor-Melvin, nor Schwarzschild-Melvin spacetimes are asymptotically flat. Recently it has been proven that the Bonnor-Melvin spacetime is stable under linear radial perturbation \cite{Tahamtan-Melvin}.   

In the next section, we study the effect of an external magnetic field and a lack of spherical symmetry on the photon surfaces and epicyclic frequencies for both the Bonnor--Melvin and Schwarzschild-Melvin spacetimes.

\begin{figure}
	\includegraphics[width=0.5\textwidth]{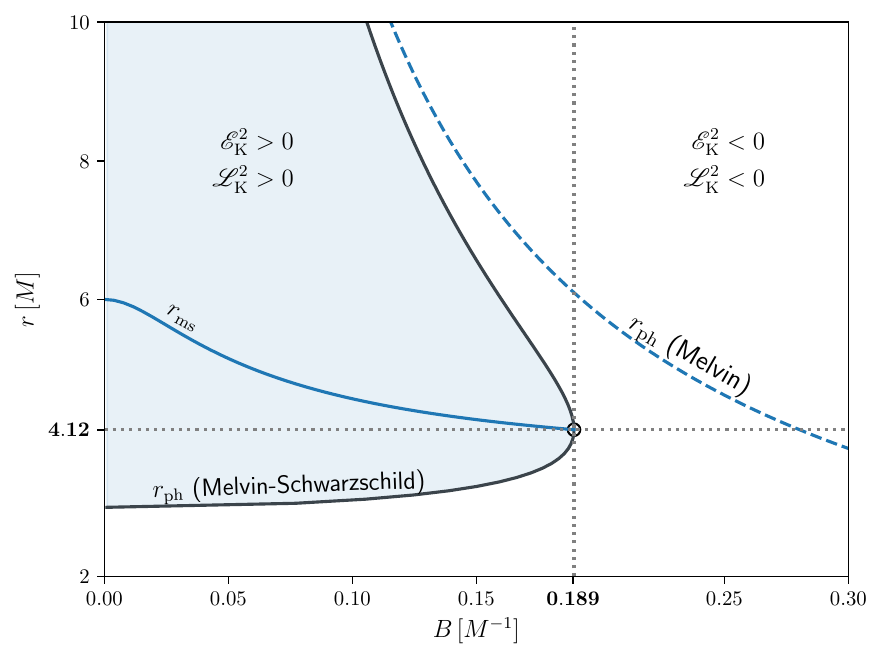}
     \caption{Positions of the circular photon orbits in theSchwarzschild-Melvin spacetime as a function of the magnetic field strength (black solid line). In the absence of the magnetic field ($B=0$), the spacetime coincides with the Schwarzschild solution, where there is a single photon orbit at $3M$. In the absence of the central black hole ($M=0$), the spacetime is described by the Bonnor-Melvin solution with a single photon orbit at $(2/\sqrt{3})B^{-1}$ (dashed blue line). For nonzero $M$ and $B$, there are two photon orbits whose radial positions are solutions of equation (\ref{eq:Melvin-rph}). Time-like circular geodesics exist only in the blue-shaded area for $B<B_\mathrm{ph}^\ast(M)\approx 0.189M^{-1}$. The solid blue line inside this region shows the position of the marginally stable orbit $r_\mathrm{ms}$ as a function of the magnetic field strength.}
    \label{fig:ms-photon-orbits}
\end{figure}

\subsection{Photon orbits}
First, we use equation (\ref{eq:radius-of-gyration}) to find the circumferential radius in the equatorial plane,
\begin{equation}
    \tilde{r} = \left(1-\frac{2M}{r}\right)^{-1/2}
    \left(1 + \frac{B^2r^2}{4}\right)^{-2} r.
    \label{eq:melwin-radius-of-gyration}
\end{equation}
When $B\neq 0$, it vanishes when $r\rightarrow\infty$ and when $r\rightarrow 0$. In addition, when $M\neq 0$, it diverges as $r\rightarrow 2M$. 

The squared energy and angular momentum at the Keplerian circular orbits per unit particle mass are given by expressions (\ref{eq:EKLK}). We obtain
\begin{equation}
    \energy_\mathrm{K}^2 = 
    \frac{\left(4-B^2r^2\right)\left(4+B^2r^2\right)^2\left(r-2M\right)^2}
    {32r\,\mathscr{D}(r)}
    \label{eq:Melvin-EK}
\end{equation}
and
\begin{equation}
    \angmom_\mathrm{K}^2 = 
    \frac{8r^2\left[4M+B^2r^2\left(2r-3M\right)\right]}
    {\left(4+B^2r^2\right)^2\mathscr{D}(r)}
    \label{eq:Melvin-LK}
\end{equation}
with
\begin{equation}
    \mathscr{D}(r) = -\frac{B^2r^2}{4}\left(3r-5M\right) + r - 3M.
    \label{eq:Melvin-Delta}
\end{equation}
Zeroes of $\mathscr{D}(r)$ correspond to photon orbits. Clearly, there is a single ``Schwarzschild'', unstable photon orbit at $r_\mathrm{ph,Schw} = 3M$ when there is no magnetic field present.  In the opposite limit, when $M=0$, there exists another, ``Melvin'', stable photon orbit located at $r_\mathrm{ph,Mel}=(2/\sqrt{3})B^{-1}$. For nonzero $M$ and $B$, positions of the photon orbits follow from solving the equation
\begin{equation}
    \left(MB\right)^2 = \frac{4M^2}{r^2}
    \frac{(r-3M)}{(3r-5M)}.
    \label{eq:Melvin-rph}
\end{equation}
The function on the right-hand side vanishes at $r=3M$ and as $r\rightarrow\infty$. In between, it reaches a maximum at $r=r_\mathrm{ph}^\ast$, $B=B_\mathrm{ph}^\ast$ with
\begin{align}
    r_\mathrm{ph}^\ast &= \frac{M}{3} \left(8 + \sqrt{19}\right)\approx4.12M\,,
    \nonumber
    \\
    MB_\mathrm{ph}^\ast &= \frac{2}{5} 
    \sqrt{\frac{1}{15} \left(169-38 \sqrt{19}\right)}\approx0.189\,.
    \nonumber
\end{align}
When $B<B_\mathrm{ph}^\ast$, the equation (\ref{eq:Melvin-rph}) has two solutions corresponding to two photon orbits. With increasing $B$, the inner ``Schwarzschild'' photon orbit moves toward higher radii. On the other hand, with increasing $M$, the outer ``Melvin'' orbit moves toward lower radii. The two photon orbits merge at $r_\mathrm{ph}^\ast$ when $B=B_\mathrm{ph}^\ast$. For a higher magnetic field no photon orbits exist. Time-like circular geodesics that correspond to positive $\energy_\mathrm{K}^2$ and $\angmom_\mathrm{K}^2$, exist only for $B<B_\mathrm{ph}^\ast$ in a limited range of radii between the two photon orbits. This situation is illustrated in Fig.~\ref{fig:ms-photon-orbits}. 

\subsection{Time-like circular orbits}
The specific Keplerian angular momentum is given by the ratio of $\angmom_\mathrm{K}$ and $\energy_\mathrm{K}$. Using expressions (\ref{eq:Melvin-EK}) and (\ref{eq:Melvin-LK}), we arrive at
\begin{align}
    \ell_\mathrm{K}^2 = 
    \frac{256\,r^3\left[B^2r^2\left(2r-3M\right)+4M\right]}
    {\left(4-B^2r^2\right)\left(4+B^2r^2\right)^4\left(r-2M\right)^2}.
    \label{lk-Melvin-Schw}
\end{align}

\begin{figure*}
	\includegraphics[width=\textwidth]{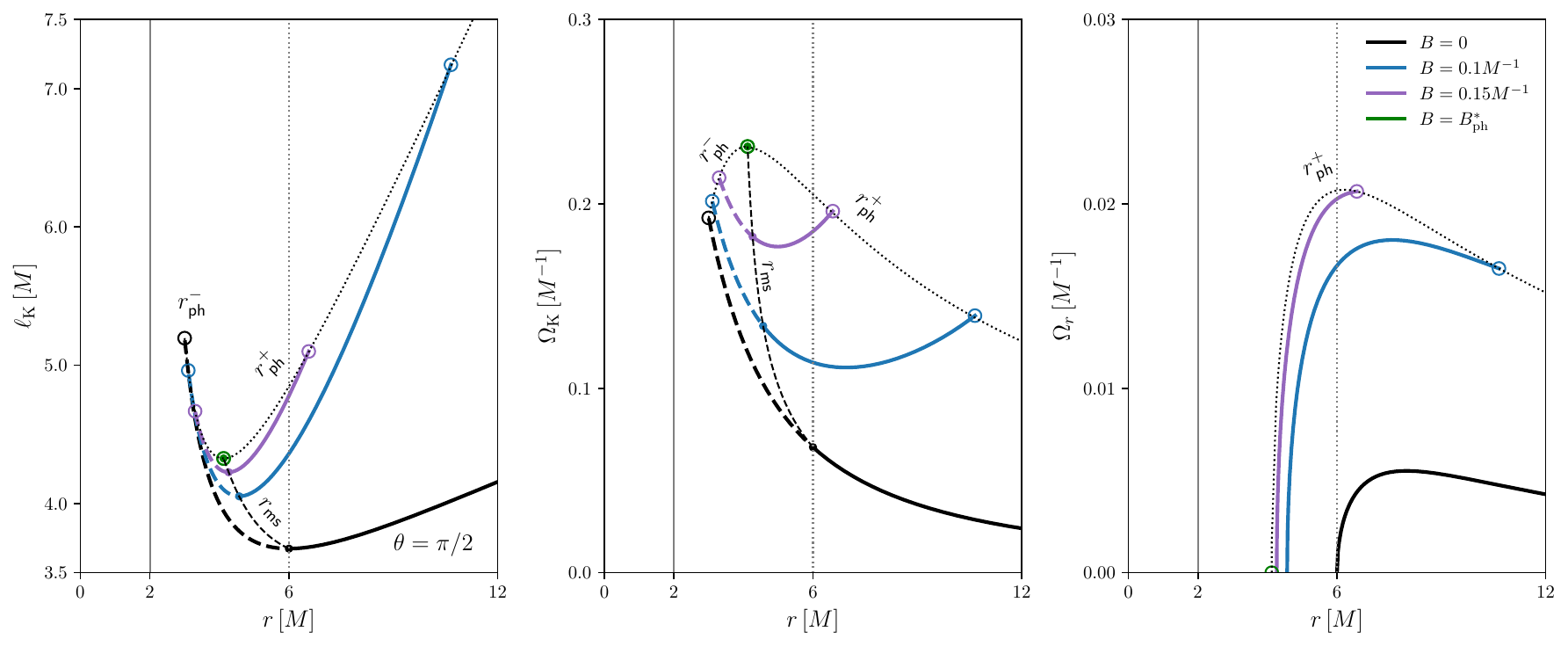}
    \\
     \includegraphics[width=\textwidth]{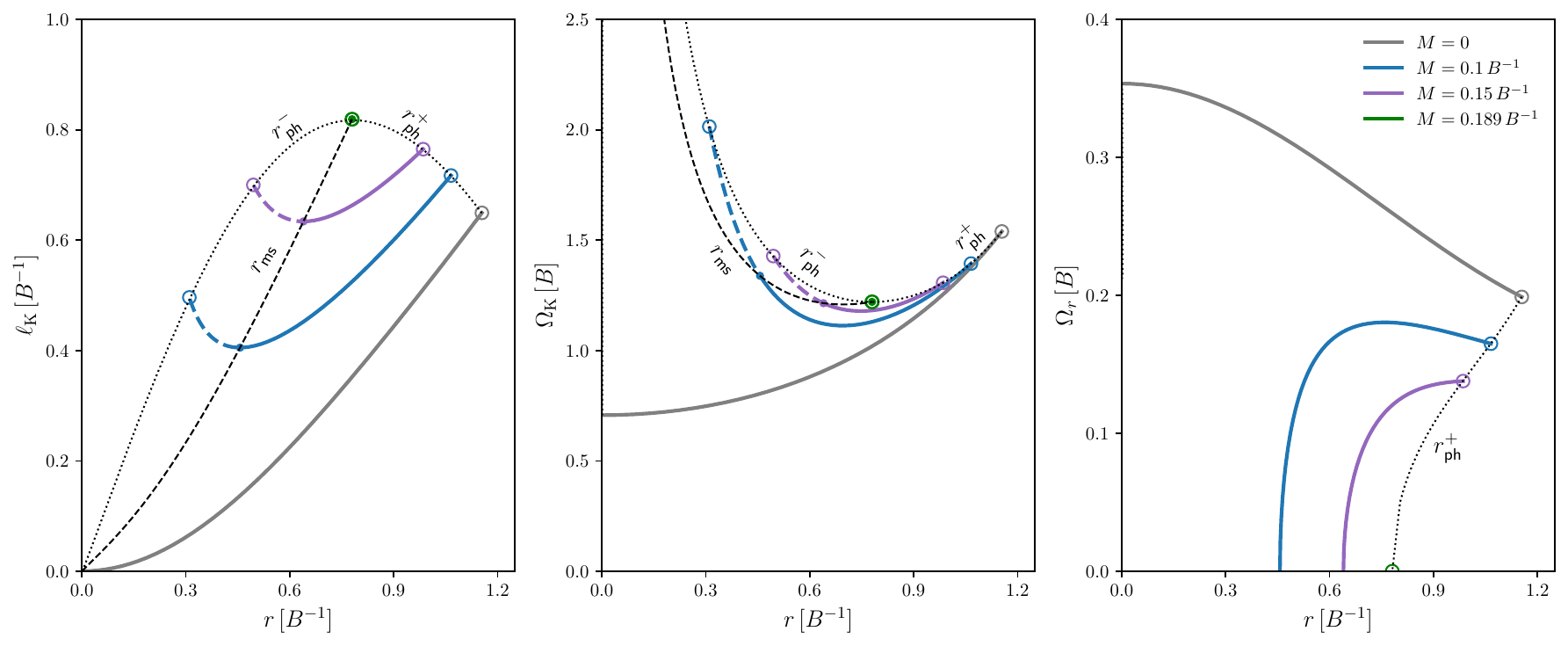}
    \caption{Keplerian angular momentum $\ell_\mathrm{K}$ (\textit{left} panels), Keplerian angular velocity $\Omega_\mathrm{K}$ (\textit{middle} panels) and the radial epicyclic frequency $\Omega_r$ (\textit{right} panels) as functions of the 
    radial coordinate $r$. The \textit{upper} panels shows several cases that correspond to a fixed black-hole mass and different magnetic field strengths. The vanishing magnetic field corresponds to Schwarzschild spacetime and is shown by the black line. The \textit{lower} panels show cases with a fixed magnetic field strength and different black-hole masses. The vanishing black-hole mass, shown by gray lines, thus corresponds to the Bonnor-Melvin spacetime. The dashed and dotted lines trace angular momenta and frequencies at the marginally-stable and photon orbits.}  
    \label{fig:schw-melvin}
\end{figure*}

Profiles of Keplerian angular momentum for different values of magnetic field and black hole mass are shown in the left panels of Fig.~\ref{fig:schw-melvin}. While the upper panel shows a sequence of curves for a fixed black-hole mass and magnetic field in the range of $0\leq B\leq B_\mathrm{ph}^\ast$ (approach from the ``Schwarzschild'' side), the lower panel shows a sequence of curves for fixed magnetic field and different masses in the range of $0\leq M\leq 1/B_\mathrm{ph}^\ast$ (approach from the ``Melvin'' side). The black curve in the upper panel corresponds to Schwarzschild spacetime where circular Keplerian motion is possible everywhere above the photon orbit at $3M$, being stable above marginally stable orbit at $6M$. The global magnetic field brings an upper bound on the radius of the time-like circular geodesics, the outer stable photon orbit, and slightly modifies positions of the inner photon orbit and the marginally stable orbit. As it increases, the outer photon orbit progressively moves in, while the inner one moves out, keeping the marginally stable orbit in between. The region of Keplerian orbits gradually shrinks and at $B=B_\mathrm{ph}^\ast$ disappears completely. Nevertheless, before that the region always contains both stable and unstable orbits, and the marginally stable orbit and the two photon orbits merge all together at $r=r_\mathrm{ph}^\ast$ when $B=B_\mathrm{ph}^\ast$. 

The lower panel shows Keplerian angular momenta for the Schwarzschild-Melvin spacetime when approached from the `Melvin side', i.e. keeping $B$ constant and gradually increasing the black hole mass. Consequently, both $r$ and $\ell_\mathrm{K}$ are normalized by $1/B$.  In the case of the Bonnor-Melvin universe ($M=0$), the angular momentum increases monotonically from zero at the origin with increasing $r$. The origin itself is a regular point entirely surrounded by stable circular orbits. This situation changes dramatically when $M>0$ as the gravity of the central black hole introduces another photon orbit that limits the radii of time-like circular geodesics and changes the stability of orbits in its neighborhood. The Keplerian angular momentum is not a monotonic function of the radius anymore, the region of unstable orbits corresponds to initially decreasing parts of the $\ell_\mathrm{K}$ curves. Again, the two photon orbits merge with the marginally stable orbit $r=r_\mathrm{ph}^\ast$ when $B=B_\mathrm{ph}^\ast$. 

\begin{figure}
    \centering
    \includegraphics[width=0.45\textwidth]{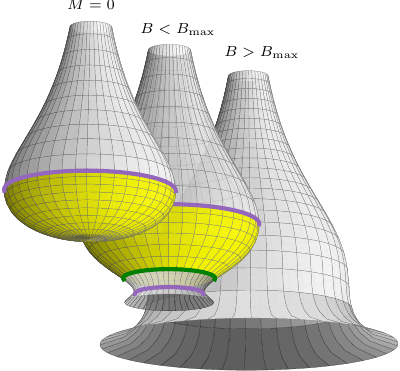}
    \caption{Embedding diagrams of the optical spaces corresponding to the Schwarzschild-Melvin spacetimes with different masses and magnetic fields. Both, the radial coordinate $r$ and the geodesic radius $\hat{r}$ increase upward. \textit{Front}: the Bonnor-Melvin universe that coincides with the Schwarzschild-Melvin spacetime with $M=0$. The spacetime hosts stable Keplerian orbits near the origin in the yellow-shaded region, up to the radius of the stable photon orbit shown by the purple line. \textit{Middle}: the Schwarzschild-Melvin spacetime with spacetime with $B<B_\mathrm{ph}^\ast$. This spacetime hosts stable Keplerian orbits in between the marginally stable orbit (green line) and the outer, stable photon orbit (the upper purple line). The inner unstable photon orbit closer to black hole is shown by the purple line at the bottom of the embedded surface. \textit{Back}: the Schwarzschild-Melvin spacetime with $B>B_\mathrm{ph}^\ast$. Such a spacetime does not host any time-like circular geodesics.}
   \label{fig:embedding-Melvin}
\end{figure}

Using expression (\ref{eq:melwin-radius-of-gyration}) and formula (\ref{eq:ell-Omega-relation}), we find the expression for the Keplerian angular frequency,
\begin{align}
    \Omega_\mathrm{K}^2 = \frac{\left(4+B^2r^2\right)^4
    \left[4M + B^2r^2\left(2r-3M\right)\right]}{256\,r^3\left(4-B^2r^2\right)}\,.     
    \label{Ok-Melvin-Schw1}
\end{align}
In the $B=0$ limit, we recover the Schwarzschild formula, $\Omega_\mathrm{K}^2=M/r^3$. In the opposite limit of $M=0$, we obtain 
\begin{equation}
    \Omega_\mathrm{K}^2 = 
    \frac{B^2}{128}\frac{\left(4 + B^2r^2\right)^4}{\left(4-B^2r^2\right)}\,,
\end{equation}
the expression for the angular frequency at the circular time-like geodesics in the Melvin universe. Similar limits also describe the behavior of $\Omega_\mathrm{K}$ at small and large radii. While at small radii the gravitational attraction of the black hole dominates, and we find approximately $\Omega_\mathrm{K}\sim\sqrt{M/r^3}$, the dynamics at large distances is mostly governed by gravity induced by the magnetic field.  Plots displaying $\Omega_\mathrm{K}$ as a function of $r$ are shown in the middle panels of Fig.~\ref{fig:schw-melvin}. Again, the black curves in the upper and lower panels correspond to the Schwarzschild and Bonnor-Melvin cases. Interplay between the black-hole gravity and magnetic field discussed above leads to a non-monotonic profile of the Keplerian angular velocity.

\subsection{Epicyclic frequencies}
The presence of the large-scale magnetic field breaks down the spherical symmetry of the Schwarzschild-Melvin spacetime. Therefore, the vertical epicyclic frequency, $\Omega_\theta$,  is no longer equal to the Keplerian angular frequency $\Omega_\mathrm{K}$ as it was in all the previous cases. Using expression (\ref{eq:Omegax}) for $x=\theta$, we obtain a very simple result
\begin{equation}
    \Omega_{\theta}=\sqrt{\frac{M}{r^3}}.
\end{equation}

Similarly, the radial epicyclic frequency $\Omega_r$ follows from \eqref{eq:Omegax} with $r=x$, 
\begin{eqnarray}
    \Omega_r&=&\frac{1}{r^2(B^2r^2+4)^2\sqrt{4-B^2r^2}} \times \left\{ \right. \nonumber \\
    && \left. M^2(30B^6r^6-200B^4r^4+672B^2r^2-384) \right.\nonumber \\
    &&\left.-M(37B^6r^7
    -204B^4r^5+624B^2r^3-64r) \right.\nonumber \\
    &&\left.+4B^2r^4(3B^4r^4-12B^2r^2+32)
    \right\}^{\frac{1}{2}}.
\end{eqnarray}

Interestingly, the vertical epicyclic motion does not feel any influence of the magnetic-field gravity. The behavior of $\Omega_{r}$ is shown in the right panels of Fig.~\ref{fig:schw-melvin}. Again, the black curves in the upper and lower panels correspond to the Schwarzschild and Bonnor-Melvin cases. In the latter case, the origin is a regular point surrounded by stable time-like circular orbits and the radial epicyclic frequency has a finite limit. As soon as the black hole gravity is introduced and the stable Keplerian orbits exist only outside the marginally stable orbit, where the epicyclic frequency vanishes. The frequency is reduced even in the stable region, which itself also shrinks a bit from the outer side. The Melvin universe differs from the Schwarzschild spacetime in that instead of having only a maximum, it has both, a minimum and a maximum. On the other hand, when the black hole mass is fixed the magnetic field increases the stability of Keplerian orbits near the black hole. The radial epicyclic frequency increases due to the gravity of the magnetic field and the marginally stable orbit slightly moves in. On the other side, the magnetic field gravity forbids time-like geodesics at large distances as noted earlier.

The existence and stability of Keplerian orbits can be easily understood from the embedding diagram of the optical spaces shown in Fig.~\ref{fig:embedding-Melvin}.  The figure shows geometries equatorial cuts of the Bonor-Melvin universe ($M=0$) and two Schwarzshild-Melwin spacetimes with $B<B_\mathrm{ph}^\ast$ and $B>B_\mathrm{ph}^\ast$. In the plot, the geodesic radius $\hat{r}$ is measured on the embedded surface in the upward direction while the circumferential radius $\tilde{r}$ corresponds to the distance from the symmetry axis of the surface. The time-like circular geodesics exist only in regions where the circumferential radius increases with the geodesic radius. This happens below the stable photon orbit in the Bonnor-Melvin case and in between the two photon orbits in the Schwarzschild-Melvin case with $B<B_\mathrm{ph}^\ast$. For the Schwarzschild-Melvin case with $B>B_\mathrm{ph}^\ast$, $\tilde{r}$ is a decreasing function of $\hat{r}$ everywhere outside the black hole horizon. The stability of the orbits follows from the behavior of the curvature radius $\mathcal{R}$ -- when it increases with increasing $\hat{r}$, the orbit is stable. In the case of the Bonnor-Melvin spacetime, the origin corresponds to $\mathcal{R}=0$. On the other hand, the curvature radius of the photon orbit is infinite. In between both $\tilde{r}$ and $\mathcal{R}$ increase with $\hat{r}$ and the Keplerian orbits are thus stable.

\section{Discussion and conclusions}
\label{sec:conclusion}
We have explored the behavior of epicyclic frequencies in several static spacetimes, including those representing naked singularities and black holes, where scalar electromagnetic fields are important sources of gravity. Our analysis provides several insights into the dynamics of orbital motion of uncharged test particles and its relationship to spacetime geometry.

For the JNW and JNWM solutions studied in Secs.~\ref{sec:scalar-field} and \ref{sec:Einstein-Maxwell-scalar-field}, we have demonstrated that a strong scalar field is the main ingredient determining not only the nature of the central object but also the existence and stability of time-like circular orbits in its vicinity. Although even an arbitrarily weak scalar field removes horizons around central singularities, for low strengths the situation looks similar to the corresponding black hole solution without it. In particular, for a weak scalar field, the singularity is still surrounded by a region where only space-like circular geodesics may exist. Above the unstable photon orbit, the circular geodesics become time-like but remain unstable. Stable Keplerian orbits exist only outside the marginally stable orbit, which is located farther away. There is only one photon orbit and one marginally stable orbit in these spacetimes. With increasing scalar field strength, both of them move inward; nevertheless, the same effect might also be achieved by increasing other types of fields (e.g., by increasing the electric charge, as in the case of the JNWM solution demonstrated in Sec.~\ref{sec:Einstein-Maxwell-scalar-field}, see Fig.~\ref{fig:jnwm-degeneracy}).

An important qualitative change occurs when the scalar field strength reaches the critical value $C_0=\sqrt{3/2}M$. In the case of the JNW spacetime, the photon orbit reaches the central singularity, but the region of unstable Keplerian orbits still persists up to $C_0=\sqrt{8/4}M$, being limited by another marginally stable orbit that appears near the singularity. In the case of the JNWM spacetime, the spacetime starts to host another photon orbit, and in that case, even the region of space-like circular geodesics persists for higher scalar field strengths. Nevertheless, in both cases, the naked singularity becomes surrounded by a narrow region of stable time-like circular geodesics, where the Keplerian angular momentum gradually decreases to zero. This change might have important consequences for accretion processes on such objects, as will be shown in a separate subsection.

In subsequent two sections, we dealt with orbital motion in spacetimes with a strong electromagnetic field. The particle circular orbits and the radial epicyclic frequency have been explored in the famous square-root model of nonlinear electrodynamics in Section~\ref{sec:nonlinear-electrodynamics}. The effects of a very strong ordered magnetic field on particle orbits have been investigated in Section~\ref{sec:magnetic-field}. None of these two spacetimes is optically flat. In addition, the latter one is neither spherically symmetric, resulting in a difference between orbital and vertical epicyclic frequency.

\subsection{Optical geometry}
For a qualitative understanding of properties of the orbital motion, a view offered by optical geometry introduced by Abramowicz, Carter and Lasota \cite{Abramowicz+Carter+Lasota1988} has proven to be particularly useful, as it allows one to identify key purely geometrical quantities characterizing the orbit.

The concept of optical geometry provides a natural explanation for degeneracy between the Keplerian and radial epicyclic frequency at large distances in asymptotically flat spacetimes. Far from the central object, the circumferential, curvature, and geodesic radii of orbits measured in almost flat optical spaces nearly coincide, $\tilde{r} \approx \mathcal{R} \approx \hat{r}$, giving equality of both frequencies in expression (\ref{eq:Abramowicz-Kluzniak-radii}). One example of a spacetime that is not asymptotically flat is provided by the square-root model of nonlinear electrodynamics studied in Sec.~\ref{sec:nonlinear-electrodynamics}. The asymptotically constant factor between the circumferential radius and the other two, $\tilde{r}/\mathcal{R} \approx \tilde{r}/\hat{r} \approx \sqrt{\alpha}$, is directly linked to the solid-angle deficit $\Delta\Omega \approx 4\pi(1-\alpha)$ of the spacetime at large distances and naturally explains why $\Omega_r \approx \sqrt{\alpha}\Omega_\mathrm{K}$. Another example is represented by the Schwarzschild-Melvin spacetimes, where the circular time-like geodesics are limited by the outer photon-orbit if it exists at all.

\begin{figure}
   \includegraphics[width=0.49\textwidth]{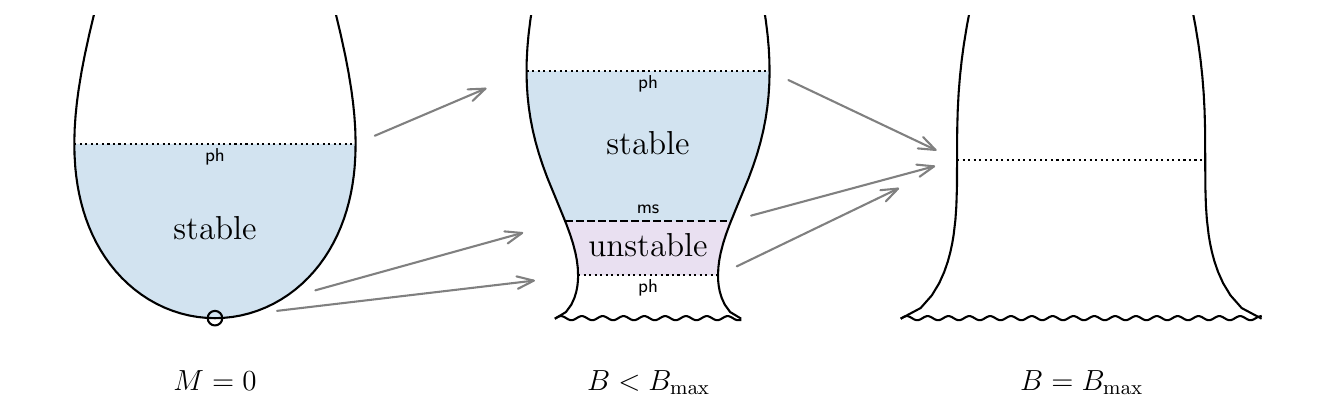}\\
   \includegraphics[width=0.49\textwidth]{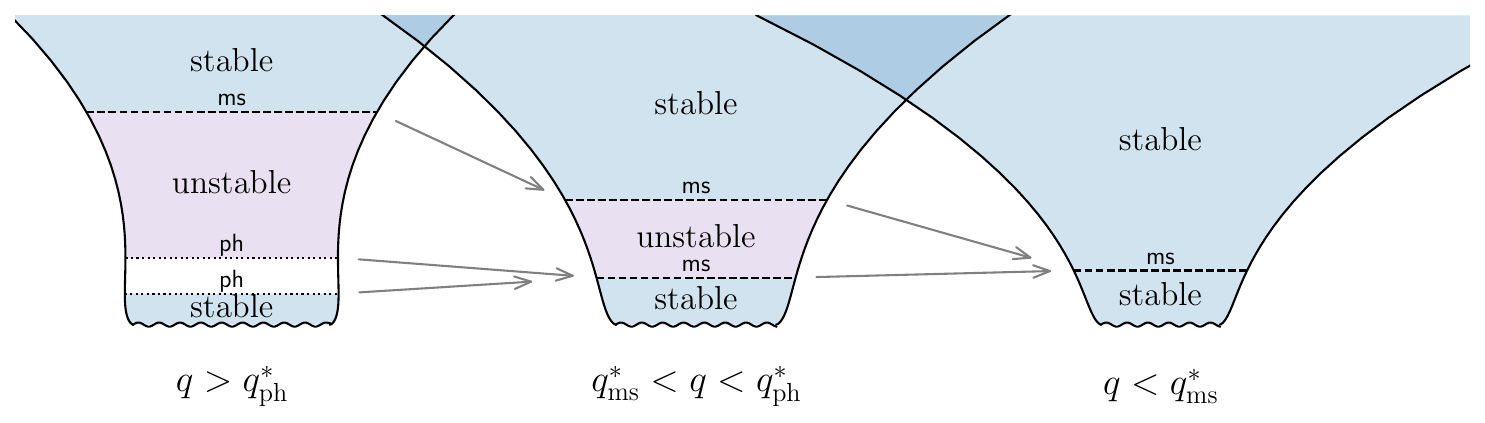}   
   \caption{Relation between properties of Keplerian circular motion and behavior of circumferential ($\tilde{r}$) and curvature ($\mathcal{R}$) radii of the orbits with increasing geodesics radius ($\hat{r}$) shown in the embedding diagrams of the Schwarzschild-Melvin (\textit{top}) and JNWM (\textit{bottom}) spacetime. In all cases both the geodesic radius and radial coordinate increase upward. While the geodesic radius $\hat{R}$ is measured on the embedded surface, the circumferential radius corresponds to a horizontal distance from the symmetry axis. Keplerian orbits exist in regions where the circumferential radius increases with increasing geodesic radius. At the same time, the orbit is stable or unstable when $\mathcal{R}$ decreases or increases with increasing $\hat{r}$. The top panel shows a sequence of embedding diagram corresponding to the Schwarzschild-Melvin spacetimes with increasing black-hole mass. The lower panel shows a similar sequence for the JNWM spacetimes with $C_0=1.3M$ and decreasing charge. The blue/purple-shaded areas indicate regions of stable/unstable orbits, while unshaded areas are regions where only space-like circular geodesic exist. The wavy lines at bottoms of the embedding diagram show boundary of the embeddable domains.}
   \label{fig:embedded_orbits}
\end{figure}

Likewise, this concept is useful for understanding behavior of photon and marginally stable orbits when parameters of the spacetime continuously change. The regions of spacetime where time-like circular orbits may exist are bounded by photon circular geodesics that correspond to extremes of the circumferential radius $\tilde{r}$ as a function of the geodesic radius $\hat{r}$. In a similar way, the regions of time-like stable circular orbits are separated by the marginally stable orbits that correspond to extremes of the curvature radius $\mathcal{R}(\hat{r})$. While a minimum of the circumferential radius reflects the presence of an unstable photon orbit, its maximum points to the presence of a stable photon orbit. Stable photon orbits separate regions of particle stable circular motion, where both $\tilde{r}$ and $\mathcal{R}$ increase with $\hat{r}$, from those where particle circular orbits are not possible because all circular geodesics are space-like, corresponding to decreasing $\tilde{r}(\hat{r})$. Likewise, the unstable photon orbits separate regions of particle unstable circular orbits [increasing $\tilde{r}(\hat{r})$ and decreasing $\mathcal{R}(\hat{r})$] from those where only space-like circular geodesics exist [decreasing $\tilde{r}(\hat{r})$]. Consequently, in between two neighboring photon orbits there is either a region where Keplerian orbits do not exist (as in the case of the JNWM spacetime with $C_0\geq\sqrt{3/2}M$ and $q>q_\mathrm{ph}^\ast$), or there are both regions of stable and unstable orbits separated by a marginally stable orbit in between (as in the case of the Schwarzschild-Melvin spacetime with $M>0$ and $B<B_\mathrm{ph}^\ast$).

Fig.~11 illustrates these points on a sequence of embedding diagrams corresponding to two spacetimes studied in the paper. The top panel shows the case of the Schwarzschild-Melvin solutions with increasing black hole mass, starting with the Melvin universe ($M=0$) with a single stable photon orbit separating the region of stable Keplerian motion near the origin and the region where circular geodesic motion is not possible further away. The curvature radius increases with increasing $\hat{r}$ from zero at the origin (shown as a black circle) up to $\mathcal{R}=\infty$ at the photon orbit. Above the photon orbit, $\mathcal{R}$ starts to increase again starting from $\mathcal{R}=-\infty$. As soon as $M$ becomes nonzero, the initially spherical shape of the embedding diagram gets perturbed by a 'throat' of dimension $\sim M$ near the origin, where black hole gravity dominates that of the magnetic field. Consequently, another, unstable photon orbit emerges at the minimum of the circumferential radius at $r=3M$, together with a region of unstable orbits between $3M$ and $6M$. As $M$ further increases, the region of black hole dominance gets larger, pushing the inner photon orbit towards the outer one, squeezing the regions of stable and unstable particle circular orbits as well as the location of the marginally stable orbit between them. Finally, at $B=B_\mathrm{ph}^\ast(M)$, when two photon orbits and the marginally stable orbit merge at the same location, spacetime becomes hostile for Keplerian circular motion.

The bottom panel of Fig.~\ref{fig:embedded_orbits} shows a sequence of the JNWM solutions with $C_0=1.3M$ and decreasing charge from $q>q_\mathrm{ph}^\ast$ down to $q<q_\mathrm{ms}^\ast$. Contrary to the previous case, the spacetime initially contains the inner unstable photon orbit, surrounded by the outer stable one. The two orbits separate the region where time-like circular geodesics do not exist from the innermost region with stable orbits and the outer region of both stable and unstable orbits. The circumferential radius is a decreasing function of $\hat{r}$ between the photon orbits, while it increases everywhere else. When $q$ reaches $q_\mathrm{ph}^\ast$, the two photon orbits merge, and $\tilde{r}(\hat{r})$ becomes monotonically increasing everywhere, except for the merging point, where the derivative $\dd\tilde{r}/\dd\hat{r}$ vanishes and $|\mathcal{R}|$ is infinite. After merging, the two photon orbits form a marginally stable orbit separating regions of stable Keplerian geodesics that were initially inside the inner photon orbit and unstable geodesics that were outside the outer one. The circumferential radius is monotonically increasing throughout the spacetime, but the curvature radius is still a decreasing function of $r$ in between the marginally stable orbits. Decreasing the charge down to $q=q_\mathrm{ms}^\ast$, these two orbits also merge, leaving the spacetime entirely stable for Keplerian orbital motion with both $\tilde{r}$ and $\mathcal{R}$ being increasing functions of $\hat{r}$.

\subsection{Impact of scalar field on accretion process}
As we have shown in Secs.~\ref{sec:scalar-field} and \ref{sec:Einstein-Maxwell-scalar-field}, the orbital dynamics is for a weak scalar field ($C_0<\sqrt{3/2}M$) qualitatively similar to the one in Schwarzschild spacetime. The spacetime singularity, with horizon or not, is surrounded by an unstable photon sphere followed by the marginally stable orbit, outside of which the  Keplerian circular orbits are stable. If the object accretes matter from its surroundings in a thin accretion disk, where the accreted material gradually losses its angular momentum and converts its mechanical energy to radiation, disk inner edge is located near the marginally stable orbit, where the Keplerian angular momentum takes its minimal value. Inside this radius, the angular momentum is already low enough to allow matter to reach the central object along freely-falling trajectories. 

\begin{figure}
   \includegraphics[width=0.49\textwidth]{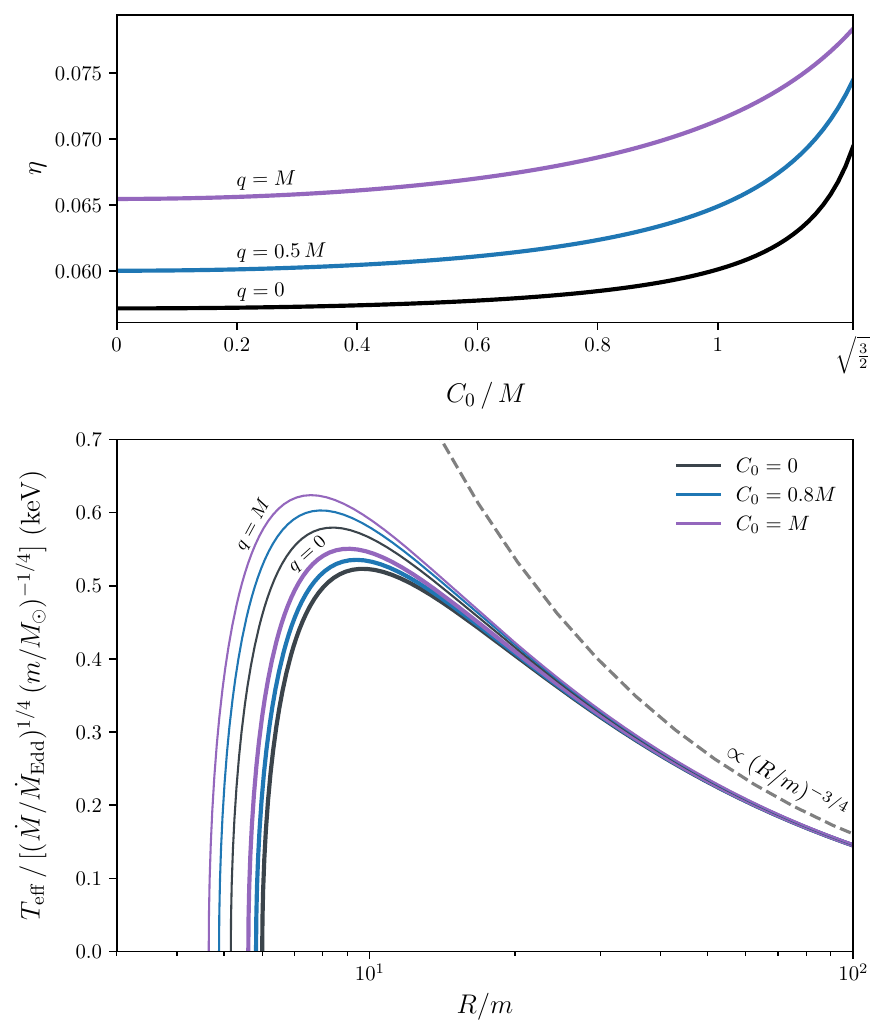}
   \caption{\textit{Top:} Efficiency of a thin accretion disk in the JNWM spacetime as a function of the scalar-field strength and several values of the electric charge. The value $C_0=0$ characterizes Reisner-N\"{o}rdstrom black hole, higher values correspond to naked singularities. The overall properties are qualitatively similar as in the Schwarzschild spacetime as long as $C_0<\sqrt{3/2}M$; the disk terminates at the marginally-stable orbit, whose position is only slightly modified by the scalar and electromagnetic fields. Since both fields tend to push its location closer to the singularity, the disk converts larger amount of the particle mechanical energy to radiation. This effect is also seen in the \textit{bottom} panel showing radial profiles of the disk effective temperature corresponding to three scalar field strengths (indicated by colors) and two values of the charge (indicated by line thickness). The temperature is normalized to the Eddington accretion rate and to one solar mass for the ADM mass of the central object. The metric function $R$ has been chosen as the radial coordinate. The proportionality $T_\mathrm{eff}\propto (R/m)^{-3/4}$ following from Newtonian calculations is shown by the dashed line.}
   \label{fig:jnwm-accretion}
\end{figure}

In the standard thin-disk scenario \cite{Shakura+Sunyaev1973, Novikov+Thorne1973}, the accretion flow follows nearly geodesic lines. Far from the central object, where the Keplerian circular orbits are stable, the trajectories of the particles are approximately Keplerian circular geodesics with additional small subsonic radial velocity, caused by outward angular-momentum transport due to turbulent stresses and energy losses converted to emerging radiation. On the other hand, near the central object, where the Keplerian circular orbits are unstable, the matter spirals inward, closely following non-circular geodesics with a specific energy and angular momentum corresponding to the marginally stable orbit, $\energy = \energy_\mathrm{K}(r_\mathrm{ms})$, $\angmom = \angmom_\mathrm{K}(r_\mathrm{ms})$. Transition between the two regimes is assumed to occur exactly at the marginally stable orbit, neglecting influence of additional stresses (pressure and turbulent stresses) that may push it slightly away from this location \cite{Abramowicz+2010, Lasota+Abramowicz2024}. 

\begin{figure*}
   \includegraphics[width=\textwidth]{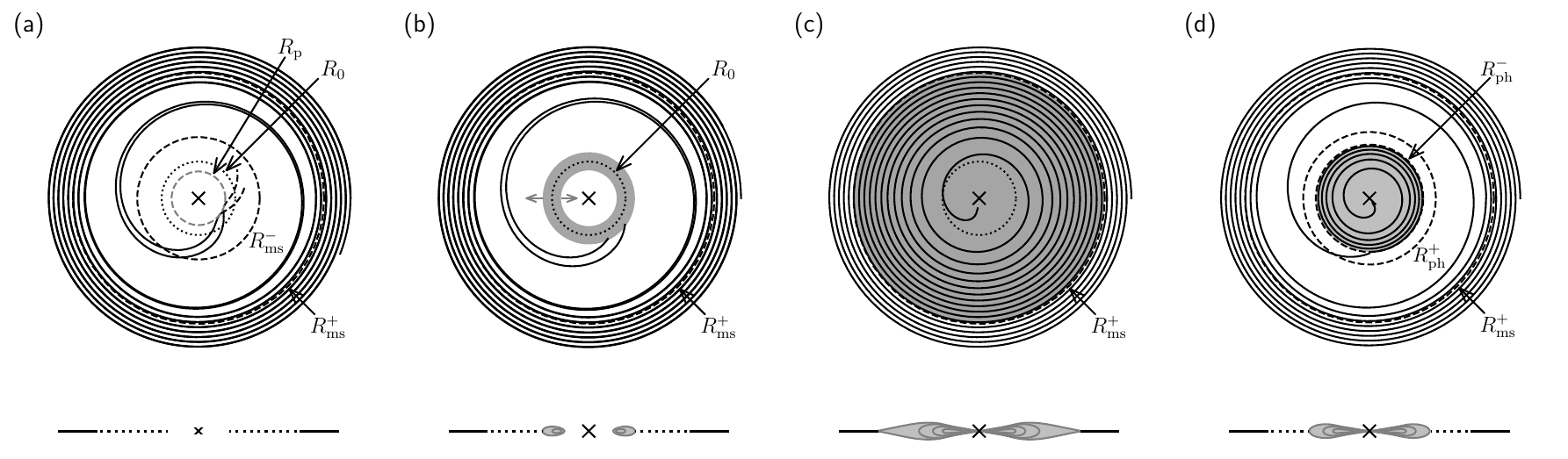}
   \caption{Accretion scenario in the case of the JNW and JNWM spacetimes, whose parameters give a non-monotonic behavior of the Keplerian angular momentum $\ell_\mathrm{K}(r)$, vanishing at the singularity. Panels (a)-(c) correspond to the JNW spacetime, the radial coordinate is the metric function $R$. (a) The flow initially plunges in from the outer marginally stable orbit $R_\mathrm{ms}^{+}$, being reflected by the infinite effective potential barrier near the singularity and colliding with the inflowing material. (b) As the kinetic energy of the radial motion dissipates, the matter settles down at the Keplerian stable circular orbit at $R_0$, whose angular momentum matches that at $R_\mathrm{ms}^{+}$ and develops a pressure-supported toroidal structure. (c) The torus is gradually spread as more matter is being accumulated and the angular momentum being transported from the inner to outer parts. Eventually, it fills the gap and connects to the outer accretion disk. The matter then follows a quasi-circular orbit everywhere outside the singularity. (d) In the case of the JNWM spacetime with $q>q_\mathrm{ph}^\ast$, the size of the torus is limited by the inner photon orbit $R_\mathrm{ph}^{-}$, outside of which even non-geodesic circular motion is not permitted. Accretion flow is in this case necessarily interrupted by a plunging region where the flow is diluted and has a significant radial velocity.}
   \label{fig:jnw-plunging}
\end{figure*}

Most of the radiation is released in the disk region outside the marginally stable orbit. Since the specific angular momentum is sub-Keplerian everywhere in the plunging region as $\ell_\mathrm{K}(r_\mathrm{ms}) < \ell_\mathrm{K}(r)$, the matter can reach the central object without need of additional energy and angular momentum losses. Particles therefore reach the central object with energy $\energy\approx\energy_\mathrm{K}(r_\mathrm{ms})$ and the plunging region  contributes negligibly to the total radiation produced during accretion. The flow in the plunging region is very diluted compared to the disk region and the radial velocity becomes supersonic almost immediately after leaving the marginally-stable orbit, not only due to increase of radial velocity but also due to sharp decrease of density and thus the sound speed of the fluid.  

The accretion efficiency is defined as the fraction of the accreted energy that is converted to radiation. Since for thin accretion disk the mechanical energy is the main contribution to the total energy of the flow and other forms (internal and magnetic energies) are assumed to be negligible, the efficiency can be estimated by a difference of the test particle energies at the inner and outer edge of the disk, $\eta=\energy_\mathrm{K}(r_\mathrm{out}) - \energy_\mathrm{K}(r_\mathrm{in})$. Assuming that the outer edge is at large radii, $\energy_\mathrm{K}(r_\mathrm{out})\approx 1$, and the inner edge coincides with the position of the marginally stable orbit, this can be easily evaluated to $\eta = 1 - \energy_\mathrm{K}(r_\mathrm{ms})$. In particular, for the Schwarzschild spacetime, we have $\eta=1 - \sqrt{8}/3\approx 0.0572$. 

The upper panel of Fig.~\ref{fig:jnwm-accretion} shows efficiencies of thin disks accreting onto a compact object described by the JNWM metric. The bottommost black curve corresponds to the JNW spacetime and its lower limit $C_0=0$ to the Schwarzschild solution. As both the scalar and electromagnetic charges push the marginally stable orbit closer to the compact object where the mechanical energy of a freely-orbiting particle is lower, both charges tent to increase the accretion efficiency. Nevertheless, this change is rather modest and the resulting efficiency hardly exceeds 10\%. The bottom panel of Fig.~\ref{fig:jnwm-accretion} shows radial profiles of the effective temperature of the accreted material as measured by the local co-rotating observer for three scalar field strengths and two electric charges. When calculating the effective temperature we have adopted a standard assumption of vanishing accretion torque at the marginally stable orbit. For a non-vanishing torque, e.g. due to a large-scale magnetic field, the temperature at the inner edge of the disk would be non-zero. The thick black line corresponds to accretion onto the Schwarzschild black hole, the dashed line shows the Newtonian limit $T_\mathrm{eff}\propto(R/m)^{-3/4}$ for accretion onto a point-mass source of gravity with potential $\Phi=-GM/R$. Clearly, the effect of the additional charges is pronounced only in a vicinity of the compact object ($r\lesssim 100m$), giving at most a 10\% difference from the Schwarzschild case. 

These circumstances change significantly when the scalar charge exceeds the critical limit $C_0=\sqrt{3/2}M$. In this case both the JNW and JNWM spacetimes admit a stable circular motion also immediately close to the singularity. When $C_0 > \sqrt{8/5}M$ (and, in addition, $q < q_\mathrm{ms}^\ast$ in the case of the JNWM spacetime, corresponding to the dashed region in Fig.~\ref{fig:jnwm-parameter-space}), the Keplerian angular momentum is monotonically increasing with increasing $r$, corresponding to stable circular geodesics everywhere outside the singularity. Therefore, accretion disks may in principle extend down to the singularity giving nearly a 100 percent efficiency, since $\energy_\mathrm{K}\rightarrow 0$ as $r\rightarrow M$ in both spacetimes. Near the singularity the effective temperature formally diverges and so does the flux and the frequency of the emitted black-body radiation. Nevertheless, the total energy released near the singularity (flux integrated over the top and bottom surfaces of the disk) indeed has a finite limit. 

An interesting situation happens when the region of stable orbital motion is interrupted by a gap of unstable Keplerian circular orbits, or even by the region where particle circular motion (whether geodesic or not) is not possible. The first case happens in the JNW spacetime when $\sqrt{3/2}M < C_0 < \sqrt{8/5}M$, or in the JNWM spacetime with $C_0>\sqrt{3/2}M$ and $q$ between $q_\mathrm{ms}^\ast$ and $q_\mathrm{ph}^\ast$ (blue region in Fig~\ref{fig:jnwm-parameter-space}). The gap is filled by unstable orbits, limited by the two marginally stable orbits. The second case occurs in the JNWM spacetime with $C_0 > \sqrt{3/2}M$ and electric charge greater than $q_\mathrm{ms}^\ast$ (purple region in Fig~\ref{fig:jnwm-parameter-space}). The gap is limited by the outer marginally stable orbit and inner photon orbit and consists of the outer region of unstable orbits and the inner region where circular geodesics are spacelike. The two regions are separated by the inner photon orbit.

Let us first consider an accretion process under these circumstances in the JNW spacetime as sketched out in panels (a)--(c) of Fig~\ref{fig:jnw-plunging}. Initially the picture is the same as in the case of $C_0<\sqrt{3/2}M$. Far from the singularity, the matter closely follows the stable Keplerian circular orbits with additional subsonic radial drift caused by a turbulent transport of the angular momentum and dissipation of the mechanical energy. After reaching the outer marginally stable orbit at $r=r_\mathrm{ms}^{+}$, the rotation does not provide any more support against free fall of the matter towards the singularity. The flow then follows sub-Keplerian quasi-elliptic geodesics, roughly conserving the energy and angular momentum from the marginally stable orbit, $\energy\approx\energy_\mathrm{K}(r_\mathrm{ms}^{+})\equiv \energy_\mathrm{ms}^{+}$, $\ell\approx \angmom_\mathrm{K}(r_\mathrm{ms}^{+})/\energy_\mathrm{K}(r_\mathrm{ms}^{+})\equiv \ell_\mathrm{ms}^{+}$. At the same time, the Keplerian angular momentum $\ell_\mathrm{K}(r)$ increases with decreasing $r$. As in the previous case, the inflow becomes quickly supersonic after leaving the marginally-stable orbit.

So far the picture is similar to the standard black-hole accretion. The difference arises due to behavior inside the inner marginally stable orbit at $r=r_\mathrm{ms}^{-}$, where $\ell_\mathrm{K}$ starts to decrease again towards zero at the singularity. Obviously, inside $r_\mathrm{ms}^{-}$, there is a location ($r=r_0$) of a stable Keplerian circular orbit whose specific angular momentum matches that of the infalling matter, $\ell_\mathrm{K}(r_0) = \ell_\mathrm{ms}^{+}$. Further inside, the flow becomes increasingly super-Keplerian and its rotation prevents matter from falling to the singularity, slowing the radial motion and bouncing the matter at some point $r_\mathrm{p} < r_0$ back to larger radii [see panel (a) of Fig~\ref{fig:jnw-plunging}]. If we considered motion of an isolated particle, the particle would finally return back to $r_\mathrm{ms}^{+}$ and continue in oscillating between $r_\mathrm{p}$ and $r_\mathrm{ms}^{+}$ in the effective-potential well. In this case, however, the returning flowcollides with the supersonic stream of infalling matter, developing a shock, in which the flow kinetic energy is dissipated, and finally settling down at the Keplerian orbit at the effective-potential minimum.   

As more material is being accreted, the matter further accumulates near $r_0$, developing pressure gradients that push the fluid away from $r_0$ and creating a toroidal structure that gradually fills the effective-potential well [see panel (b) of Fig~\ref{fig:jnw-plunging}]. Since the fluid rotates faster closer than further away from the singularity, the turbulent stresses cause outward angular momentum transport leading to gradual steepening of the angular momentum distribution. This helps the material in the inner regions of the torus to descent towards smaller radii. At the same time, it further increases the spreading of the the outer parts of the torus through out the gap.

In the case of the JNW spacetime, the gap might in principle be completely filled by the torus, smoothly connecting at $r_\mathrm{ms}^{+}$ to the outer thin disk [see panel (c) of Fig.~\ref{fig:jnw-plunging}]. The angular momentum from the innermost regions of the torus would then be transported outward through the outer disk, allowing a steady accretion onto the singularity. At the same time, the torus would torque the innermost parts of the outer thin disk, further modifying its dissipation profile in the transition region. Although the time-like circular geodesics are unstable in the gap region, the fluid in the torus follows non-geodesic sub-Keplerian circular orbits where the matter is accelerated by the outward pressure gradient. Since the fluid angular momentum in the torus is never decreases with increasing $r$, such a configuration would be linearly stable according to the Rayleigh criterion. 

More puzzling situation occurs in the case of the JNWM spacetime with $q>q_\mathrm{ph}^\ast$, where the two photon orbits bound a region where neither non-geodesic circular motion is possible. In that case, the inner photon orbit would serve as the outer limit for the radial extension of the torus. The fact that circular orbits are space-like above this orbit alone prevents the flow from closing the gap as it does in the previous case of the JNW singularity. The material from the outer disk would therefore plunge in over a whole region between the (outer) marginally stable orbit and the photon orbit [see panel (d) of Fig~\ref{fig:jnw-plunging}]. Since the mater has to lose all its angular momentum before being accreted onto the naked singularity, the accretion from the inner torus requires a significant outward flux of the angular momentum. It is difficult to imagine that this flux might be mediated to the outer disk over the plunging region where the matter moves radially with a supersonic velocity by ordinary viscous or turbulent stresses. Additional processes of the angular momentum transport/removal must therefore be involved in the accretion process. The efficient transport of angular momentum can be achieved by ordered large-scale magnetic fields as shown in Ref.~\cite{Gammie1999} for the case of accretion on Kerr black hole. A significant amount might also be removed by the thermal radiation since the matter near the photon orbit moves close to speed of light. Another option are outflows that frequently accompany the accretion process. Although this question is rather of academic interest and well beyond the scope of this paper, we believe that it is interesting enough to deserve more attention in the future.

\section*{Acknowledgments}
We are grateful for helpful discussions with O. Sv\'{\i}tek.  J. H.  is supported by GA\v{C}R  21-06825X and T. T. is supported by  GA\v{C}R 23-07457S grant of the Czech Science Foundation.

\appendix

\section{Misner Sharp Mass} 
\label{MS}
Misner and Sharp in 1964 \cite{Misner+Sharp1964} used a well-defined coordinate independent quantity for spherically symmetric spacetimes, the circumferential radius $r_C$, in order to define mass contained inside a given radius, the so-called Misner--Sharp mass,
\begin{equation}
    M_\mathrm{MS}=\frac{r_C}{2}\left(1-\left|\boldsymbol{\nabla} r_C\right|^2\right)\,.
\end{equation}
The Misner-Sharp mass and ADM mass are equivalent at spatial infinity in asymptotically flat and spherically symmetric spacetimes. Since the solutions used in the paper have a complicated form and the circumferential radius is not linear in $r$, it would be interesting to explore the corresponding Misner-Sharp masses. 

\begin{figure}
	\includegraphics[width=0.47\textwidth]{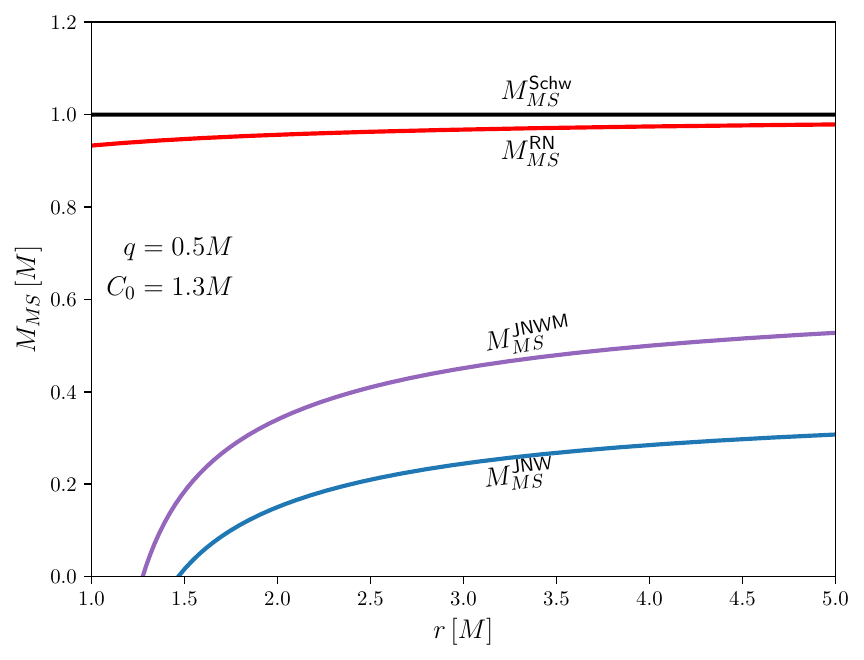}
    \caption{The Misner-Sharp mass calculated for the Schwarzschild, Reissner-Nordstr\"om, JNW and JNWM spacetimes with all the event horizons coordinates shifted to $r=M$ for parameters $C_0=1.3M$ and $q=1/2$. }
    \label{fig:MSmass}
\end{figure}

The Misner-Sharp mass for the JNW spacetime, namely for metric solutions \eqref{JNW}, reads
\begin{equation}
    M_\mathrm{MS} = \frac{M\left[\mu(2r-\mu\,M) -M\right]}
    {2\,\mu^2\sqrt{f(r)\,\left(r^2-M^2\right)}}\,.
    \label{MS-JNW}
\end{equation}
The above expression monotonically increases, as is depicted in Fig.~\ref{fig:MSmass}. The asymptotic behavior of this function would be $M_\mathrm{MS}\rightarrow M/\mu$ as $r\rightarrow\infty$, which is exactly the same as the ADM mass \eqref{ADM-JNW}. 

In order to obtain the Misner-Sharp mass for the JNWM spacetime, we use \eqref{JNW-Maxwell} and \eqref{JNW-Maxwell-R}. The result takes the form
\begin{align}
    \label{MS-JNWM}
    M_\mathrm{MS} = & -\frac{M^2}{2\sqrt{f(r)(r^2 - M^2)}}\Big[1 + 
    \nonumber\\ 
    & \nu\frac{r}{M}\left(\frac{C_1x + C_2}{C_1x - C2}\right) +
    \frac{\nu^2}{4}\left(\frac{C_1x + C_2}{C_1x - C2}\right)^2
    \Big].
\end{align}
Also, here the mass monotonically increases. The plot is shown in Fig. \ref{fig:MSmass}. Asymptotically, the mass would be $M_\mathrm{MS}\approx\sqrt{(\nu M/2)^2 + q^2}$ as $r\rightarrow\infty$, which is exactly the same as the ADM mass \eqref{eq:JNWM-AMD}. 

In Fig.~\ref{fig:MSmass}, the comparison  of the Misner-Sharp mass for the Schwarzschild, Reissner-Nordstr\"om, JNW and JNWM with all the event horizons shifted to $r=M$ is shown.

\bibliography{epicycles}

\begin{thebibliography}{92}
\expandafter\ifx\csname natexlab\endcsname\relax\def\natexlab#1{#1}\fi
\expandafter\ifx\csname bibnamefont\endcsname\relax
  \def\bibnamefont#1{#1}\fi
\expandafter\ifx\csname bibfnamefont\endcsname\relax
  \def\bibfnamefont#1{#1}\fi
\expandafter\ifx\csname citenamefont\endcsname\relax
  \def\citenamefont#1{#1}\fi
\expandafter\ifx\csname url\endcsname\relax
  \def\url#1{\texttt{#1}}\fi
\expandafter\ifx\csname urlprefix\endcsname\relax\def\urlprefix{URL }\fi
\providecommand{\bibinfo}[2]{#2}
\providecommand{\eprint}[2][]{\url{#2}}

\bibitem[{\citenamefont{{van der Klis}}(2000)}]{Klis2000}
\bibinfo{author}{\bibfnamefont{M.}~\bibnamefont{{van der Klis}}},
  \bibinfo{journal}{\araa} \textbf{\bibinfo{volume}{38}}, \bibinfo{pages}{717}
  (\bibinfo{year}{2000}), \eprint{astro-ph/0001167}.

\bibitem[{\citenamefont{{Remillard} and
  {McClintock}}(2006)}]{Remillard+McClintock2006}
\bibinfo{author}{\bibfnamefont{R.~A.} \bibnamefont{{Remillard}}}
  \bibnamefont{and} \bibinfo{author}{\bibfnamefont{J.~E.}
  \bibnamefont{{McClintock}}}, \bibinfo{journal}{\araa}
  \textbf{\bibinfo{volume}{44}}, \bibinfo{pages}{49} (\bibinfo{year}{2006}),
  \eprint{astro-ph/0606352}.

\bibitem[{\citenamefont{{Ingram} and {Motta}}(2019)}]{Ingram+Motta2019}
\bibinfo{author}{\bibfnamefont{A.~R.} \bibnamefont{{Ingram}}} \bibnamefont{and}
  \bibinfo{author}{\bibfnamefont{S.~E.} \bibnamefont{{Motta}}},
  \bibinfo{journal}{\nar} \textbf{\bibinfo{volume}{85}}, \bibinfo{eid}{101524}
  (\bibinfo{year}{2019}), \eprint{2001.08758}.

\bibitem[{\citenamefont{{Gilfanov} et~al.}(2000)\citenamefont{{Gilfanov},
  {Churazov}, and {Revnivtsev}}}]{2000MNRAS.316..923G}
\bibinfo{author}{\bibfnamefont{M.}~\bibnamefont{{Gilfanov}}},
  \bibinfo{author}{\bibfnamefont{E.}~\bibnamefont{{Churazov}}},
  \bibnamefont{and}
  \bibinfo{author}{\bibfnamefont{M.}~\bibnamefont{{Revnivtsev}}},
  \bibinfo{journal}{\mnras} \textbf{\bibinfo{volume}{316}},
  \bibinfo{pages}{923} (\bibinfo{year}{2000}), \eprint{astro-ph/0001450}.

\bibitem[{\citenamefont{{Abramowicz} et~al.}(2004)\citenamefont{{Abramowicz},
  {Klu{\'z}niak}, {McClintock}, and {Remillard}}}]{2004ApJ...609L..63A}
\bibinfo{author}{\bibfnamefont{M.~A.} \bibnamefont{{Abramowicz}}},
  \bibinfo{author}{\bibfnamefont{W.}~\bibnamefont{{Klu{\'z}niak}}},
  \bibinfo{author}{\bibfnamefont{J.~E.} \bibnamefont{{McClintock}}},
  \bibnamefont{and} \bibinfo{author}{\bibfnamefont{R.~A.}
  \bibnamefont{{Remillard}}}, \bibinfo{journal}{\apjl}
  \textbf{\bibinfo{volume}{609}}, \bibinfo{pages}{L63} (\bibinfo{year}{2004}),
  \eprint{astro-ph/0402012}.

\bibitem[{\citenamefont{{Goluchov{\'a}}
  et~al.}(2019)\citenamefont{{Goluchov{\'a}}, {T{\"o}r{\"o}k},
  {{\v{S}}r{\'a}mkov{\'a}}, {Abramowicz}, {Stuchl{\'\i}k}, and
  {Hor{\'a}k}}}]{Gol-etal:2019}
\bibinfo{author}{\bibfnamefont{K.}~\bibnamefont{{Goluchov{\'a}}}},
  \bibinfo{author}{\bibfnamefont{G.}~\bibnamefont{{T{\"o}r{\"o}k}}},
  \bibinfo{author}{\bibfnamefont{E.}~\bibnamefont{{{\v{S}}r{\'a}mkov{\'a}}}},
  \bibinfo{author}{\bibfnamefont{M.~A.} \bibnamefont{{Abramowicz}}},
  \bibinfo{author}{\bibfnamefont{Z.}~\bibnamefont{{Stuchl{\'\i}k}}},
  \bibnamefont{and}
  \bibinfo{author}{\bibfnamefont{J.}~\bibnamefont{{Hor{\'a}k}}},
  \bibinfo{journal}{A\&A} \textbf{\bibinfo{volume}{622}}, \bibinfo{eid}{L8}
  (\bibinfo{year}{2019}), \eprint{1901.05419}.

\bibitem[{\citenamefont{{Belloni} et~al.}(2005)\citenamefont{{Belloni},
  {M{\'e}ndez}, and {Homan}}}]{Belloni+2005}
\bibinfo{author}{\bibfnamefont{T.}~\bibnamefont{{Belloni}}},
  \bibinfo{author}{\bibfnamefont{M.}~\bibnamefont{{M{\'e}ndez}}},
  \bibnamefont{and} \bibinfo{author}{\bibfnamefont{J.}~\bibnamefont{{Homan}}},
  \bibinfo{journal}{\aap} \textbf{\bibinfo{volume}{437}}, \bibinfo{pages}{209}
  (\bibinfo{year}{2005}), \eprint{astro-ph/0501186}.

\bibitem[{\citenamefont{{Abramowicz}
  et~al.}(2003{\natexlab{a}})\citenamefont{{Abramowicz}, {Bulik}, {Bursa}, and
  {Klu{\'z}niak}}}]{Abramowicz+Bulik+Bursa+Kluzniak2003}
\bibinfo{author}{\bibfnamefont{M.~A.} \bibnamefont{{Abramowicz}}},
  \bibinfo{author}{\bibfnamefont{T.}~\bibnamefont{{Bulik}}},
  \bibinfo{author}{\bibfnamefont{M.}~\bibnamefont{{Bursa}}}, \bibnamefont{and}
  \bibinfo{author}{\bibfnamefont{W.}~\bibnamefont{{Klu{\'z}niak}}},
  \bibinfo{journal}{\aap} \textbf{\bibinfo{volume}{404}}, \bibinfo{pages}{L21}
  (\bibinfo{year}{2003}{\natexlab{a}}), \eprint{astro-ph/0206490}.

\bibitem[{\citenamefont{{T{\"o}r{\"o}k}
  et~al.}(2008)\citenamefont{{T{\"o}r{\"o}k}, {Abramowicz}, {Bakala}, {Bursa},
  {Hor{\'a}k}, {Kluzniak}, {Rebusco}, and {Stuchlik}}}]{Torok+2008b}
\bibinfo{author}{\bibfnamefont{G.}~\bibnamefont{{T{\"o}r{\"o}k}}},
  \bibinfo{author}{\bibfnamefont{M.~A.} \bibnamefont{{Abramowicz}}},
  \bibinfo{author}{\bibfnamefont{P.}~\bibnamefont{{Bakala}}},
  \bibinfo{author}{\bibfnamefont{M.}~\bibnamefont{{Bursa}}},
  \bibinfo{author}{\bibfnamefont{J.}~\bibnamefont{{Hor{\'a}k}}},
  \bibinfo{author}{\bibfnamefont{W.}~\bibnamefont{{Kluzniak}}},
  \bibinfo{author}{\bibfnamefont{P.}~\bibnamefont{{Rebusco}}},
  \bibnamefont{and}
  \bibinfo{author}{\bibfnamefont{Z.}~\bibnamefont{{Stuchlik}}},
  \bibinfo{journal}{\actaa} \textbf{\bibinfo{volume}{58}}, \bibinfo{pages}{15}
  (\bibinfo{year}{2008}), \eprint{0802.4070}.

\bibitem[{\citenamefont{{Torok} et~al.}(2008)\citenamefont{{Torok}, {Bakala},
  {Stuchlik}, and {Cech}}}]{Torok+2008a}
\bibinfo{author}{\bibfnamefont{G.}~\bibnamefont{{Torok}}},
  \bibinfo{author}{\bibfnamefont{P.}~\bibnamefont{{Bakala}}},
  \bibinfo{author}{\bibfnamefont{Z.}~\bibnamefont{{Stuchlik}}},
  \bibnamefont{and} \bibinfo{author}{\bibfnamefont{P.}~\bibnamefont{{Cech}}},
  \bibinfo{journal}{\actaa} \textbf{\bibinfo{volume}{58}}, \bibinfo{pages}{1}
  (\bibinfo{year}{2008}).

\bibitem[{\citenamefont{{Boutelier} et~al.}(2010)\citenamefont{{Boutelier},
  {Barret}, {Lin}, and {T{\"o}r{\"o}k}}}]{Boutelier+2010}
\bibinfo{author}{\bibfnamefont{M.}~\bibnamefont{{Boutelier}}},
  \bibinfo{author}{\bibfnamefont{D.}~\bibnamefont{{Barret}}},
  \bibinfo{author}{\bibfnamefont{Y.}~\bibnamefont{{Lin}}}, \bibnamefont{and}
  \bibinfo{author}{\bibfnamefont{G.}~\bibnamefont{{T{\"o}r{\"o}k}}},
  \bibinfo{journal}{\mnras} \textbf{\bibinfo{volume}{401}},
  \bibinfo{pages}{1290} (\bibinfo{year}{2010}), \eprint{0909.2990}.

\bibitem[{\citenamefont{{Belloni} et~al.}(2012)\citenamefont{{Belloni},
  {Sanna}, and {M{\'e}ndez}}}]{Belloni+2012}
\bibinfo{author}{\bibfnamefont{T.~M.} \bibnamefont{{Belloni}}},
  \bibinfo{author}{\bibfnamefont{A.}~\bibnamefont{{Sanna}}}, \bibnamefont{and}
  \bibinfo{author}{\bibfnamefont{M.}~\bibnamefont{{M{\'e}ndez}}},
  \bibinfo{journal}{\mnras} \textbf{\bibinfo{volume}{426}},
  \bibinfo{pages}{1701} (\bibinfo{year}{2012}), \eprint{1207.2311}.

\bibitem[{\citenamefont{{Motta} et~al.}(2014)\citenamefont{{Motta},
  {Munoz-Darias}, {Sanna}, {Fender}, {Belloni}, and {Stella}}}]{Motta+2014}
\bibinfo{author}{\bibfnamefont{S.~E.} \bibnamefont{{Motta}}},
  \bibinfo{author}{\bibfnamefont{T.}~\bibnamefont{{Munoz-Darias}}},
  \bibinfo{author}{\bibfnamefont{A.}~\bibnamefont{{Sanna}}},
  \bibinfo{author}{\bibfnamefont{R.}~\bibnamefont{{Fender}}},
  \bibinfo{author}{\bibfnamefont{T.}~\bibnamefont{{Belloni}}},
  \bibnamefont{and} \bibinfo{author}{\bibfnamefont{L.}~\bibnamefont{{Stella}}},
  \bibinfo{journal}{\mnras} \textbf{\bibinfo{volume}{439}},
  \bibinfo{pages}{L65} (\bibinfo{year}{2014}), \eprint{1312.3114}.

\bibitem[{\citenamefont{{Varniere} and
  {Rodriguez}}(2018)}]{Varniere+Rodriguez2018}
\bibinfo{author}{\bibfnamefont{P.}~\bibnamefont{{Varniere}}} \bibnamefont{and}
  \bibinfo{author}{\bibfnamefont{J.}~\bibnamefont{{Rodriguez}}},
  \bibinfo{journal}{\apj} \textbf{\bibinfo{volume}{865}}, \bibinfo{eid}{113}
  (\bibinfo{year}{2018}), \eprint{1808.06823}.

\bibitem[{\citenamefont{{Stella} and {Vietri}}(1999)}]{Stella+Vietri1999}
\bibinfo{author}{\bibfnamefont{L.}~\bibnamefont{{Stella}}} \bibnamefont{and}
  \bibinfo{author}{\bibfnamefont{M.}~\bibnamefont{{Vietri}}},
  \bibinfo{journal}{\prl} \textbf{\bibinfo{volume}{82}}, \bibinfo{pages}{17}
  (\bibinfo{year}{1999}), \eprint{astro-ph/9812124}.

\bibitem[{\citenamefont{{Rezzolla} et~al.}(2003)\citenamefont{{Rezzolla},
  {Yoshida}, {Maccarone}, and {Zanotti}}}]{Rezzolla+2003}
\bibinfo{author}{\bibfnamefont{L.}~\bibnamefont{{Rezzolla}}},
  \bibinfo{author}{\bibfnamefont{S.}~\bibnamefont{{Yoshida}}},
  \bibinfo{author}{\bibfnamefont{T.~J.} \bibnamefont{{Maccarone}}},
  \bibnamefont{and}
  \bibinfo{author}{\bibfnamefont{O.}~\bibnamefont{{Zanotti}}},
  \bibinfo{journal}{\mnras} \textbf{\bibinfo{volume}{344}},
  \bibinfo{pages}{L37} (\bibinfo{year}{2003}), \eprint{astro-ph/0307487}.

\bibitem[{\citenamefont{{Blaes} et~al.}(2006)\citenamefont{{Blaes}, {Arras},
  and {Fragile}}}]{Blaes+2006}
\bibinfo{author}{\bibfnamefont{O.~M.} \bibnamefont{{Blaes}}},
  \bibinfo{author}{\bibfnamefont{P.}~\bibnamefont{{Arras}}}, \bibnamefont{and}
  \bibinfo{author}{\bibfnamefont{P.~C.} \bibnamefont{{Fragile}}},
  \bibinfo{journal}{\mnras} \textbf{\bibinfo{volume}{369}},
  \bibinfo{pages}{1235} (\bibinfo{year}{2006}), \eprint{astro-ph/0601379}.

\bibitem[{\citenamefont{{Abramowicz} et~al.}(2006)\citenamefont{{Abramowicz},
  {Blaes}, {Hor{\'a}k}, {Kluzniak}, and {Rebusco}}}]{Abramowicz+2006}
\bibinfo{author}{\bibfnamefont{M.~A.} \bibnamefont{{Abramowicz}}},
  \bibinfo{author}{\bibfnamefont{O.~M.} \bibnamefont{{Blaes}}},
  \bibinfo{author}{\bibfnamefont{J.}~\bibnamefont{{Hor{\'a}k}}},
  \bibinfo{author}{\bibfnamefont{W.}~\bibnamefont{{Kluzniak}}},
  \bibnamefont{and}
  \bibinfo{author}{\bibfnamefont{P.}~\bibnamefont{{Rebusco}}},
  \bibinfo{journal}{Classical and Quantum Gravity}
  \textbf{\bibinfo{volume}{23}}, \bibinfo{pages}{1689} (\bibinfo{year}{2006}),
  \eprint{astro-ph/0511375}.

\bibitem[{\citenamefont{{Abramowicz} and
  {Klu{\'z}niak}}(2001)}]{Abramowicz+Kluzniak2001}
\bibinfo{author}{\bibfnamefont{M.~A.} \bibnamefont{{Abramowicz}}}
  \bibnamefont{and}
  \bibinfo{author}{\bibfnamefont{W.}~\bibnamefont{{Klu{\'z}niak}}},
  \bibinfo{journal}{\aap} \textbf{\bibinfo{volume}{374}}, \bibinfo{pages}{L19}
  (\bibinfo{year}{2001}), \eprint{astro-ph/0105077}.

\bibitem[{\citenamefont{{Abramowicz}
  et~al.}(2003{\natexlab{b}})\citenamefont{{Abramowicz}, {Karas}, {Kluzniak},
  {Lee}, and {Rebusco}}}]{Abramowicz+2003}
\bibinfo{author}{\bibfnamefont{M.~A.} \bibnamefont{{Abramowicz}}},
  \bibinfo{author}{\bibfnamefont{V.}~\bibnamefont{{Karas}}},
  \bibinfo{author}{\bibfnamefont{W.}~\bibnamefont{{Kluzniak}}},
  \bibinfo{author}{\bibfnamefont{W.~H.} \bibnamefont{{Lee}}}, \bibnamefont{and}
  \bibinfo{author}{\bibfnamefont{P.}~\bibnamefont{{Rebusco}}},
  \bibinfo{journal}{\pasj} \textbf{\bibinfo{volume}{55}}, \bibinfo{pages}{467}
  (\bibinfo{year}{2003}{\natexlab{b}}), \eprint{astro-ph/0302183}.

\bibitem[{\citenamefont{{Kluzniak} and
  {Abramowicz}}(2002)}]{Kluzniak+Abramowicz2002}
\bibinfo{author}{\bibfnamefont{W.}~\bibnamefont{{Kluzniak}}} \bibnamefont{and}
  \bibinfo{author}{\bibfnamefont{M.~A.} \bibnamefont{{Abramowicz}}},
  \bibinfo{journal}{arXiv e-prints} \bibinfo{eid}{astro-ph/0203314}
  (\bibinfo{year}{2002}), \eprint{astro-ph/0203314}.

\bibitem[{\citenamefont{{Hor{\'a}k}}(2008)}]{Horak2008}
\bibinfo{author}{\bibfnamefont{J.}~\bibnamefont{{Hor{\'a}k}}},
  \bibinfo{journal}{\aap} \textbf{\bibinfo{volume}{486}}, \bibinfo{pages}{1}
  (\bibinfo{year}{2008}), \eprint{0805.2059}.

\bibitem[{\citenamefont{{Kato}}(2001)}]{Kato2001}
\bibinfo{author}{\bibfnamefont{S.}~\bibnamefont{{Kato}}},
  \bibinfo{journal}{\pasj} \textbf{\bibinfo{volume}{53}}, \bibinfo{pages}{1}
  (\bibinfo{year}{2001}).

\bibitem[{\citenamefont{{Lai} and {Tsang}}(2009)}]{Lai+Tsang2009}
\bibinfo{author}{\bibfnamefont{D.}~\bibnamefont{{Lai}}} \bibnamefont{and}
  \bibinfo{author}{\bibfnamefont{D.}~\bibnamefont{{Tsang}}},
  \bibinfo{journal}{\mnras} \textbf{\bibinfo{volume}{393}},
  \bibinfo{pages}{979} (\bibinfo{year}{2009}), \eprint{0810.0203}.

\bibitem[{\citenamefont{{Hor{\'a}k} and {Lai}}(2013)}]{Horak+Lai2013}
\bibinfo{author}{\bibfnamefont{J.}~\bibnamefont{{Hor{\'a}k}}} \bibnamefont{and}
  \bibinfo{author}{\bibfnamefont{D.}~\bibnamefont{{Lai}}},
  \bibinfo{journal}{\mnras} \textbf{\bibinfo{volume}{434}},
  \bibinfo{pages}{2761} (\bibinfo{year}{2013}), \eprint{1307.8077}.

\bibitem[{\citenamefont{{Kato}}(2004)}]{Kato2004}
\bibinfo{author}{\bibfnamefont{S.}~\bibnamefont{{Kato}}},
  \bibinfo{journal}{\pasj} \textbf{\bibinfo{volume}{56}}, \bibinfo{pages}{905}
  (\bibinfo{year}{2004}), \eprint{astro-ph/0409051}.

\bibitem[{\citenamefont{{Ferreira} and {Ogilvie}}(2008)}]{Ferreira+Ogilvie2008}
\bibinfo{author}{\bibfnamefont{B.~T.} \bibnamefont{{Ferreira}}}
  \bibnamefont{and} \bibinfo{author}{\bibfnamefont{G.~I.}
  \bibnamefont{{Ogilvie}}}, \bibinfo{journal}{\mnras}
  \textbf{\bibinfo{volume}{386}}, \bibinfo{pages}{2297} (\bibinfo{year}{2008}),
  \eprint{0803.1671}.

\bibitem[{\citenamefont{{Okazaki} et~al.}(1987)\citenamefont{{Okazaki}, {Kato},
  and {Fukue}}}]{Okazaki+1987}
\bibinfo{author}{\bibfnamefont{A.~T.} \bibnamefont{{Okazaki}}},
  \bibinfo{author}{\bibfnamefont{S.}~\bibnamefont{{Kato}}}, \bibnamefont{and}
  \bibinfo{author}{\bibfnamefont{J.}~\bibnamefont{{Fukue}}},
  \bibinfo{journal}{\pasj} \textbf{\bibinfo{volume}{39}}, \bibinfo{pages}{457}
  (\bibinfo{year}{1987}).

\bibitem[{\citenamefont{{Nowak} and {Wagoner}}(1992)}]{Nowak+Wagoner1992}
\bibinfo{author}{\bibfnamefont{M.~A.} \bibnamefont{{Nowak}}} \bibnamefont{and}
  \bibinfo{author}{\bibfnamefont{R.~V.} \bibnamefont{{Wagoner}}},
  \bibinfo{journal}{\apj} \textbf{\bibinfo{volume}{393}}, \bibinfo{pages}{697}
  (\bibinfo{year}{1992}).

\bibitem[{\citenamefont{{Wagoner} et~al.}(2001)\citenamefont{{Wagoner},
  {Silbergleit}, and {Ortega-Rodr{\'\i}guez}}}]{Wagoner+2001}
\bibinfo{author}{\bibfnamefont{R.~V.} \bibnamefont{{Wagoner}}},
  \bibinfo{author}{\bibfnamefont{A.~S.} \bibnamefont{{Silbergleit}}},
  \bibnamefont{and}
  \bibinfo{author}{\bibfnamefont{M.}~\bibnamefont{{Ortega-Rodr{\'\i}guez}}},
  \bibinfo{journal}{\apjl} \textbf{\bibinfo{volume}{559}}, \bibinfo{pages}{L25}
  (\bibinfo{year}{2001}), \eprint{astro-ph/0107168}.

\bibitem[{\citenamefont{{T{\"o}r{\"o}k} and
  {Stuchl{\'\i}k}}(2005)}]{Torok+Stuchlik2005}
\bibinfo{author}{\bibfnamefont{G.}~\bibnamefont{{T{\"o}r{\"o}k}}}
  \bibnamefont{and}
  \bibinfo{author}{\bibfnamefont{Z.}~\bibnamefont{{Stuchl{\'\i}k}}},
  \bibinfo{journal}{\aap} \textbf{\bibinfo{volume}{437}}, \bibinfo{pages}{775}
  (\bibinfo{year}{2005}), \eprint{astro-ph/0502127}.

\bibitem[{\citenamefont{{Kotrlov{\'a}}
  et~al.}(2008)\citenamefont{{Kotrlov{\'a}}, {Stuchl{\'\i}k}, and
  {T{\"o}r{\"o}k}}}]{Kotrlova+2008}
\bibinfo{author}{\bibfnamefont{A.}~\bibnamefont{{Kotrlov{\'a}}}},
  \bibinfo{author}{\bibfnamefont{Z.}~\bibnamefont{{Stuchl{\'\i}k}}},
  \bibnamefont{and}
  \bibinfo{author}{\bibfnamefont{G.}~\bibnamefont{{T{\"o}r{\"o}k}}},
  \bibinfo{journal}{Classical and Quantum Gravity}
  \textbf{\bibinfo{volume}{25}}, \bibinfo{eid}{225016} (\bibinfo{year}{2008}),
  \eprint{0812.0720}.

\bibitem[{\citenamefont{{Rayimbaev} et~al.}(2021)\citenamefont{{Rayimbaev},
  {Abdujabbarov}, and {Wen-Biao}}}]{Rayimbaev+2021}
\bibinfo{author}{\bibfnamefont{J.}~\bibnamefont{{Rayimbaev}}},
  \bibinfo{author}{\bibfnamefont{A.}~\bibnamefont{{Abdujabbarov}}},
  \bibnamefont{and}
  \bibinfo{author}{\bibfnamefont{H.}~\bibnamefont{{Wen-Biao}}},
  \bibinfo{journal}{\prd} \textbf{\bibinfo{volume}{103}}, \bibinfo{eid}{104070}
  (\bibinfo{year}{2021}).

\bibitem[{\citenamefont{{Deligianni}
  et~al.}(2021{\natexlab{a}})\citenamefont{{Deligianni}, {Kunz}, {Nedkova},
  {Yazadjiev}, and {Zheleva}}}]{Deligianni+2021a}
\bibinfo{author}{\bibfnamefont{E.}~\bibnamefont{{Deligianni}}},
  \bibinfo{author}{\bibfnamefont{J.}~\bibnamefont{{Kunz}}},
  \bibinfo{author}{\bibfnamefont{P.}~\bibnamefont{{Nedkova}}},
  \bibinfo{author}{\bibfnamefont{S.}~\bibnamefont{{Yazadjiev}}},
  \bibnamefont{and}
  \bibinfo{author}{\bibfnamefont{R.}~\bibnamefont{{Zheleva}}},
  \bibinfo{journal}{\prd} \textbf{\bibinfo{volume}{104}}, \bibinfo{eid}{024048}
  (\bibinfo{year}{2021}{\natexlab{a}}), \eprint{2103.13504}.

\bibitem[{\citenamefont{{Deligianni}
  et~al.}(2021{\natexlab{b}})\citenamefont{{Deligianni}, {Kleihaus}, {Kunz},
  {Nedkova}, and {Yazadjiev}}}]{Deligianni+2021b}
\bibinfo{author}{\bibfnamefont{E.}~\bibnamefont{{Deligianni}}},
  \bibinfo{author}{\bibfnamefont{B.}~\bibnamefont{{Kleihaus}}},
  \bibinfo{author}{\bibfnamefont{J.}~\bibnamefont{{Kunz}}},
  \bibinfo{author}{\bibfnamefont{P.}~\bibnamefont{{Nedkova}}},
  \bibnamefont{and}
  \bibinfo{author}{\bibfnamefont{S.}~\bibnamefont{{Yazadjiev}}},
  \bibinfo{journal}{\prd} \textbf{\bibinfo{volume}{104}}, \bibinfo{eid}{064043}
  (\bibinfo{year}{2021}{\natexlab{b}}), \eprint{2107.01421}.

\bibitem[{\citenamefont{{Stuchl{\'\i}k} and
  {Vrba}}(2021{\natexlab{a}})}]{Stuchlik+Vrba2021a}
\bibinfo{author}{\bibfnamefont{Z.}~\bibnamefont{{Stuchl{\'\i}k}}}
  \bibnamefont{and} \bibinfo{author}{\bibfnamefont{J.}~\bibnamefont{{Vrba}}},
  \bibinfo{journal}{Universe} \textbf{\bibinfo{volume}{7}},
  \bibinfo{pages}{279} (\bibinfo{year}{2021}{\natexlab{a}}),
  \eprint{2108.09562}.

\bibitem[{\citenamefont{{Stuchl{\'\i}k} and
  {Vrba}}(2021{\natexlab{b}})}]{Stuchlik+Vrba2021b}
\bibinfo{author}{\bibfnamefont{Z.}~\bibnamefont{{Stuchl{\'\i}k}}}
  \bibnamefont{and} \bibinfo{author}{\bibfnamefont{J.}~\bibnamefont{{Vrba}}},
  \bibinfo{journal}{European Physical Journal Plus}
  \textbf{\bibinfo{volume}{136}}, \bibinfo{eid}{1127}
  (\bibinfo{year}{2021}{\natexlab{b}}), \eprint{2110.10569}.

\bibitem[{\citenamefont{{Rayimbaev} et~al.}(2022)\citenamefont{{Rayimbaev},
  {Bokhari}, and {Ahmedov}}}]{Rayimbaev+2022}
\bibinfo{author}{\bibfnamefont{J.}~\bibnamefont{{Rayimbaev}}},
  \bibinfo{author}{\bibfnamefont{A.~H.} \bibnamefont{{Bokhari}}},
  \bibnamefont{and}
  \bibinfo{author}{\bibfnamefont{B.}~\bibnamefont{{Ahmedov}}},
  \bibinfo{journal}{Classical and Quantum Gravity}
  \textbf{\bibinfo{volume}{39}}, \bibinfo{eid}{075021} (\bibinfo{year}{2022}).

\bibitem[{\citenamefont{{De Falco} et~al.}(2023)\citenamefont{{De Falco},
  {Bajardi}, {D'Agostino}, {Benetti}, and {Capozziello}}}]{DeFalco+2023}
\bibinfo{author}{\bibfnamefont{V.}~\bibnamefont{{De Falco}}},
  \bibinfo{author}{\bibfnamefont{F.}~\bibnamefont{{Bajardi}}},
  \bibinfo{author}{\bibfnamefont{R.}~\bibnamefont{{D'Agostino}}},
  \bibinfo{author}{\bibfnamefont{M.}~\bibnamefont{{Benetti}}},
  \bibnamefont{and}
  \bibinfo{author}{\bibfnamefont{S.}~\bibnamefont{{Capozziello}}},
  \bibinfo{journal}{European Physical Journal C} \textbf{\bibinfo{volume}{83}},
  \bibinfo{eid}{456} (\bibinfo{year}{2023}), \eprint{2305.04695}.

\bibitem[{\citenamefont{{Rayimbaev} et~al.}(2023)\citenamefont{{Rayimbaev},
  {Dialektopoulos}, {Sarikulov}, and {Abdujabbarov}}}]{Rayimbaev+2023}
\bibinfo{author}{\bibfnamefont{J.}~\bibnamefont{{Rayimbaev}}},
  \bibinfo{author}{\bibfnamefont{K.~F.} \bibnamefont{{Dialektopoulos}}},
  \bibinfo{author}{\bibfnamefont{F.}~\bibnamefont{{Sarikulov}}},
  \bibnamefont{and}
  \bibinfo{author}{\bibfnamefont{A.}~\bibnamefont{{Abdujabbarov}}},
  \bibinfo{journal}{European Physical Journal C} \textbf{\bibinfo{volume}{83}},
  \bibinfo{eid}{572} (\bibinfo{year}{2023}), \eprint{2307.03019}.

\bibitem[{\citenamefont{Shahzadi et~al.}(2024)\citenamefont{Shahzadi,
  Kolo\v{s}, Saleem, and Stuchl\'\i{}k}}]{Shahzadi+2023a}
\bibinfo{author}{\bibfnamefont{M.}~\bibnamefont{Shahzadi}},
  \bibinfo{author}{\bibfnamefont{M.}~\bibnamefont{Kolo\v{s}}},
  \bibinfo{author}{\bibfnamefont{R.}~\bibnamefont{Saleem}}, \bibnamefont{and}
  \bibinfo{author}{\bibfnamefont{Z.}~\bibnamefont{Stuchl\'\i{}k}},
  \bibinfo{journal}{Class. Quant. Grav.} \textbf{\bibinfo{volume}{41}},
  \bibinfo{pages}{075014} (\bibinfo{year}{2024}), \eprint{2309.09712}.

\bibitem[{\citenamefont{{Semer{\'a}k} and
  {Z{\'a}cek}}(2000)}]{Semerak+Zacek2000a}
\bibinfo{author}{\bibfnamefont{O.}~\bibnamefont{{Semer{\'a}k}}}
  \bibnamefont{and}
  \bibinfo{author}{\bibfnamefont{M.}~\bibnamefont{{Z{\'a}cek}}},
  \bibinfo{journal}{Classical and Quantum Gravity}
  \textbf{\bibinfo{volume}{17}}, \bibinfo{pages}{1613} (\bibinfo{year}{2000}).

\bibitem[{\citenamefont{{Semer{\'a}k} and
  {{\v{Z}}{\'a}{\v{c}}ek}}(2000)}]{Semerak+Zacek2000b}
\bibinfo{author}{\bibfnamefont{O.}~\bibnamefont{{Semer{\'a}k}}}
  \bibnamefont{and}
  \bibinfo{author}{\bibfnamefont{M.}~\bibnamefont{{{\v{Z}}{\'a}{\v{c}}ek}}},
  \bibinfo{journal}{Publ. Astron. Soc. Japan} \textbf{\bibinfo{volume}{52}},
  \bibinfo{pages}{1067} (\bibinfo{year}{2000}).

\bibitem[{\citenamefont{Tahamtan}(2020)}]{Tahamtan2020}
\bibinfo{author}{\bibfnamefont{T.}~\bibnamefont{Tahamtan}},
  \bibinfo{journal}{Phys. Rev. D} \textbf{\bibinfo{volume}{101}},
  \bibinfo{pages}{124023} (\bibinfo{year}{2020}), \eprint{2006.02810}.

\bibitem[{\citenamefont{Tahamtan}(2018-2019-2020)}]{Tahamtan-RAG19}
\bibinfo{author}{\bibfnamefont{T.}~\bibnamefont{Tahamtan}}, in
  \emph{\bibinfo{booktitle}{Proceedings of RAGtime 20–22:Workshops on black
  holes and neutron stars}}, edited by \bibinfo{editor}{\bibfnamefont{G.~T.}
  \bibnamefont{Z.~Stuchlík}} \bibnamefont{and}
  \bibinfo{editor}{\bibfnamefont{V.}~\bibnamefont{Karas}}
  (\bibinfo{publisher}{Institute of Physics in Opava No. 9},
  \bibinfo{year}{2018-2019-2020}), ISBN \bibinfo{isbn}{978-80-7510-433-5},
  \urlprefix\url{https://proceedings.physics.cz/wp-content/uploads/2021/04/tah.pdf}.

\bibitem[{\citenamefont{{Shirokov}}(1973)}]{Shikorov1973}
\bibinfo{author}{\bibfnamefont{M.~F.} \bibnamefont{{Shirokov}}},
  \bibinfo{journal}{General Relativity and Gravitation}
  \textbf{\bibinfo{volume}{4}}, \bibinfo{pages}{131} (\bibinfo{year}{1973}).

\bibitem[{\citenamefont{{Aliev} and {Galtsov}}(1981)}]{Aliev+Galtsov1981}
\bibinfo{author}{\bibfnamefont{A.~N.} \bibnamefont{{Aliev}}} \bibnamefont{and}
  \bibinfo{author}{\bibfnamefont{D.~V.} \bibnamefont{{Galtsov}}},
  \bibinfo{journal}{General Relativity and Gravitation}
  \textbf{\bibinfo{volume}{13}}, \bibinfo{pages}{899} (\bibinfo{year}{1981}).

\bibitem[{\citenamefont{{Kato}}(1990)}]{Kato1990}
\bibinfo{author}{\bibfnamefont{S.}~\bibnamefont{{Kato}}},
  \bibinfo{journal}{\pasj} \textbf{\bibinfo{volume}{42}}, \bibinfo{pages}{99}
  (\bibinfo{year}{1990}).

\bibitem[{\citenamefont{{Abramowicz} and
  {Klu{\'z}niak}}(2005)}]{Abramowicz+Kluzniak2005}
\bibinfo{author}{\bibfnamefont{M.~A.} \bibnamefont{{Abramowicz}}}
  \bibnamefont{and}
  \bibinfo{author}{\bibfnamefont{W.}~\bibnamefont{{Klu{\'z}niak}}},
  \bibinfo{journal}{APSS} \textbf{\bibinfo{volume}{300}}, \bibinfo{pages}{127}
  (\bibinfo{year}{2005}), \eprint{astro-ph/0411709}.

\bibitem[{\citenamefont{{Kerner} et~al.}(2001)\citenamefont{{Kerner}, {van
  Holten}, and {Colistete}}}]{Kerner+2001}
\bibinfo{author}{\bibfnamefont{R.}~\bibnamefont{{Kerner}}},
  \bibinfo{author}{\bibfnamefont{J.~W.} \bibnamefont{{van Holten}}},
  \bibnamefont{and}
  \bibinfo{author}{\bibfnamefont{J.}~\bibnamefont{{Colistete}},
  \bibfnamefont{R.}}, \bibinfo{journal}{Classical and Quantum Gravity}
  \textbf{\bibinfo{volume}{18}}, \bibinfo{pages}{4725} (\bibinfo{year}{2001}),
  \eprint{gr-qc/0102099}.

\bibitem[{\citenamefont{{Abramowicz} et~al.}(1993)\citenamefont{{Abramowicz},
  {Miller}, and {Stuchl{\'\i}k}}}]{Abramowicz+1993}
\bibinfo{author}{\bibfnamefont{M.~A.} \bibnamefont{{Abramowicz}}},
  \bibinfo{author}{\bibfnamefont{J.~C.} \bibnamefont{{Miller}}},
  \bibnamefont{and}
  \bibinfo{author}{\bibfnamefont{Z.}~\bibnamefont{{Stuchl{\'\i}k}}},
  \bibinfo{journal}{\prd} \textbf{\bibinfo{volume}{47}}, \bibinfo{pages}{1440}
  (\bibinfo{year}{1993}).

\bibitem[{\citenamefont{{Abramowicz}}(1971)}]{Abramowicz1971}
\bibinfo{author}{\bibfnamefont{M.~A.} \bibnamefont{{Abramowicz}}},
  \bibinfo{journal}{\actaa} \textbf{\bibinfo{volume}{21}}, \bibinfo{pages}{81}
  (\bibinfo{year}{1971}).

\bibitem[{\citenamefont{{Abramowicz} and
  {Klu{\'z}niak}}(2003)}]{Abramowicz+Kluzniak2003}
\bibinfo{author}{\bibfnamefont{M.~A.} \bibnamefont{{Abramowicz}}}
  \bibnamefont{and}
  \bibinfo{author}{\bibfnamefont{W.}~\bibnamefont{{Klu{\'z}niak}}},
  \bibinfo{journal}{General Relativity and Gravitation}
  \textbf{\bibinfo{volume}{35}}, \bibinfo{pages}{69} (\bibinfo{year}{2003}),
  \eprint{gr-qc/0206063}.

\bibitem[{\citenamefont{{Abramowicz} et~al.}(1988)\citenamefont{{Abramowicz},
  {Carter}, and {Lasota}}}]{Abramowicz+Carter+Lasota1988}
\bibinfo{author}{\bibfnamefont{M.~A.} \bibnamefont{{Abramowicz}}},
  \bibinfo{author}{\bibfnamefont{B.}~\bibnamefont{{Carter}}}, \bibnamefont{and}
  \bibinfo{author}{\bibfnamefont{J.~P.} \bibnamefont{{Lasota}}},
  \bibinfo{journal}{General Relativity and Gravitation}
  \textbf{\bibinfo{volume}{20}}, \bibinfo{pages}{1173} (\bibinfo{year}{1988}).

\bibitem[{\citenamefont{{Keir}}(2016)}]{Keir2016}
\bibinfo{author}{\bibfnamefont{J.}~\bibnamefont{{Keir}}},
  \bibinfo{journal}{Classical and Quantum Gravity}
  \textbf{\bibinfo{volume}{33}}, \bibinfo{eid}{135009} (\bibinfo{year}{2016}),
  \eprint{1404.7036}.

\bibitem[{\citenamefont{Janis et~al.}(1968)\citenamefont{Janis, Newman, and
  Winicour}}]{JNW1}
\bibinfo{author}{\bibfnamefont{A.~I.} \bibnamefont{Janis}},
  \bibinfo{author}{\bibfnamefont{E.~T.} \bibnamefont{Newman}},
  \bibnamefont{and} \bibinfo{author}{\bibfnamefont{J.}~\bibnamefont{Winicour}},
  \bibinfo{journal}{Phys. Rev. Lett.} \textbf{\bibinfo{volume}{20}},
  \bibinfo{pages}{878} (\bibinfo{year}{1968}).

\bibitem[{\citenamefont{Fisher}(1948)}]{Fisher}
\bibinfo{author}{\bibfnamefont{I.~Z.} \bibnamefont{Fisher}},
  \bibinfo{journal}{Zh. Eksp. Teor. Fiz.} \textbf{\bibinfo{volume}{18}},
  \bibinfo{pages}{636} (\bibinfo{year}{1948}), \eprint{gr-qc/9911008}.

\bibitem[{\citenamefont{Sv\'\i{}tek et~al.}(2020)\citenamefont{Sv\'\i{}tek,
  Tahamtan, and Zampeli}}]{Svitek+2020}
\bibinfo{author}{\bibfnamefont{O.}~\bibnamefont{Sv\'\i{}tek}},
  \bibinfo{author}{\bibfnamefont{T.}~\bibnamefont{Tahamtan}}, \bibnamefont{and}
  \bibinfo{author}{\bibfnamefont{A.}~\bibnamefont{Zampeli}},
  \bibinfo{journal}{Annals Phys.} \textbf{\bibinfo{volume}{418}},
  \bibinfo{pages}{168195} (\bibinfo{year}{2020}), \eprint{1606.05635}.

\bibitem[{\citenamefont{Virbhadra and Ellis}(2002)}]{JNW-lensing1}
\bibinfo{author}{\bibfnamefont{K.~S.} \bibnamefont{Virbhadra}}
  \bibnamefont{and} \bibinfo{author}{\bibfnamefont{G.~F.~R.}
  \bibnamefont{Ellis}}, \bibinfo{journal}{Phys. Rev. D}
  \textbf{\bibinfo{volume}{65}}, \bibinfo{pages}{103004}
  (\bibinfo{year}{2002}),
  \urlprefix\url{https://link.aps.org/doi/10.1103/PhysRevD.65.103004}.

\bibitem[{\citenamefont{Virbhadra et~al.}(1998)\citenamefont{Virbhadra,
  Narasimha, and Chitre}}]{JNW-lensing2}
\bibinfo{author}{\bibfnamefont{K.~S.} \bibnamefont{Virbhadra}},
  \bibinfo{author}{\bibfnamefont{D.}~\bibnamefont{Narasimha}},
  \bibnamefont{and} \bibinfo{author}{\bibfnamefont{S.~M.}
  \bibnamefont{Chitre}}, \bibinfo{journal}{Astronomy and Astrophysics}
  \textbf{\bibinfo{volume}{337}}, \bibinfo{pages}{1} (\bibinfo{year}{1998}),
  \eprint{astro-ph/9801174}.

\bibitem[{\citenamefont{Chowdhury et~al.}(2012)\citenamefont{Chowdhury, Patil,
  Malafarina, and Joshi}}]{JNW-accretion1}
\bibinfo{author}{\bibfnamefont{A.~N.} \bibnamefont{Chowdhury}},
  \bibinfo{author}{\bibfnamefont{M.}~\bibnamefont{Patil}},
  \bibinfo{author}{\bibfnamefont{D.}~\bibnamefont{Malafarina}},
  \bibnamefont{and} \bibinfo{author}{\bibfnamefont{P.~S.} \bibnamefont{Joshi}},
  \bibinfo{journal}{Phys. Rev. D} \textbf{\bibinfo{volume}{85}},
  \bibinfo{pages}{104031} (\bibinfo{year}{2012}),
  \urlprefix\url{https://link.aps.org/doi/10.1103/PhysRevD.85.104031}.

\bibitem[{\citenamefont{Ghosh et~al.}(2015)\citenamefont{Ghosh, Sarkar, and
  Bhadra}}]{JNW-accretion2}
\bibinfo{author}{\bibfnamefont{S.}~\bibnamefont{Ghosh}},
  \bibinfo{author}{\bibfnamefont{T.}~\bibnamefont{Sarkar}}, \bibnamefont{and}
  \bibinfo{author}{\bibfnamefont{A.}~\bibnamefont{Bhadra}},
  \bibinfo{journal}{Phys. Rev. D} \textbf{\bibinfo{volume}{92}},
  \bibinfo{pages}{083010} (\bibinfo{year}{2015}),
  \urlprefix\url{https://link.aps.org/doi/10.1103/PhysRevD.92.083010}.

\bibitem[{\citenamefont{Gyulchev et~al.}(2019)\citenamefont{Gyulchev, Nedkova,
  Vetsov, and Yazadjiev}}]{Gyulchev+2019}
\bibinfo{author}{\bibfnamefont{G.}~\bibnamefont{Gyulchev}},
  \bibinfo{author}{\bibfnamefont{P.}~\bibnamefont{Nedkova}},
  \bibinfo{author}{\bibfnamefont{T.}~\bibnamefont{Vetsov}}, \bibnamefont{and}
  \bibinfo{author}{\bibfnamefont{S.}~\bibnamefont{Yazadjiev}},
  \bibinfo{journal}{Phys. Rev. D} \textbf{\bibinfo{volume}{100}},
  \bibinfo{pages}{024055} (\bibinfo{year}{2019}),
  \urlprefix\url{https://link.aps.org/doi/10.1103/PhysRevD.100.024055}.

\bibitem[{\citenamefont{Sau et~al.}(2020)\citenamefont{Sau, Banerjee, and
  SenGupta}}]{JNW-shadow}
\bibinfo{author}{\bibfnamefont{S.}~\bibnamefont{Sau}},
  \bibinfo{author}{\bibfnamefont{I.}~\bibnamefont{Banerjee}}, \bibnamefont{and}
  \bibinfo{author}{\bibfnamefont{S.}~\bibnamefont{SenGupta}},
  \bibinfo{journal}{Phys. Rev. D} \textbf{\bibinfo{volume}{102}},
  \bibinfo{pages}{064027} (\bibinfo{year}{2020}),
  \urlprefix\url{https://link.aps.org/doi/10.1103/PhysRevD.102.064027}.

\bibitem[{\citenamefont{Mirza et~al.}(2023)\citenamefont{Mirza, Kangazi, and
  Sadeghi}}]{Mirza+2023}
\bibinfo{author}{\bibfnamefont{B.}~\bibnamefont{Mirza}},
  \bibinfo{author}{\bibfnamefont{P.~K.} \bibnamefont{Kangazi}},
  \bibnamefont{and} \bibinfo{author}{\bibfnamefont{F.}~\bibnamefont{Sadeghi}},
  \bibinfo{journal}{Eur. Phys. J. C} \textbf{\bibinfo{volume}{83}},
  \bibinfo{pages}{1161} (\bibinfo{year}{2023}), \eprint{2307.13588}.

\bibitem[{\citenamefont{Azizallahi et~al.}(2024)\citenamefont{Azizallahi,
  Mirza, Hajibarat, and Anjomshoa}}]{Azizallahi+2024}
\bibinfo{author}{\bibfnamefont{A.}~\bibnamefont{Azizallahi}},
  \bibinfo{author}{\bibfnamefont{B.}~\bibnamefont{Mirza}},
  \bibinfo{author}{\bibfnamefont{A.}~\bibnamefont{Hajibarat}},
  \bibnamefont{and}
  \bibinfo{author}{\bibfnamefont{H.}~\bibnamefont{Anjomshoa}},
  \bibinfo{journal}{Nuclear Physics B} \textbf{\bibinfo{volume}{998}},
  \bibinfo{pages}{116414} (\bibinfo{year}{2024}), ISSN
  \bibinfo{issn}{0550-3213},
  \urlprefix\url{https://www.sciencedirect.com/science/article/pii/S0550321323003413}.

\bibitem[{\citenamefont{Janis et~al.}(1969)\citenamefont{Janis, Robinson, and
  Winicour}}]{JNW2}
\bibinfo{author}{\bibfnamefont{A.~I.} \bibnamefont{Janis}},
  \bibinfo{author}{\bibfnamefont{D.~C.} \bibnamefont{Robinson}},
  \bibnamefont{and} \bibinfo{author}{\bibfnamefont{J.}~\bibnamefont{Winicour}},
  \bibinfo{journal}{Phys. Rev.} \textbf{\bibinfo{volume}{186}},
  \bibinfo{pages}{1729} (\bibinfo{year}{1969}).

\bibitem[{\citenamefont{Penney}(1969)}]{Penney1969}
\bibinfo{author}{\bibfnamefont{R.}~\bibnamefont{Penney}},
  \bibinfo{journal}{Phys. Rev.} \textbf{\bibinfo{volume}{182}},
  \bibinfo{pages}{1383} (\bibinfo{year}{1969}),
  \urlprefix\url{https://link.aps.org/doi/10.1103/PhysRev.182.1383}.

\bibitem[{\citenamefont{Teixeira et~al.}(1976)\citenamefont{Teixeira, Wolk, and
  Som}}]{Teixeira1976}
\bibinfo{author}{\bibfnamefont{A.~F.~F.} \bibnamefont{Teixeira}},
  \bibinfo{author}{\bibfnamefont{I.}~\bibnamefont{Wolk}}, \bibnamefont{and}
  \bibinfo{author}{\bibfnamefont{M.~M.} \bibnamefont{Som}},
  \bibinfo{journal}{Journal of Physics A: Mathematical and General}
  \textbf{\bibinfo{volume}{9}}, \bibinfo{pages}{53} (\bibinfo{year}{1976}).

\bibitem[{\citenamefont{Banerjee and Choudhury}(1977)}]{Banerjee}
\bibinfo{author}{\bibfnamefont{A.}~\bibnamefont{Banerjee}} \bibnamefont{and}
  \bibinfo{author}{\bibfnamefont{S.~B.~D.} \bibnamefont{Choudhury}},
  \bibinfo{journal}{Phys. Rev. D} \textbf{\bibinfo{volume}{15}},
  \bibinfo{pages}{3062} (\bibinfo{year}{1977}),
  \urlprefix\url{https://link.aps.org/doi/10.1103/PhysRevD.15.3062}.

\bibitem[{\citenamefont{Sorokin}(2022)}]{Sorokin2021}
\bibinfo{author}{\bibfnamefont{D.~P.} \bibnamefont{Sorokin}},
  \bibinfo{journal}{Fortsch. Phys.} \textbf{\bibinfo{volume}{70}},
  \bibinfo{pages}{2200092} (\bibinfo{year}{2022}), \eprint{2112.12118}.

\bibitem[{\citenamefont{Boillat}(1970)}]{Boillat1970}
\bibinfo{author}{\bibfnamefont{G.}~\bibnamefont{Boillat}}, \bibinfo{journal}{J.
  Math. Phys.} \textbf{\bibinfo{volume}{11}}, \bibinfo{pages}{941}
  (\bibinfo{year}{1970}).

\bibitem[{\citenamefont{Born and Infeld}(1934)}]{BornInfeld}
\bibinfo{author}{\bibfnamefont{M.}~\bibnamefont{Born}} \bibnamefont{and}
  \bibinfo{author}{\bibfnamefont{L.}~\bibnamefont{Infeld}},
  \bibinfo{journal}{Proceedings of the Royal Society of London. Series A,
  Containing Papers of a Mathematical and Physical Character}
  \textbf{\bibinfo{volume}{144}}, \bibinfo{pages}{425} (\bibinfo{year}{1934}).

\bibitem[{\citenamefont{Letelier}(1979)}]{Letelier-1979}
\bibinfo{author}{\bibfnamefont{P.~S.} \bibnamefont{Letelier}},
  \bibinfo{journal}{Phys. Rev. D} \textbf{\bibinfo{volume}{20}},
  \bibinfo{pages}{1294} (\bibinfo{year}{1979}),
  \urlprefix\url{https://link.aps.org/doi/10.1103/PhysRevD.20.1294}.

\bibitem[{\citenamefont{Barriola and Vilenkin}(1989)}]{Barriola}
\bibinfo{author}{\bibfnamefont{M.}~\bibnamefont{Barriola}} \bibnamefont{and}
  \bibinfo{author}{\bibfnamefont{A.}~\bibnamefont{Vilenkin}},
  \bibinfo{journal}{Phys. Rev. Lett.} \textbf{\bibinfo{volume}{63}},
  \bibinfo{pages}{341} (\bibinfo{year}{1989}),
  \urlprefix\url{https://link.aps.org/doi/10.1103/PhysRevLett.63.341}.

\bibitem[{\citenamefont{Harari and Loust\'o}(1990)}]{global1}
\bibinfo{author}{\bibfnamefont{D.}~\bibnamefont{Harari}} \bibnamefont{and}
  \bibinfo{author}{\bibfnamefont{C.}~\bibnamefont{Loust\'o}},
  \bibinfo{journal}{Phys. Rev. D} \textbf{\bibinfo{volume}{42}},
  \bibinfo{pages}{2626} (\bibinfo{year}{1990}),
  \urlprefix\url{https://link.aps.org/doi/10.1103/PhysRevD.42.2626}.

\bibitem[{\citenamefont{Tahamtan and Sv{i}tek}(2014)}]{Tahamtan-quantum:2014}
\bibinfo{author}{\bibfnamefont{T.}~\bibnamefont{Tahamtan}} \bibnamefont{and}
  \bibinfo{author}{\bibfnamefont{O.}~\bibnamefont{Sv{i}tek}},
  \bibinfo{journal}{Eur. Phys. J.} \textbf{\bibinfo{volume}{C74}},
  \bibinfo{pages}{2987} (\bibinfo{year}{2014}), \eprint{1312.7806}.

\bibitem[{\citenamefont{Svitek and Tahamtan}(2016)}]{Tahamtan-Boosting}
\bibinfo{author}{\bibfnamefont{O.}~\bibnamefont{Svitek}} \bibnamefont{and}
  \bibinfo{author}{\bibfnamefont{T.}~\bibnamefont{Tahamtan}},
  \bibinfo{journal}{Gen. Rel. Grav.} \textbf{\bibinfo{volume}{48}},
  \bibinfo{pages}{22} (\bibinfo{year}{2016}), \eprint{1406.6334}.

\bibitem[{\citenamefont{Novello et~al.}(2000)\citenamefont{Novello, De~Lorenci,
  Salim, and Klippert}}]{Novello:1999pg}
\bibinfo{author}{\bibfnamefont{M.}~\bibnamefont{Novello}},
  \bibinfo{author}{\bibfnamefont{V.~A.} \bibnamefont{De~Lorenci}},
  \bibinfo{author}{\bibfnamefont{J.~M.} \bibnamefont{Salim}}, \bibnamefont{and}
  \bibinfo{author}{\bibfnamefont{R.}~\bibnamefont{Klippert}},
  \bibinfo{journal}{Phys. Rev. D} \textbf{\bibinfo{volume}{61}},
  \bibinfo{pages}{045001} (\bibinfo{year}{2000}), \eprint{gr-qc/9911085}.

\bibitem[{\citenamefont{Ernst}(1976)}]{Ernst}
\bibinfo{author}{\bibfnamefont{F.~J.} \bibnamefont{Ernst}},
  \bibinfo{journal}{Journal of Mathematical Physics}
  \textbf{\bibinfo{volume}{17}}, \bibinfo{pages}{54} (\bibinfo{year}{1976}),
  \eprint{https://doi.org/10.1063/1.522781},
  \urlprefix\url{https://doi.org/10.1063/1.522781}.

\bibitem[{\citenamefont{Stephani et~al.}(2003)\citenamefont{Stephani, Kramer,
  MacCallum, Hoenselaers, and Herlt}}]{Stephanietal:book}
\bibinfo{author}{\bibfnamefont{H.}~\bibnamefont{Stephani}},
  \bibinfo{author}{\bibfnamefont{D.}~\bibnamefont{Kramer}},
  \bibinfo{author}{\bibfnamefont{M.}~\bibnamefont{MacCallum}},
  \bibinfo{author}{\bibfnamefont{C.}~\bibnamefont{Hoenselaers}},
  \bibnamefont{and} \bibinfo{author}{\bibfnamefont{E.}~\bibnamefont{Herlt}},
  \emph{\bibinfo{title}{Exact Solutions of Einstein's Field Equations}},
  Cambridge Monographs on Mathematical Physics (\bibinfo{publisher}{Cambridge
  University Press}, \bibinfo{year}{2003}), \bibinfo{edition}{2nd} ed.

\bibitem[{\citenamefont{Ortaggio}(2004)}]{Marcello}
\bibinfo{author}{\bibfnamefont{M.}~\bibnamefont{Ortaggio}},
  \bibinfo{journal}{Phys. Rev. D} \textbf{\bibinfo{volume}{69}},
  \bibinfo{pages}{064034} (\bibinfo{year}{2004}),
  \urlprefix\url{https://link.aps.org/doi/10.1103/PhysRevD.69.064034}.

\bibitem[{\citenamefont{Bonnor}(1954)}]{Bonor-Melvin}
\bibinfo{author}{\bibfnamefont{W.~B.} \bibnamefont{Bonnor}},
  \bibinfo{journal}{Proc. Phys. Soc. A} \textbf{\bibinfo{volume}{67}},
  \bibinfo{pages}{225} (\bibinfo{year}{1954}).

\bibitem[{\citenamefont{Melvin}(1964)}]{Melvin1}
\bibinfo{author}{\bibfnamefont{M.}~\bibnamefont{Melvin}},
  \bibinfo{journal}{Physics Letters} \textbf{\bibinfo{volume}{8}},
  \bibinfo{pages}{65} (\bibinfo{year}{1964}), ISSN \bibinfo{issn}{0031-9163},
  \urlprefix\url{https://www.sciencedirect.com/science/article/pii/0031916364908017}.

\bibitem[{\citenamefont{{Melvin} and {Wallingford}}(1966)}]{Melvin2}
\bibinfo{author}{\bibfnamefont{M.~A.} \bibnamefont{{Melvin}}} \bibnamefont{and}
  \bibinfo{author}{\bibfnamefont{J.~S.} \bibnamefont{{Wallingford}}},
  \bibinfo{journal}{Journal of Mathematical Physics}
  \textbf{\bibinfo{volume}{7}}, \bibinfo{pages}{333} (\bibinfo{year}{1966}).

\bibitem[{\citenamefont{Tahamtan and Halilsoy}(2013)}]{Tahamtan-Melvin}
\bibinfo{author}{\bibfnamefont{T.}~\bibnamefont{Tahamtan}} \bibnamefont{and}
  \bibinfo{author}{\bibfnamefont{M.}~\bibnamefont{Halilsoy}},
  \bibinfo{journal}{Astrophys. Space Sci.} \textbf{\bibinfo{volume}{343}},
  \bibinfo{pages}{435} (\bibinfo{year}{2013}), \eprint{1104.3401}.

\bibitem[{\citenamefont{{Shakura} and {Sunyaev}}(1973)}]{Shakura+Sunyaev1973}
\bibinfo{author}{\bibfnamefont{N.~I.} \bibnamefont{{Shakura}}}
  \bibnamefont{and} \bibinfo{author}{\bibfnamefont{R.~A.}
  \bibnamefont{{Sunyaev}}}, \bibinfo{journal}{\aap}
  \textbf{\bibinfo{volume}{24}}, \bibinfo{pages}{337} (\bibinfo{year}{1973}).

\bibitem[{\citenamefont{{Novikov} and {Thorne}}(1973)}]{Novikov+Thorne1973}
\bibinfo{author}{\bibfnamefont{I.~D.} \bibnamefont{{Novikov}}}
  \bibnamefont{and} \bibinfo{author}{\bibfnamefont{K.~S.}
  \bibnamefont{{Thorne}}}, in \emph{\bibinfo{booktitle}{Black Holes (Les Astres
  Occlus)}}, edited by
  \bibinfo{editor}{\bibfnamefont{C.}~\bibnamefont{{Dewitt}}} \bibnamefont{and}
  \bibinfo{editor}{\bibfnamefont{B.~S.} \bibnamefont{{Dewitt}}}
  (\bibinfo{year}{1973}), pp. \bibinfo{pages}{343--450}.

\bibitem[{\citenamefont{{Abramowicz} et~al.}(2010)\citenamefont{{Abramowicz},
  {Jaroszy{\'n}ski}, {Kato}, {Lasota}, {R{\'o}{\.z}a{\'n}ska}, and
  {S{\k{a}}dowski}}}]{Abramowicz+2010}
\bibinfo{author}{\bibfnamefont{M.~A.} \bibnamefont{{Abramowicz}}},
  \bibinfo{author}{\bibfnamefont{M.}~\bibnamefont{{Jaroszy{\'n}ski}}},
  \bibinfo{author}{\bibfnamefont{S.}~\bibnamefont{{Kato}}},
  \bibinfo{author}{\bibfnamefont{J.~P.} \bibnamefont{{Lasota}}},
  \bibinfo{author}{\bibfnamefont{A.}~\bibnamefont{{R{\'o}{\.z}a{\'n}ska}}},
  \bibnamefont{and}
  \bibinfo{author}{\bibfnamefont{A.}~\bibnamefont{{S{\k{a}}dowski}}},
  \bibinfo{journal}{\aap} \textbf{\bibinfo{volume}{521}}, \bibinfo{eid}{A15}
  (\bibinfo{year}{2010}), \eprint{1003.3887}.

\bibitem[{\citenamefont{{Lasota} and
  {Abramowicz}}(2024)}]{Lasota+Abramowicz2024}
\bibinfo{author}{\bibfnamefont{J.-P.} \bibnamefont{{Lasota}}} \bibnamefont{and}
  \bibinfo{author}{\bibfnamefont{M.}~\bibnamefont{{Abramowicz}}},
  \bibinfo{journal}{arXiv e-prints} \bibinfo{eid}{arXiv:2410.06200}
  (\bibinfo{year}{2024}), \eprint{2410.06200}.

\bibitem[{\citenamefont{{Gammie}}(1999)}]{Gammie1999}
\bibinfo{author}{\bibfnamefont{C.~F.} \bibnamefont{{Gammie}}},
  \bibinfo{journal}{\apjl} \textbf{\bibinfo{volume}{522}}, \bibinfo{pages}{L57}
  (\bibinfo{year}{1999}), \eprint{astro-ph/9906223}.

\bibitem[{\citenamefont{Misner and Sharp}(1964)}]{Misner+Sharp1964}
\bibinfo{author}{\bibfnamefont{C.~W.} \bibnamefont{Misner}} \bibnamefont{and}
  \bibinfo{author}{\bibfnamefont{D.~H.} \bibnamefont{Sharp}},
  \bibinfo{journal}{Phys. Rev.} \textbf{\bibinfo{volume}{136}},
  \bibinfo{pages}{B571} (\bibinfo{year}{1964}),
  \urlprefix\url{https://link.aps.org/doi/10.1103/PhysRev.136.B571}.

\end{thebibliography}

\end{document}